 \documentclass[final,3p,10pt,times]{elsarticle}

\usepackage{amsmath, amssymb, amsthm}

\usepackage{geometry}
\usepackage{graphicx} % 插入图片
\usepackage{caption}  % 自定义 caption
\usepackage{subcaption} % 子图环境
\usepackage{algorithm}
\usepackage{algpseudocode}
\usepackage{threeparttable}
\usepackage{tikz}               % 绘图
\usepackage{xcolor}             % 颜色
\usepackage{setspace}           % 行距设置
\usepackage{epstopdf}           % 将 eps 转换为 pdf
\usepackage{enumitem}           % 自定义列表
\usepackage{booktabs}           % 绘制专业表格
\usepackage{multirow}           % 表格跨行
\usepackage{makecell}           % 表格内换行
\usepackage{float}              % 浮动体位置控制
\usepackage{csquotes}           % 引用
\usepackage{appendix}           % 附录
\usepackage{multicol}           % 多栏布局
\usepackage{hyperref}           % 超链接 (通常放在最后加载)
\usepackage{tikz}
\usetikzlibrary{shapes, arrows.meta, positioning, shadows}

\makeatletter

\newcommand{\Rmnum}[1]{\expandafter\@slowromancap\romannumeral #1@}
\makeatother

\newcommand{\RomanNumeral}[1]{\uppercase\expandafter{\romannumeral #1\relax}}

\journal{Signal Processing}

\begin{document}

\begin{frontmatter}



\title{A Unified Framework for 2D Nonseparable Fractional Fourier Transform: From Geometric Completeness to Applications}

 \author[1]{Daxiang Li}
 \author[1,2,3,4]{Zhichao Zhang\corref{cor1}}
 \ead{zzc910731@163.com}
 \author[1]{Wei Yao}

\cortext[cor1]{Corresponding author; Tel: +86-13376073017.}

\address[1]{School of Mathematics and Statistics, Nanjing University of Information Science and Technology, Nanjing 210044, China}
\address[2]{Hubei Key Laboratory of Applied Mathematics, Hubei University, Wuhan 430062, China}
\address[3]{Key Laboratory of System Control and Information Processing, Ministry of Education, Shanghai Jiao Tong University, Shanghai 200240, China}
\address[4]{Key Laboratory of Computational Science and Application of Hainan Province, Hainan Normal University, Haikou 571158, China}

\tnotetext[mytitlenote]{This work was supported in part by the Open Foundation of Hubei Key Laboratory of Applied Mathematics (Hubei University) under Grant HBAM202404; in part by the Foundation of Key Laboratory of System Control and Information Processing, Ministry of Education under Grant Scip20240121; in part by the Foundation of Key Laboratory of Computational Science and Application of Hainan Province under Grant JSKX202401; and in part by the Foundation of Key Laboratory of Numerical Simulation of Sichuan Provincial Universities under Grant KLNS--2024SZFZ005.}

\begin{abstract}
	The one-dimensional (1D) fractional Fourier transform (FRFT) generalizes the Fourier transform, offering significant advantages in the time-frequency analysis of non-stationary signals. While various 2D extensions exist, such as the 2D separable FRFT (SFRFT), gyrator transform (GT), coupled FRFT (CFRFT), and earlier nonseparable definitions, they suffer from fragmented theoretical frameworks and a fundamental lack of geometric consistency with the 2D Wigner distribution (WD). Addressing these limitations, we propose a unified 2D nonseparable FRFT (NSFRFT) framework. Theoretically derived from the intersection of the symplectic and special orthogonal groups (isomorphic to the unitary group $\mathrm{U}(2)$), this transform inherently possesses four degrees of freedom and mathematically incorporates the 2D SFRFT, GT, and CFRFT as special cases. Unlike prior algebraic generalizations, it strictly preserves the rigid 4D rotational geometry of the 2D WD, ensuring geometric consistency and numerical stability. We derive its essential properties and develop efficient discrete algorithms with a computational complexity of $O(N^{2}\log N)$. Numerical simulations validate the superiority of the 2D NSFRFT in analyzing coupled chirp signals and demonstrate its robustness in filtering and image encryption and decryption applications.
\end{abstract}



\end{frontmatter}
\section{Introduction}
%%第一段
 
Fractional signal processing has rapidly evolved into a prominent research hotspot in modern signal analysis, owing to its superior adaptability in characterizing non-stationary signals compared with conventional integer-order signal processing frameworks. Within this burgeoning research paradigm, the fractional Fourier transform (FRFT) \cite{1,2,3,4,5} and the fractional wavelet transform stand out as two fundamental analytical tools, and their applications have permeated a broad spectrum of academic and engineering fields. These span from classic domains including optical system modeling and analysis \cite{6,7,8,9,10,11}, optimal filter design and implementation \cite{12,13,14,15,16}, analytical and numerical solutions of differential equations \cite{1,17,18,19}, phase retrieval for coherent imaging systems \cite{20,21,22,23}, and pattern recognition with high-dimensional feature extraction \cite{24,25,26,27,28}, to cutting-edge applications in biomedical engineering such as electrocardiogram signal denoising and feature mining, as well as the development of intelligent automated healthcare systems \cite{a1,a2,a3,a4,a5}. Between these two pivotal transforms, the FRFT is distinguished by its unique geometric attribute: an intrinsic rotational relationship with the Wigner distribution (WD). Specifically, the one-dimensional (1D) FRFT can be rigorously interpreted as a rigid rotation operation of the 1D WD in the  time-frequency plane \cite{29,30,31}. This distinctive property endows the FRFT with the capability to flexibly manipulate the time-frequency aggregation of signals, thereby rendering it highly effective for a series of critical signal processing tasks, such as noise suppression, high-efficiency signal compression, and blind separation of mixed multi-component signals.

To extend these benefits to higher-dimensional signals, the definition and property analysis of 2D transforms have become a focal point of research \cite{32,33,33.1,33.2,34,35,36}. From a rigorous mathematical perspective, the 2D nonseparable linear canonical transform (NSLCT) \cite{45} serves as the most generalized framework for processing 2D non-stationary signals. 
% --- 针对 Reviewer 5：插入 Pei (2011) 对比 ---
{The theoretical properties of the general 2D NSLCT, such as eigenfunctions and self-imaging phenomena, have been extensively studied by Pei \textit{et al.} \cite{45.5}.} Theoretically, the 2D NSLCT corresponds to generalized affine transformations in the 4D phase space. While this allows it to model complex signal couplings (e.g., shearing and scaling), this generality incurs a cost: it obscures the intuitive rotational geometry characteristic of the FRFT family. Therefore, a critical theoretical gap remains: determining the specific parameterization of the rigid rotation subgroup within the general 2D NSLCT framework that strictly preserves the WD's geometry without shearing. We identify the need for a transform that inherits the 2D NSLCT's capability to handle nonseparable coupling but is strictly constrained to represent a generalized rotation rather than a general affine deformation.

Within the FRFT family, several 2D variants have been developed to address this need, but they face inherent limitations. The 2D separable FRFT (SFRFT) \cite{32} extends the 1D FRFT by independently applying it along two orthogonal directions. While simple to implement, it is defined by the tensor product of two 1D FRFTs and thus fails to capture 2D signals with nonseparable (coupled) terms. To overcome the limitations of separability, Sahin \textit{et al.} \cite{33} pioneered the nonseparable 2D FRFT by introducing coordinate affine transformations. While this was a significant milestone, Sahin's definition is fundamentally based on coordinate affine transformations, which correspond to shearing and scaling operations in the phase space rather than rigid rotation. Consequently, it fails to maintain strict 4D rotational geometry with the 2D WD. As explicitly noted by the authors, this transform is "not a fractional operator in the strict sense" and lacks the index additivity property essential for rigorous fractional signal processing. This geometric inconsistency leads to unavoidable waveform distortions when processing coupled signals that require precise rotational focusing.

\begin{figure}[h!]
    \centering
    \subfloat[]{\includegraphics[width=0.28\textwidth]{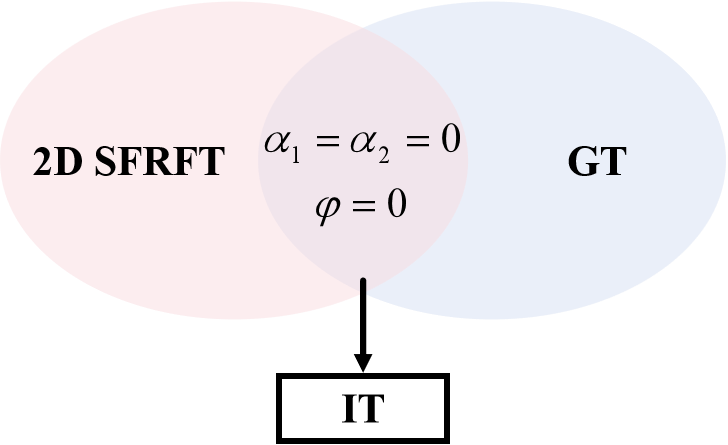}}\hspace{0.5em}
    \subfloat[]{\includegraphics[width=0.28\textwidth]{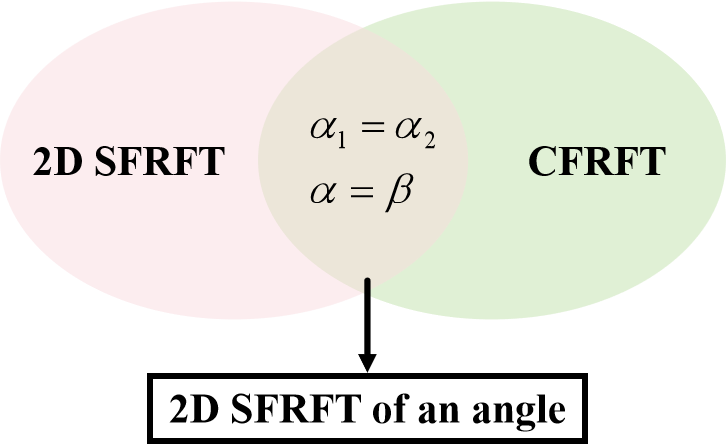}}\hspace{0.5em}
    \subfloat[]{\includegraphics[width=0.28\textwidth]{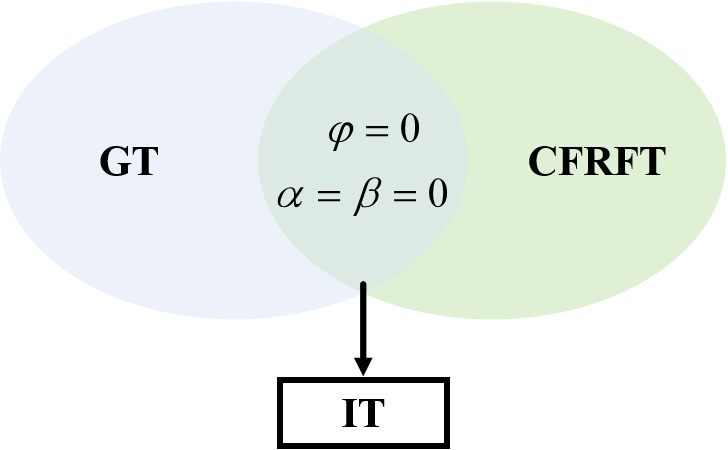}}
    \caption{The relationship between the 2D SFRFT,GT, and CFRFT. Here, $\alpha_1$, $\alpha_2$ are the parameters of the 2D SFRFT, $\varphi$ is the parameter of the GT, and $\alpha$, $\beta$ are the parameters of the CFRFT. The abbreviation \enquote{IT} denotes the 2D identity transform.}
    \label{fig:1}
\end{figure}

Subsequently, more advanced transforms were proposed to restore geometric consistency. In 2007, Rodrigo \textit{et al.} introduced the gyrator transform (GT) \cite{33.1,33.2}, capable of inducing rotations in the twisted space/spatial-frequency planes. Later, in 2017, Zayed introduced the coupled FRFT (CFRFT) \cite{34} by leveraging a new family of Hermite functions. While both the GT and CFRFT exhibit a 4D rotational relationship with the 2D WD, resolving the consistency issue, they essentially represent different subspaces of the fractional domain. Neither includes the 2D SFRFT as a special case; instead, their relationship is intersecting rather than inclusive (see Fig. \ref{fig:1}), resulting in a fragmented theoretical landscape that limits their utility as universal tools. This lack of uniformity limits their applicability as a universal tool for diverse 2D signal processing scenarios.

To bridge the gap between the general coupling capability of the 2D NSLCT and the geometric rotation property of the 2D FRFT, we propose a more general 2D nonseparable FRFT (NSFRFT).
% --- 新增关于 Obstacles 的描述 ---
The primary challenges in constructing this transform lie in (1) theoretically unifying the fragmented definitions of existing 2D FRFTs, (2) strictly preserving the 4D rotational relationship with the 2D WD while allowing for nonseparable coupling, and (3) overcoming the high computational complexity ($O(N^4)$) of the direct implementation.
% ---------------------------------
By addressing these obstacles, the proposed 2D NSFRFT is constructed as a distinctive subset of the 2D NSLCT family. Unlike the generic 2D NSLCT which permits arbitrary phase space deformations, our transform is mathematically constrained to strictly preserve the 4D rotational structure of the 2D WD, thereby ensuring both geometric consistency and the ability to handle nonseparable terms.

The primary contributions and novelties of this study are summarized as follows:
\begin{itemize}
    \item \textbf{Theoretical Unification and Completeness:} We introduce the 2D NSFRFT, a unified framework with four degrees of freedom. It mathematically incorporates the existing 2D SFRFT, GT, and CFRFT as special cases, resolving the lack of theoretical uniformity. By deriving the transform from the intersection of the symplectic group and the special orthogonal group, we ensure a geometrically complete description of rigid rotations in 4D phase space.
    
    \item \textbf{Geometric Consistency:} We prove that the proposed transform maintains a generalized 4D rotational relationship with the 2D WD. This ensures the geometric consistency required for rigorous time-frequency analysis, distinguishing it from earlier nonseparable attempts.

    \item \textbf{Capability for Coupled Signals:} The 2D NSFRFT demonstrates a distinct ability to effectively process 2D chirp signals with nonseparable terms. We demonstrate that while conventional transforms fail to focus such coupled signals, the proposed structure successfully handles the coupling effects.

    \item \textbf{Efficient Algorithms \& Applications:} We develop discrete algorithms with complexity $O(N^2 \log N)$ and validate the transform's superiority in  filtering and image encryption applications through rigorous performance metrics (MSE, PSNR, SSIM).
\end{itemize}

The remaining sections of this paper are organized as follows. Section \ref{Section 2} presents the relevant preliminary concepts, including the general 2D NSLCT framework and existing 2D FRFTs. Section \ref{Section 3} defines the 2D NSFRFT, explores its properties, and provides its geometric interpretation. Section \ref{Section 4} details the discrete algorithms derived from matrix decomposition. Section \ref{Section 5} covers practical applications, including filtering and image encryption and decryption. Finally, Section \ref{Section 6} summarizes the findings and outlines potential directions for future research.

\section{Preliminaries}
\label{Section 2}

In this section, we establish the mathematical foundation for the proposed transform. We begin with the general framework of the 2D NSLCT, followed by existing 2D FRFTs. We then introduce the WD as a generalized time-frequency analysis tool and discuss the mathematical decomposition of 4D rotations.

\subsection{The General Framework: 2D NSLCT}
The 2D NSLCT serves as a generalized framework for integral transforms and optical system analysis. It represents the most general form of linear integral transforms, corresponding to arbitrary affine transformations in the 4D phase space.

\textit{Definition 1:} The 2D NSLCT of a signal $f(x,y)$ is defined by \cite{45}
\begin{equation}
\mathcal{L}^\mathbf{M} [f](u,v) = \frac{1}{2\pi\sqrt{-\det(\mathbf{B})}} \int_{\mathbb{R}^2} f(x,y) K_{(\mathbf{A},\mathbf{B},\mathbf{C},\mathbf{D})}(x, y, u, v)  \mathrm{d}x \, \mathrm{d}y,
\label{eq:2D NSLCT}
\end{equation}
where $K_{(\mathbf{A},\mathbf{B},\mathbf{C},\mathbf{D})}(x, y, u, v)$ is the kernel function of the 2D NSLCT, given by
%\begin{equation}
%K_{(\mathbf{A},\mathbf{B},\mathbf{C},\mathbf{D})}(x,y,u,v) = \frac{1}{2\pi\sqrt{-\det(\mathbf{B})}} \exp\left[{\mathrm{j}\frac{k_1 u^2 + k_2 uv + k_3 v^2}{2\det(\mathbf{B})}}\right] \exp\left[{\mathrm{j}\frac{(-b_{22}u + b_{12}v)x + (b_{21}u - b_{11}v)y}{\det(\mathbf{B})}}\right] \\ \times \exp\left[{\mathrm{j}\frac{p_1 x^2 + p_2 xy + p_3 y^2}{2\det(\mathbf{B})}}\right],
%\end{equation}
\begin{equation}
	\begin{split}
		K_{(\mathbf{A},\mathbf{B},\mathbf{C},\mathbf{D})}(x,y,u,v) &= \frac{1}{2\pi\sqrt{-\det(\mathbf{B})}} 
		\exp\left[{\mathrm{j}\frac{k_1 u^2 + k_2 uv + k_3 v^2}{2\det(\mathbf{B})}}\right] \exp\left[{\mathrm{j}\frac{(-b_{22}u + b_{12}v)x + (b_{21}u - b_{11}v)y}{\det(\mathbf{B})}}\right] \\
		&\quad \times \exp\left[{\mathrm{j}\frac{p_1 x^2 + p_2 xy + p_3 y^2}{2\det(\mathbf{B})}}\right],
	\end{split}
\end{equation}
and 
\begin{equation}
k_1 = d_{11}b_{22} - d_{12}b_{21},\ k_2 = 2\bigl(-d_{11}b_{12} + d_{12}b_{11}\bigr), k_3 = -d_{21}b_{12} + d_{22}b_{11},
\end{equation}
\begin{equation}
    \ p_1 = a_{11}b_{22} - a_{21}b_{12}, p_2 = 2\bigl(a_{12}b_{22} - a_{22}b_{12}\bigr),\ p_3 = -a_{12}b_{21} + a_{22}b_{11}.
\end{equation}
The symplectic matrix $\mathbf{M}$ of the transform in Eq. (\ref{eq:2D NSLCT}) is defined by
\begin{equation}
\mathbf{M} = \begin{bmatrix} \mathbf{A} & \mathbf{B} \\ \mathbf{C} & \mathbf{D} \end{bmatrix} = \begin{bmatrix}
a_{11} & a_{12} & b_{11} & b_{12} \\
a_{21} & a_{22} & b_{21} & b_{22} \\
c_{11} & c_{12} & d_{11} & d_{12} \\
c_{21} & c_{22} & d_{21} & d_{22}
\end{bmatrix},
\label{eq:n5}
\end{equation}
where the sub-matrices $\mathbf{A}, \mathbf{B}, \mathbf{C}, \mathbf{D} \in \mathbb{R}^{2 \times 2}$, and $\det(\mathbf{B}) \neq 0$, i.e. $\mathbf{B}$ is invertible. In addition, sub-matrices $\mathbf{A}, \mathbf{B}, \mathbf{C}, \mathbf{D}$ satisfy the following three constraint conditions:
\begin{equation}
    \mathbf{A}\mathbf{B}^T=\mathbf{B}\mathbf{A}^T, \mathbf{C}\mathbf{D}^T=\mathbf{D}\mathbf{C}^T,
    \mathbf{A}\mathbf{D}^T-\mathbf{B}\mathbf{C}^T=\mathbf{I},
\label{eq:n6}
\end{equation}
 or equivalently
\begin{equation}
    \mathbf{A}^T\mathbf{C}=\mathbf{C}^T\mathbf{A}, \mathbf{B}^T\mathbf{D}=\mathbf{D}^T\mathbf{B},
    \mathbf{A}^T\mathbf{D}-\mathbf{C}^T\mathbf{B}=\mathbf{I},
\end{equation}
where $\mathbf{I}$ is a 2 × 2 identity matrix. Linear Eq. (\ref{eq:n6}) can be written as the following six constraints:
\begin{equation}
\begin{split}
&a_{11}b_{21} + a_{12}b_{22} = a_{21}b_{11} + a_{22}b_{12}, \\
&c_{11}d_{21} + c_{12}d_{22} = c_{21}d_{11} + c_{22}d_{12}, \\
&a_{11}d_{11} + a_{12}d_{12} - \bigl(b_{11}c_{11} + b_{12}c_{12}\bigr) = 1, \\
&a_{21}d_{21} + a_{22}d_{22} - \bigl(b_{21}c_{21} + b_{22}c_{22}\bigr) = 1, \\
&a_{11}d_{21} + a_{12}d_{22} = b_{11}c_{21} + b_{12}c_{22}, \\
&a_{21}d_{11} + a_{22}d_{12} = b_{21}c_{11} + b_{22}c_{12}.
\end{split}
\end{equation}

It can be seen from Eq. (\ref{eq:n5}) that there are 16 parameters in matrix $\mathbf{M}$ for the 2D NSLCT. With the above six constraint conditions, the number of independent free parameters is 10.

\subsection{Related 2D FRFTs}
As special cases within the 2D NSLCT framework, several 2D transforms have been defined to realize time-frequency rotation.

Before discussing the 2D case, we briefly recall the 1D FRFT, which serves as the fundamental building block.
\textit{Definition 2:} The 1D FRFT of a signal $f(x)$ is defined by
\begin{equation}
\mathcal{F}_\alpha [f](u) = {F}_\alpha (u)=\int_{\mathbb{R}} f(x) K_\alpha(u, x) \, \mathrm{d}x,
\label{eq:1}
\end{equation}
where
\begin{equation}
K_\alpha (u, x) = 
\begin{cases} 
A_\alpha \exp \left[ \mathrm{j} \left( \frac{\cot \alpha}{2} (x^2 + u^2) - \csc \alpha \, u x \right) \right], & \alpha \neq p\pi \\
\delta(u - x), & \alpha = 2p\pi \\
\delta(u + x), & \alpha = (2p \pm 1)\pi 
\end{cases}
\label{eq:2}
\end{equation}
is the kernel of the 1D FRFT with $A_\alpha = \sqrt{\frac{1 - \mathrm{j} \cot \alpha}{2\pi}}$, $\alpha = \frac{p\pi}{2}$, where $p$ is an integer and $\delta$(·) denotes the Dirac delta
function.

\textit{Definition 3:} The 2D SFRFT  is defined by the tensor product of two 1D FRFTs, namely,
\begin{equation}
\mathcal{F}_{\alpha_1, \alpha_2}[f](u, v) = {F}_{\alpha_1, \alpha_2}(u, v)=\int_{\mathbb{R}^2}  f(x, y) K_{\alpha_1}(u, x)K_{\alpha_2}(v, y) \, \mathrm{d}x \, \mathrm{d}y.
\label{eq:3}
\end{equation}
This definition corresponds to independent rotations in the $(x, u)$ and $(y, v)$ planes but lacks coupling between the $x$ and $y$ dimensions.

\textit{Definition 4:} The GT of a signal $f(x,y)$ is defined by

\begin{equation}
\begin{aligned}
\mathcal{F}_{\varphi}[f](u, v) = F_{\varphi}(u, v) =  \frac{\left| \csc \varphi \right|}{2\pi}
\int_{\mathbb{R}^2}
\exp\big\{
    \frac{\mathrm{j}(uv+xy)}{\tan \varphi}
    - \frac{\mathrm{j}(uy+vx)}{\sin \varphi}
\big\}f(x,y) \, \mathrm{d}x\,\mathrm{d}y.
\end{aligned}
\label{eq:4}
\end{equation}

\textit{Definition 5:} The CFRFT of a signal $f(x,y)$ is defined by 
\begin{equation}
\begin{aligned}
\mathcal{F}_{\alpha, \beta}[f](u, v) 
& = F_{\alpha, \beta}(u, v) = \int_{\mathbb{R}^2} f(x, y) K_{\alpha, \beta}(x, y, u, v) \, \mathrm{d}x \, \mathrm{d}y,
\end{aligned}
\label{eq:5}
\end{equation}
where
\begin{equation}
\begin{aligned}
K_{\alpha, \beta}(x, y, u, v) &= d(\gamma) \exp \big\{a(\gamma)(x^2 + y^2 + u^2 + v^2) - b(\gamma, \delta)(ux + vy) - c(\gamma, \delta)(vx - uy) \big\}
\end{aligned}
\label{eq:6}
\end{equation}
with $a(\gamma) = \dfrac{\mathrm{j} \cot \gamma}{2}$, 
$b(\gamma, \delta) = \dfrac{\mathrm{j} \cos \delta}{\sin \gamma}$, 
$c(\gamma, \delta) = \dfrac{\mathrm{j} \sin \delta}{\sin \gamma}$, 
$d(\gamma) = - \dfrac{\mathrm{j} \exp(\mathrm{j}\gamma)}{2 \pi \sin \gamma}$,
and where $\gamma = \frac{\alpha+\beta}{2}$ and $\delta = \frac{\alpha-\beta}{2}$.

While the GT and CFRFT introduce coupling, they represent distinct rotational subspaces within the 2D NSLCT framework and do not include the 2D SFRFT as a special case.

\subsection{Phase Space Analysis and Geometric Interpretation}
To explicitly reveal the geometric meaning of the transforms defined in the preceding subsections, we introduce the WD. The WD provides a powerful phase-space tool \cite{37,38,39,40,41,42} where the 2D NSLCTs correspond to specific affine operations (e.g., rotation, shearing). This geometric perspective enables us to characterize the specific affine operations of different transforms, distinguishing the rigid rotation of the 2D FRFT family from the shearing or scaling of general 2D NSLCTs.

%The WD is one of the most important and widely used time-frequency analysis tools. It is closely related to the FRFT and has numerous applications in quantum mechanics, signal analysis and optics .

\textit{Definition 6:} The 1D WD of a signal $f(x)$ is defined by
\begin{equation}
W_{f}(x, u) = \int_{\mathbb{R}} f\left(x + \frac{\tau}{2}\right) f^*\left(x - \frac{\tau}{2}\right) \exp({-\mathrm{j}u\tau}) \, \mathrm{d}\tau,
\label{eq:7}
\end{equation}
where * denotes the complex conjugate operator.

\textit{Definition 7:} The 2D WD of a signal $f(x,y)$ is generally defined by
\begin{equation}
W_f(x, y, u, v) = \int_{\mathbb{R}^2} f\left(x+\frac{\tau}{2}, y+\frac{\eta}{2}\right) f^*\left(x-\frac{\tau}{2}, y-\frac{\eta}{2}\right) \exp\left[{-\mathrm{j}(u\tau + v\eta)}\right] \, \mathrm{d}\tau \, \mathrm{d}\eta.
\label{eq:WD2D}
\end{equation}

The fundamental connection between the transforms defined in the preceding subsections and the WD is governed by the symplectic covariance property. Specifically, if a signal $f$ undergoes a general 2D NSLCT parameterized by a symplectic matrix $\mathbf{M}$, its 2D WD follows a coordinate affine transformation $W_{\mathcal{L}^\mathbf{M} [f]}(\mathbf{w}) = W_f(\mathbf{M}^{-1} \mathbf{w})$, where $\mathbf{w}=[x, y, u, v]^T$ represents the 4D phase-space coordinates.
Crucially, while a general 2D NSLCT, such as the affine-based nonseparable transform \cite{33}, typically induces shearing or scaling (leading to geometric distortions), the FRFT family is distinct in that it strictly corresponds to rigid rotation, which preserves the isometric structure of the signal distribution.

To explicitly illustrate this rotational characteristic, we first examine the 1D case. The 1D WD $W_{F_{\alpha}}$ of the 1D FRFT $F_{\alpha}$ of a function $f$ corresponds to a 2D rotation of the 1D WD $W_f$ of the function $f$, and the corresponding 2D rotation matrix $\mathbf{R_{2D}}$ is

\begin{equation}
	\mathbf{R_{2D}} =
	\left[
	\begin{matrix}
		\cos \alpha & -\sin \alpha \\
		\sin \alpha &  \cos \alpha
	\end{matrix}
	\right].
	\label{eq:8}
\end{equation}

Similarly, the 2D SFRFT, GT, and CFRFT all exhibit rotational transformation relationships with the 2D WD, the corresponding 4D rotation matrices 
$\mathbf{R_{4D}^\mathrm{S}}$, $\mathbf{R_{4D}^\mathrm{G}}$ and $\mathbf{R_{4D}^\mathrm{C}}$ are
\begin{equation}
	\mathbf{R_{4D}^\mathrm{S}} = 
	\left[
	\begin{matrix}
		\cos \alpha_1 & 0 & -\sin \alpha_1 & 0 \\
		0 & \cos \alpha_2 & 0 & -\sin \alpha_2 \\
		\sin \alpha_1 & 0 & \cos \alpha_1 & 0 \\
		0 & \sin \alpha_2 & 0 & \cos \alpha_2
	\end{matrix}
	\right],
	\label{eq:9}
\end{equation}

\begin{equation}
	\mathbf{R_{4D}^\mathrm{G}} = 
	\left[
	\begin{matrix}
		\cos \varphi & 0 & 0 & -\sin \varphi \\
		0 & \cos \varphi & -\sin \varphi & 0 \\
		0 & \sin \varphi & \cos \varphi & 0 \\
		\sin \varphi & 0 & 0 & \cos \varphi
	\end{matrix}
	\right],
	\label{eq:10}
\end{equation}
and
\begin{equation}
	\mathbf{R_{4D}^\mathrm{C}} = 
	\begin{bmatrix}
		\cos\gamma\cos\delta & \cos\gamma\sin\delta & -\sin\gamma\cos\delta & -\sin\gamma\sin\delta \\
		-\cos\gamma\sin\delta & \cos\gamma\cos\delta & \sin\gamma\sin\delta & -\sin\gamma\cos\delta \\
		\sin\gamma\cos\delta & \sin\gamma\sin\delta & \cos\gamma\cos\delta & \cos\gamma\sin\delta \\
		-\sin\gamma\sin\delta & \sin\gamma\cos\delta & -\cos\gamma\sin\delta & \cos\gamma\cos\delta
	\end{bmatrix},
	\label{eq:11}
\end{equation}
respectively.

The aforementioned rotational relationships demonstrate that the 2D SFRFT, GT, and CFRFT can all be regarded as specific rigid rotations (subgroups of the symplectic group) in the 4D joint time-frequency phase space. This geometric rigidity distinguishes them from general affine transforms and provides a solid theoretical foundation for the proposed unified framework.

\subsection{4D Rotation Matrix Decomposition}
To construct a unified transform that encompasses all aforementioned rotational properties, we utilize the geometric decomposition of 4D rotations. It has been shown in \cite{34,43} that any 4D rotation matrix can be decomposed using unit quaternions.

\textit{Proposition 1:} Any 4D rotation matrix $\mathbf{Z} \in \mathbb{R}^{4 \times 4}$ can be expressed as the product of two matrices $\mathbf{M}_L$ and $\mathbf{M}_R$, corresponding to left- and right-multiplication by unit quaternions. There exist real scalars $a, b, c, d$ and $p, q, r, s$ satisfying $a^2 + b^2 + c^2 + d^2 = 1$ and $p^2 + q^2 + r^2 + s^2 = 1$, such that
\begin{equation}
\mathbf{M}_L = 
\begin{bmatrix}
a & -b & -c & -d \\
b & a & -d & c \\
c & d & a & -b \\
d & -c & b & a
\end{bmatrix}, \quad
\mathbf{M}_R = 
\begin{bmatrix}
p & -q & -r & -s \\
q & p & s & -r \\
r & -s & p & q \\
s & r & -q & p
\end{bmatrix},
\label{eq:12}
\end{equation}
and their product $\mathbf{Z} = \mathbf{M}_L \mathbf{M}_R$ is a valid 4D rotation matrix.

Existing transforms (2D SFRFT, GT, and CFRFT) utilize only limited subsets of these parameters. This proposition provides the theoretical basis for the proposed 2D NSFRFT, which utilizes the full parameter space of $\mathbf{M}_L$ (with $\mathbf{M}_R$ determined by specific angles) to achieve a general unified definition.

\section{Definition, Properties, Geometric Interpretation and Distinctive Advantage of the 2D NSFRFT}
\label{Section 3}
In this section, we introduce the definition of the 2D NSFRFT, discuss its properties, give its geometric interpretation and  highlight its distinctive advantage. 
\subsection{Definition of the 2D NSFRFT}

In this subsection, we introduce the definition of the 2D NSFRFT. Distinct from definitions based directly on the integral kernel, we construct the transform through its symplectic structure in the phase space to highlight its geometric properties.

\textit{Definition 8:} The 2D NSFRFT with parameters ${P}=(a,b,c,d,\theta)$ is defined as the specific case of the 2D NSLCT. {Derived from the geometric decomposition of 4D rotations (see Section 2.4),} it corresponds to the orthogonal symplectic matrix $\mathbf{X} \in \mathrm{Sp}(4,\mathbb{R}) \cap \mathrm{SO}(4)$,
\begin{equation}
	\mathbf{X} = 
	\begin{bmatrix}
		\mathbf{A} &\mathbf{B}\\\mathbf{-B} &\mathbf{A}
	\end{bmatrix},
	\label{eq:X}
\end{equation}
where the sub-matrices $\mathbf{A}$ and $\mathbf{B}$ are constructed using ($a,b,c,d$) and the angle $\theta$,
% ... (接后续公式)
\begin{equation}
\mathbf{A} = 
\begin{bmatrix}
a\cos\theta - c\sin\theta & b\cos\theta - d\sin\theta  \\
-b\cos\theta - d\sin\theta & a\cos\theta + c\sin\theta 
\end{bmatrix}, 
\mathbf{B} = 
\begin{bmatrix}
a\sin\theta + c\cos\theta & b\sin\theta + d\cos\theta  \\
-b\sin\theta + d\cos\theta & a\sin\theta - c\cos\theta
\end{bmatrix},
\label{eq:22}
\end{equation}
subject to the constraints $a,b,c,d \in \mathbb{R}$ with $a^2+b^2+c^2+d^2=1$, and $\theta \in [0, 2\pi]$.

By substituting the sub-matrices into the general 2D NSLCT framework (Eq. (\ref{eq:2D NSLCT})), we derive the explicit integral expression. The 2D NSFRFT of a signal $f(x,y)$ is thus given by

% \textit{Definition:} The 2D NSFRFT of a signal $f(x,y)$ is defined by
\begin{equation}
\begin{aligned}
\mathcal{F}_{P}[f](u, v) &= F_{P}(u,v) = \int_{\mathbb{R}^2} f(x,y) 
K_{P}(x,y,u,v) \mathrm{d}x \, \mathrm{d}y,
\end{aligned}
\label{eq:13}
\end{equation}
where
\begin{equation}
\begin{aligned}
&K_{P}(x,y,u,v) = \frac{1}{2\pi\sqrt{-\mathbf{T}}} \exp\left[{\frac{\mathrm{j}(p_1x^2 + p_2xy + p_3y^2)}{2\mathbf{T}}} \right]\exp \left[ {\frac{\mathrm{j}(m_1ux + m_2vx + m_3uy + m_4vy)}{\mathbf{T}}} \right]\exp \left[ {\frac{\mathrm{j}(k_1u^2 + k_2uv + k_3v^2)}{2\mathbf{T}}}\right],  \\
\end{aligned}
\label{eq:14}
\end{equation}
and where
\begin{equation}
\mathbf{T} = a^2 \sin^2 \theta + b^2 \sin^2 \theta - c^2 \cos^2 \theta - d^2 \cos^2 \theta \neq 0,
\label{eq:15}
\end{equation}
\begin{equation}
p_1 = \sin\theta \cos\theta - ac + bd,  p_2 = -2(bc + ad),  p_3 = \sin\theta \cos\theta + ac - bd,
\label{eq:16}
\end{equation}
\begin{equation}
m_1 = -a \sin\theta + c \cos\theta, 
m_2 = b \sin\theta + d \cos\theta,
m_3 = -b \sin\theta + d \cos\theta, 
m_4 = -a \sin\theta - c \cos\theta,
\label{eq:17}
\end{equation}
\begin{equation}
k_1 = \sin\theta \cos\theta - ac - bd, 
k_2 = 2(bc-ad), 
k_3 = \sin\theta \cos\theta + ac + bd.
\label{eq:18}
\end{equation}

To maintain the readability of the main text, the detailed construction of this orthogonal symplectic matrix and the algebraic derivation of the integral definition are provided in \ref{Appendix C}.

{\textit{Remark 1 (Geometric Completeness and Group Structure):}} 
The construction of the orthogonal symplectic matrix $\mathbf{X}$ in \textit{Definition 8} is not arbitrary. It is rigorously derived from the intersection of the symplectic group $\mathrm{Sp}(4,\mathbb{R})$ (governing all 2D NSLCTs) and the special orthogonal group $\mathrm{SO}(4)$ (governing rigid rotations). This intersection is isomorphic to the unitary group $\mathrm{U}(2)$.
Therefore, unlike general 2D NSLCTs which allow shearing and scaling, the proposed 2D NSFRFT represents the complete family of rigid rotations in the 4D phase space. The parameters $a,b,c,d$ combined with the angle $\theta$, under the constraint $a^2 + b^2+c^2+d^2=1$, provide exactly 4 independent degrees of freedom (3 from the $S^3$ sphere + 1 angle $\theta$). These parameters function collectively to determine a generalized mixed rotation, enabling the transform to simultaneously model signal coupling (via cross-terms) and geometric reorientation in the phase space.

\textit{Remark 2 (Role of the Kernel Function):}
The kernel function $K_P(x,y,u,v)$ in Eq. (\ref{eq:14}) fundamentally governs the properties of the transform by acting as a unitary operator that projects the input signal onto a complete basis of nonseparable 2D chirp functions. Specifically, the cross-terms within the kernel exponent (e.g., $xy$, $uv$, $xu$) encode the signal coupling mechanism, determining precisely how the signal energy is spatially coupled and rotated in the phase space. This nonseparable structure enables the 2D NSFRFT to effectively concentrate coupled chirp signals that conventional separable kernels, which lack these critical cross-terms, fail to focus.

In particular, we define that when the parameters $P=(a, b, c, d, \theta)=(1,0,0,0,0)$, the corresponding  \(\mathbf{T}\) becomes zero, and the 2D NSFRFT is specifically defined as the IT, namely,
\begin{equation}
\mathcal{F}_{(1,0,0,0,0)}[f](u,v) = F_{(1,0,0,0,0)}(u,v) = f(x,y).
\label{eq:20}
\end{equation}

Fig. \ref{fig:2} illustrates the hierarchical relationship between the 2D NSFRFT and its degenerate cases. Specifically, apart from a constant phase factor, the 2D NSFRFT simplifies to the 2D SFRFT when $ b = d = 0 $, reduces to the GT when $ b = c = \theta = 0 $, and reduces to the CFRFT when $ c = d = 0 $.

%\begin{figure}[h!]
%    \centering
%    \includegraphics[width=0.6\textwidth]{3.1.png}
%    \caption{Relations between the 2D NSFRFT and its special cases.} 
%    \label{fig:2} 
%\end{figure}
\begin{figure}[htbp]
	\centering
	\begin{tikzpicture}[
		% 全局样式设置
		node distance=1.8cm and 2.5cm, % 垂直间距和水平间距
		% 字体样式：加粗
		font=\bfseries, 
		% 箭头样式
		>={Latex[width=2mm,length=2mm]},
		line/.style={draw=gray!80, line width=1pt, ->},
		label_text/.style={font=\small\normalfont, text=black, fill=white, inner sep=1pt},
		% 蓝色方块样式 (2D NSFRFT)
		blue_box/.style={
			rectangle, 
			draw=none, 
			fill=cyan!30!blue!60!, % 调节这个颜色可以改变蓝色的深浅
			text=white, 
			minimum width=3.5cm, 
			minimum height=1cm, 
			rounded corners=5pt,
			drop shadow
		},
		% 绿色方块样式 (GT, IT, SFRFT, CFRFT)
		green_box/.style={
			rectangle, 
			draw=none, 
			fill=green!40!black!60!, % 调节这个颜色可以改变绿色的深浅 (类似海藻绿)
			text=white, 
			minimum width=3cm, 
			minimum height=1cm, 
			rounded corners=5pt,
			drop shadow
		},
		% 灰色方块样式 (2D SFRFT of an angle)
		gray_box/.style={
			rectangle, 
			draw=none, 
			fill=gray!30, 
			text=black, 
			minimum width=3cm, 
			minimum height=1cm, 
			rounded corners=5pt,
			align=center, % 允许换行
			drop shadow
		}
		]
		
		% --- 1. 放置节点 (Nodes) ---
		
		% 顶层：2D NSFRFT (蓝色)
		\node [blue_box] (nsfrft) {2D NSFRFT};
		
		% 中层中间：GT (绿色)
		\node [green_box, below=2cm of nsfrft] (gt) {GT};
		
		% 左侧：2D SFRFT (绿色)
		\node [green_box, left=3.5cm of gt] (sfrft) {2D SFRFT};
		
		% 右侧：CFRFT (绿色)
		\node [green_box, right=3.5cm of gt] (cfrft) {CFRFT};
		
		% 下层中间：IT (绿色 - 跟你要求的一样，背景和GT一样)
		\node [gray_box, below=1.5cm of gt] (it) {IT};
		
		% 最底层：2D SFRFT of an angle (灰色)
		\node [gray_box, below=1.0cm of it] (angle) {2D SFRFT\\of an angle};

		% --- 2. 绘制连线 (Paths) ---
		
		% 线：NSFRFT -> GT
		\draw [line] (nsfrft) -- node[right, label_text] {$b=c=\theta=0$} (gt);
		
		% 线：NSFRFT -> 2D SFRFT (折线)
		\draw [line] (nsfrft.south) -- ++(0,-0.5) -| node[pos=0.25, above, label_text] {$b=d=0$} (sfrft.north);
		
		% 线：NSFRFT -> CFRFT (折线)
		\draw [line] (nsfrft.south) -- ++(0,-0.5) -| node[pos=0.25, above, label_text] {$c=d=0$} (cfrft.north);
		
		% 线：GT -> IT
		\draw [line] (gt) -- node[right, label_text] {$a=1, d=0$} (it);
		
		% 线：2D SFRFT -> Angle (斜线)
		\draw [line] (sfrft.south east) -- node[below left, label_text, pos=0.6] {$a=1, c=0$} (angle.west);
		
		% 线：CFRFT -> Angle (斜线)
		\draw [line] (cfrft.south west) -- node[below right, label_text, pos=0.6] {$a=1, b=0$} (angle.east);
		
		% 线：Angle -> IT (向上指)
		\draw [line] (angle) -- node[right, label_text] {$\theta=0$} (it);
		
	\end{tikzpicture}
	\caption{Hierarchical relationships between the 2D NSFRFT and its special cases.}
	\label{fig:2}
\end{figure}

\subsection{Properties of the 2D NSFRFT}

In this subsection, we summarize the key properties of the 2D NSFRFT including orthogonality of kernel functions, inverse formula,  energy conservation, conjugation, space reversal, differentiation, integration, time-frequency shifting and the relationship with the 2D WD. The detailed mathematical derivations are provided in \ref{Appendix B}.

\textbf{Property 1 (Orthogonality of Kernel Functions).}
The 2D NSFRFT is a unitary transformation on the Hilbert space $L^2(\mathbb{R}^2)$. This implies that its kernel functions satisfy the generalized orthogonality condition
\begin{equation}
	\int_{\mathbb{R}^2} K_{P}(x,y,u,v) K_{P}^*(x,y,\hat{u},\hat{v}) \, \mathrm{d}x \, \mathrm{d}y = \delta(u-\hat{u}, v-\hat{v}).
	\label{eq:orthogonality}
\end{equation}

\textit{Remark 1 (Mathematical Completeness):}
This property asserts that the kernel set $\{K_P\}$ forms a complete orthonormal basis in the generalized sense. The Dirac delta function implies that distinct spectral components are uncorrelated. This completeness is the fundamental prerequisite for the existence of a unique inverse transform, ensuring that the projection from the spatial domain to the 2D NSFRFT domain involves no information loss.

\textbf{Property 2 (Inverse Formula).} 
The original signal $f(x,y)$ can be perfectly reconstructed from its 2D NSFRFT spectrum $F_{P}(u,v)$ via the following inverse transformation
\begin{equation}
	f(x,y) = \int_{\mathbb{R}^2} F_{P}(u,v) K_{P}^*(x,y,u,v) \, \mathrm{d}u \, \mathrm{d}v.
	\label{eq:inverse_prop}
\end{equation}

\textit{Remark 2 (Hermitian Adjoint):} 
From an operator theory perspective, the inverse operator is simply the Hermitian adjoint of the forward operator, which is analytically implemented by the complex conjugate kernel $K_{P}^*$. Physically, this signifies that signal reconstruction is equivalent to propagating the wavefield through the corresponding optical system in reverse, effectively undoing the phase-space rotation.

\textbf{Property 3 (Energy Conservation Property).}
The 2D NSFRFT preserves the total energy of the signal, satisfying the Parseval-type relation
\begin{equation}
	\int_{\mathbb{R}^2} \left|f(x,y)\right|^2 \, \mathrm{d}x \, \mathrm{d}y = \int_{\mathbb{R}^2} \left|F_{P}(u,v)\right|^2 \, \mathrm{d}u \, \mathrm{d}v.
	\label{eq:energy_conservation}
\end{equation}

\textit{Remark 3 (Geometric Interpretation):} 
While this property reflects the unitary nature of the transform, its deeper root lies in symplectic geometry. Since the 2D NSFRFT is governed by a symplectic matrix with unit determinant ($\det(\mathbf{X})=1$), the transformation preserves the volume element in the 4D phase space (Liouville’s Theorem). Consequently, the global integral of the WD, representing signal energy, remains invariant under any rigid rotation.

\textbf{Property 4 (Conjugation Property).} 
The complex conjugate of the 2D NSFRFT kernel corresponds to the kernel of the inverse transformation with the inverse parameter set. This relationship is formally expressed as
\begin{equation}
	K_{P}^*(x,y,u,v) = \gamma \cdot K_{P_{inv}}(u,v,x,y),
	\label{eq:conjugation_prop}
\end{equation}
where $P_{inv} = (a, -b, -c, -d, -\theta)$ denotes the inverse parameter set, and the phase adjustment factor $\gamma$ is determined by the sign of the discriminant $\mathbf{T}$:
\begin{equation}
	\gamma = \mathrm{sgn}(-\mathbf{T}) = 
	\begin{cases} 
		1, & \text{if } \mathbf{T} < 0, \\
		-1, & \text{if } \mathbf{T} > 0.
	\end{cases}
\end{equation}

\textit{Remark 4 (Reciprocity and Phase Correction):} 
This property bridges the algebraic conjugation with the geometric inversion. Mathematically, the parameter set $P_{inv}$ maps directly to the inverse symplectic matrix $\mathbf{X}^{-1}$. The inclusion of the factor $\gamma$ is crucial for ensuring analytical consistency: it compensates for the phase discontinuity (branch cut) that occurs in the amplitude term when the discriminant $\mathbf{T}$ flips sign, thereby maintaining the reciprocity of the transform.

\textbf{Property 5 (Space Reversal Property).} 
The 2D NSFRFT of a space-reversed signal $f(-x, -y)$ is equivalent to the frequency-reversed spectrum of the original signal, namely,
\begin{equation}
	\mathcal{F}_P [f(-x, -y)] (u, v) = F_P (-u, -v).
\end{equation}

\textit{Remark 5 (Parity and Scaling):} 
This property highlights the parity conservation of the transformation. Since the chirp components of the kernel (Eq. (\ref{eq:14})) are quadratic forms of spatial variables (involving terms like $x^2, xy, y^2$), they remain invariant under spatial inversion (i.e., scaling by $\sigma = -1$). Consequently, the spatial reversal only affects the sign of the bilinear kernel cross-terms, which is mathematically equivalent to reversing the frequency coordinates. 
It is important to note that for a general scaling factor $\sigma$ where $|\sigma| \neq 1$, the 2D NSFRFT of $f(\sigma x, \sigma y)$ would map the signal to a generalized domain with altered parameters (becoming a general 2D NSLCT), rather than staying within the same 2D NSFRFT definition. Thus, space reversal represents a unique, closed-form symmetry within the 2D NSFRFT framework.

%\textbf{Property 6 (Differentiation Property).}
%Let $F(u,v)$ denote the 2D NSFRFT of the signal $f(x,y)$. The transformation of the partial derivatives of $f(x,y)$ with respect to the spatial variables $x$ and $y$ is given by the following linear combination of frequency domain operators:
%
%For differentiation with respect to the $x$-axis:
%\begin{equation}
%	\mathcal{F}\left[ \frac{\partial f(x,y)}{\partial x} \right](u,v) = a_{11} \frac{\partial F(u,v)}{\partial u} + a_{21} \frac{\partial F(u,v)}{\partial v} - \mathrm{j} c_{11} u F(u,v) - \mathrm{j} c_{21} v F(u,v),
%\end{equation}
%where the coefficients $a_{11}, a_{21}, c_{11}, c_{21}$ correspond to the elements of the first column of the parameter matrix $\mathbf{X}$.
%
%Similarly, for differentiation with respect to the $y$-axis:
%\begin{equation}
%	\mathcal{F}\left[ \frac{\partial f(x,y)}{\partial y} \right](u,v) = a_{12} \frac{\partial F(u,v)}{\partial u} + a_{22} \frac{\partial F(u,v)}{\partial v} - \mathrm{j} c_{12} u F(u,v) - \mathrm{j} c_{22} v F(u,v),
%\end{equation}
%where the coefficients $a_{12}, a_{22}, c_{12}, c_{22}$ correspond to the elements of the second column of the parameter matrix $\mathbf{X}$.
%
%\textit{Remark 6 :} These properties demonstrate that the differentiation operation in the spatial domain is mapped to a weighted sum of differentiation and multiplication operations in the 2D NSFRFT domain. This linear mapping is a hallmark of 2D NSLCTs, strictly determined by the columns of the symplectic matrix $\mathbf{M}$.

\textbf{Property 6 (Differentiation Property).}
Let $F_P(u,v)$ denote the 2D NSFRFT of the signal $f(x,y)$. The transformation of the partial derivatives of $f(x,y)$ with respect to the spatial variables $x$ and $y$ is given by a linear combination of frequency domain operators.

For differentiation with respect to the $x$-axis:
\begin{equation}
	\mathcal{F}_P \left[ \frac{\partial f(x,y)}{\partial x} \right](u,v) = a_{11} \frac{\partial F_P(u,v)}{\partial u} + a_{21} \frac{\partial F_P(u,v)}{\partial v} - \mathrm{j} c_{11} u F_P(u,v) - \mathrm{j} c_{21} v F_P(u,v),
\end{equation}
where the coefficients correspond to the elements of the first column of the parameter matrix $\mathbf{X}$ (specifically, the first columns of its sub-matrices $\mathbf{A}$ and $\mathbf{C}=\mathbf{-B}$).

Similarly, for differentiation with respect to the $y$-axis:
\begin{equation}
	\mathcal{F}_P \left[ \frac{\partial f(x,y)}{\partial y} \right](u,v) = a_{12} \frac{\partial F_P(u,v)}{\partial u} + a_{22} \frac{\partial F_P(u,v)}{\partial v} - \mathrm{j} c_{12} u F_P(u,v) - \mathrm{j} c_{22} v F_P(u,v),
\end{equation}
where the coefficients correspond to the elements of the second column of the parameter matrix $\mathbf{X}$.

\textit{Remark 6 (Operational Mapping):} 
These properties demonstrate that the differentiation operation in the spatial domain is mapped to a weighted sum of differentiation and multiplication operations in the 2D NSFRFT domain. This linear mapping is a hallmark of the symplectic canonical transformation, strictly determined by the columns of the transformation matrix $\mathbf{X}$. This property provides the theoretical foundation for applying the 2D NSFRFT to solve partial differential equations with variable coefficients.

\textbf{Property 7 (Integration Property).}
Let $h(x,y) = \int_{-\infty}^{x} f(\tau, y) \, \mathrm{d}\tau$ denote the integration of the signal along the $x$-axis. The 2D NSFRFT of $h(x,y)$, denoted as $H_P(u,v)$, is the solution to the following first-order linear partial differential equation:
\begin{equation} \label{eq:int_prop_x}
	\left( a_{11} \frac{\partial}{\partial u} + a_{21} \frac{\partial}{\partial v} - \mathrm{j} c_{11} u - \mathrm{j} c_{21} v \right) H_P(u,v) = F_P(u,v),
\end{equation}
where $F_P(u,v)$ is the 2D NSFRFT of the original signal $f(x,y)$, and the coefficients correspond to the elements of the first column of the parameter matrix $\mathbf{X}$.

Similarly, let $g(x,y) = \int_{-\infty}^{y} f(x, \eta) \, \mathrm{d}\eta$ be the integral along the $y$-axis. Its transform $G_P(u,v)$ satisfies:
\begin{equation} \label{eq:int_prop_y}
	\left( a_{12} \frac{\partial}{\partial u} + a_{22} \frac{\partial}{\partial v} - \mathrm{j} c_{12} u - \mathrm{j} c_{22} v \right) G_P(u,v) = F_P(u,v),
\end{equation}
where the coefficients correspond to the elements of the second column of $\mathbf{X}$.

\textit{Remark 7 (Algebraic Complexity):} 
Unlike the classical Fourier transform, where integration corresponds to a simple algebraic division by frequency variables, the integration property in the 2D NSFRFT domain manifests as a differential equation. This complexity arises from the symplectic coupling of spatial and frequency variables introduced by the transformation matrix $\mathbf{X}$, reflecting that the direction of the integral operation is effectively {rotated} relative to the canonical axes in the phase space.

\textbf{Property 8 (Shift and Modulation Property).} 
The 2D NSFRFT exhibits a unified response to signal displacement and modulation. A simultaneous spatial translation by $(m_1, m_2)$ and frequency modulation by $(n_1, n_2)$ in the input domain results in a coordinate shift and linear phase modulation in the transform domain.

Let the input signal be $g(x,y) = e^{\mathrm{j}(n_1 x + n_2 y)} f(x - m_1, y - m_2)$. Its 2D NSFRFT is given by
\begin{equation}
	G_P(u,v) = \Phi \cdot e^{\mathrm{j}(s_1 u + s_2 v)} F_P(u - r_1, v - r_2),
	\label{eq:tf_shifting}
\end{equation}
where $\Phi$ is a phase factor dependent on the shift parameters. The frequency domain shifts $(r_1, r_2)$ and phase modulation coefficients $(s_1, s_2)$ are linearly related to the input parameters $(m_1, m_2, n_1, n_2)$ via the orthogonal symplectic matrix $\mathbf{X}$
\begin{equation}
	\begin{bmatrix} r_1 \\ r_2 \\ s_1 \\ s_2 \end{bmatrix} 
	= \mathbf{X} 
	\begin{bmatrix} m_1 \\ m_2 \\ n_1 \\ n_2 \end{bmatrix}
	= \begin{bmatrix} \mathbf{A} & \mathbf{B} \\ \mathbf{-B} & \mathbf{A} \end{bmatrix}
	\begin{bmatrix} m_1 \\ m_2 \\ n_1 \\ n_2 \end{bmatrix}.
\end{equation}

\textit{Remark 8 (Centroid Propagation):} 
This property demonstrates that the orthogonal symplectic matrix $\mathbf{X}$ strictly governs the propagation of the signal's center in the 4D phase space. The input parameter vector $[m_1, m_2, n_1, n_2]^T$ is mapped to the output vector $[r_1, r_2, s_1, s_2]^T$ through the linear canonical transformation. Geometrically, this implies that the center of mass of the signal's Wigner distribution follows the rigid rotation defined by $\mathbf{X}$, preserving the relative geometry of the signal support.

\textbf{Property 9 (Relationship with the 2D WD).}
The 2D NSFRFT corresponds to a unitary rotation in the 4D phase space. Consequently, the 2D WD of the transformed signal $F_P(u,v)$ is a strictly rotated version of the 2D WD of the input signal $f(x,y)$
\begin{equation}
	W_{F_P}(x, y, u, v) = W_f(\widetilde{x}, \widetilde{y}, \widetilde{u}, \widetilde{v}),
	\label{eq:wigner_relation}
\end{equation}
where the coordinate vector is transformed via the transpose of the parameter matrix:
\begin{equation}
	\begin{bmatrix} \widetilde{x} \\ \widetilde{y} \\ \widetilde{u} \\ \widetilde{v} \end{bmatrix} 
	= \mathbf{X}^{-1} 
	\begin{bmatrix} x \\ y \\ u \\ v \end{bmatrix}
	= \mathbf{X}^{T} 
	\begin{bmatrix} x \\ y \\ u \\ v \end{bmatrix}.
	\label{eq:444}
\end{equation}
Since $\mathbf{X}$ is an orthogonal matrix ($\mathbf{X}^{-1} = \mathbf{X}^T$), the inverse transformation corresponds exactly to the matrix transpose.

\textit{Remark 9 (4D Phase Space Rotation):}
This property reveals the geometric essence of the 2D NSFRFT. Unlike general 2D NSLCTs, which typically introduce shearing or scaling distortions to the phase space support, the 2D NSFRFT performs a {rigid rotation} of the Wigner distribution. The signal's energy distribution is rotated without geometric distortion, strictly preserving the volumetric structure. This confirms that the 2D NSFRFT is a true generalization of the rotation principle of the classical FRFT to the 4D nonseparable phase space.

%\subsection{Geometric Interpretation of the 2D NSFRFT}
%
%From Eq. (\ref{eq:444}), we can conclude that the 2D WD \( W_{F_P} \) of the 2D NSFRFT \( F_P \) of a 2D function \( f \) can be obtained by applying a 4D linear transformation to the 2D WD \( W_f \) of \( f \), using the transformation matrix
%\begin{equation}
%\begin{aligned}
%\mathbf{R_{4D}^\mathrm{NS}} &=
%\begin{bmatrix}
%a\cos\theta-c\sin\theta & -b\cos\theta-d\sin\theta & -a\sin\theta-c\cos\theta & b\sin\theta-d\cos\theta \\
%b\cos\theta-d\sin\theta & a\cos\theta+c\sin\theta & -b\sin\theta-d\cos\theta & -a\sin\theta+c\cos\theta \\
%a\sin\theta+c\cos\theta & -b\sin\theta+d\cos\theta & a\cos\theta-c\sin\theta & -b\cos\theta-d\sin\theta \\
%b\sin\theta+d\cos\theta & a\sin\theta-c\cos\theta & b\cos\theta-d\sin\theta & a\cos\theta+c\sin\theta
%\end{bmatrix}.
%\end{aligned}
%\label{eq:29}
%\end{equation}
%
%Furthermore, according to Proposition 1, it can be verified that $\mathbf{R_{4D}^\mathrm{NS}}$ is a specific 4D rotation matrix. This confirms that the proposed 2D NSFRFT preserves the energy and symplectic structure of the WD. Moreover, the matrix $\mathbf{R_{4D}^\mathrm{NS}}$ provides a unified representation that includes the classical forms $\mathbf{R_{4D}^\mathrm{S}}$, $\mathbf{R_{4D}^\mathrm{G}}$, and $\mathbf{R_{4D}^\mathrm{C}}$ as special cases. Consequently, the 2D NSFRFT generalizes the 2D SFRFT, GT, and CFRFT, offering greater flexibility and broader potential in fractional signal processing applications.

\subsection{Geometric Interpretation of the 2D NSFRFT}

From Eq. (\ref{eq:444}), it is evident that the 2D WD \( W_{F_P} \) of the transformed signal is obtained by applying a 4D linear coordinate transformation to the input distribution \( W_f \). The governing transformation matrix, denoted as $\mathbf{R_{4D}^\mathrm{NS}}$, is explicitly given by
\begin{equation}
	\mathbf{R_{4D}^\mathrm{NS}} = \mathbf{X}^{-1}= \mathbf{X}^T= 
	\begin{bmatrix}
		a\cos\theta-c\sin\theta & -b\cos\theta-d\sin\theta & -a\sin\theta-c\cos\theta & b\sin\theta-d\cos\theta \\
		b\cos\theta-d\sin\theta & a\cos\theta+c\sin\theta & -b\sin\theta-d\cos\theta & -a\sin\theta+c\cos\theta \\
		a\sin\theta+c\cos\theta & -b\sin\theta+d\cos\theta & a\cos\theta-c\sin\theta & -b\cos\theta-d\sin\theta \\
		b\sin\theta+d\cos\theta & a\sin\theta-c\cos\theta & b\cos\theta-d\sin\theta & a\cos\theta+c\sin\theta
	\end{bmatrix}.
	\label{eq:rotation_matrix_4d}
\end{equation}

Based on the orthogonality derived in Proposition 1, $\mathbf{R_{4D}^\mathrm{NS}}$ is an orthogonal matrix with unit determinant ($\det(\mathbf{R_{4D}^\mathrm{NS}}) = 1$). Mathematically, this identifies the transformation as an element of the special orthogonal group $\mathrm{SO}(4)$, representing a strict rigid rotation in the 4D phase space. 

This geometric insight has two profound implications:
First, unlike general 2D NSLCTs which may introduce shearing or scaling, the proposed 2D NSFRFT preserves the Euclidean geometry of the WD, ensuring no distortion of the signal's energy support. 
Second, the matrix $\mathbf{R_{4D}^\mathrm{NS}}$ serves as a unified geometric operator. It seamlessly encapsulates the transformation kernels of the 2D SFRFT , GT , and CFRFT  as degenerate subspaces. Consequently, the 2D NSFRFT generalizes these existing transforms, offering a versatile framework for manipulating signal energy trajectories in high-dimensional phase spaces.

\subsection{Unique Advantage: Handling Specific Coupled 2D Chirp Signals}
\label{section 3.4}

In practical applications such as radar, sonar, communication, and optical imaging, 2D chirp signals characterized by complex phase structures frequently arise. These signals typically possess a full quadratic phase structure, simultaneously containing squared terms ($x^2, y^2$) representing self-chirp rates and cross-terms ($xy$) representing spatial coupling.

Existing transforms exhibit inherent structural limitations when processing such signals due to the restricted definitions of their kernel functions. Specifically,
\begin{itemize}
	\item The 2D SFRFT (Eq. (\ref{eq:3})) is defined as the tensor product of two 1D FRFT kernels. Its kernel phase is strictly separable, comprising a sum of independent quadratic terms ($\alpha x^2 + \beta y^2$). This mathematical structure inherently excludes the spatial coupling term ($xy$), rendering it incapable of matching coupled signals.
	\item The GT kernel (Eq. (\ref{eq:4})) primarily focuses on the spatial cross-terms ($xy$) to effectuate plane rotations. However, it lacks independent degrees of freedom for separate quadratic scaling ($x^2$ and $y^2$), limiting its flexibility in handling anisotropic chirps.
	\item The CFRFT kernel (Eq. (\ref{eq:6})) emphasizes radial quadratic terms ($x^2+y^2$) and angular momentum. While it handles rotation, its structure is tied to isotropic or specific radial symmetries, limiting its ability to match arbitrary rectangular coupling coefficients.
\end{itemize}

Consequently, none of these methods provides the necessary degrees of freedom to simultaneously match the diverse coefficients ($x^2, y^2, xy$) of a general coupled chirp signal.

In contrast, the proposed {2D NSFRFT} incorporates a generalized kernel function (Eq. \ref{eq:14}) that explicitly includes {both} the separable quadratic terms ($x^2, y^2$) and the nonseparable cross-terms ($xy$). This flexible structure allows the 2D NSFRFT to perfectly match the phase curvature of a specific class of coupled 2D chirp signals that existing transforms fail to focus. By tuning the parameters, the transform concentrates the energy of these signals into highly localized impulses (Dirac delta functions).

A rigorous mathematical proof demonstrating that the 2D NSFRFT can transform this specific class of nonseparable 2D chirp signals into ideal impulses is provided in {\ref{Appendix A}}. This unique focusing capability makes the proposed method particularly advantageous for parameter estimation and weak signal detection. {Crucially, it also enables precise signal filtering, allowing for the design of effective passband and stopband filters to extract or suppress coupled signals in complex interference environments.}

\section{Discrete Algorithms of the 2D NSFRFT}
\label{Section 4}

In this section, we present three discrete implementations of the 2D NSFRFT, comprising a direct method and two fast algorithms. {We adapt the established matrix decomposition frameworks of the general 2D NSLCT to the specific context of the 2D NSFRFT.} While the structural decompositions utilize existing architectures, our specific algorithmic contribution lies in deriving the explicit closed-form parameters required to instantiate these frameworks for the proposed transform. This provides the first complete and verifiable implementation path for the 2D NSFRFT.

We then conduct a comparative analysis to evaluate the computational complexity and accuracy of these methods. {Finally, to further validate the effectiveness of the proposed discrete algorithms, we demonstrate and analyze the transformation results of several typical 2D signals.}

\subsection{The Direct Method}
\label{sec4.1}
The fundamental approach to implementing the discrete 2D NSFRFT is to discretize the continuous integral definition in Eq. (\ref{eq:13}) using a direct Riemann summation approximation. Assuming an input signal of size $M \times N$, the discrete transformation is expressed as
\begin{equation}
	\begin{aligned}
		F_{P}(p \Delta_u, q \Delta_v) &= \frac{1}{2\pi \sqrt{-\mathbf{T}}} \exp\left[{\frac{\mathrm{j}(k_1 p^2 \Delta_u^2 + k_2 pq \Delta_u \Delta_v + k_3 q^2 \Delta_v^2)}{2\mathbf{T}}}\right] \\
		&\times \sum_{m=0}^{M-1} \sum_{n=0}^{N-1} f(m \Delta_x, n \Delta_y) \Delta_x \Delta_y \\
		&\times \exp\left[{\frac{\mathrm{j}(p_1 m^2 \Delta_x^2 + p_2 mn \Delta_x \Delta_y + p_3 n^2 \Delta_y^2)}{2\mathbf{T}}} \right] \\
		&\times \exp\left[{\frac{\mathrm{j}(m_1 p \Delta_u + m_2 q \Delta_v) m \Delta_x + \mathrm{j}(m_3 p \Delta_u + m_4 q \Delta_v) n \Delta_y}{\mathbf{T}}}\right],
	\end{aligned} 
	\label{eq:30}
\end{equation}
where $\Delta_x, \Delta_y$ and $\Delta_u, \Delta_v$ denote the sampling intervals in the spatial and frequency domains, respectively. The coefficients $k_i, p_i, m_i$ correspond to the expanded terms of the kernel function derived in Eqs. (\ref{eq:16}) - (\ref{eq:18}).

While this method is conceptually intuitive and theoretically accurate, it suffers from high computational cost. For an $N \times N$ image, calculating each of the $N^2$ output points requires iterating over all $N^2$ input pixels, resulting in a total computational complexity of $O(N^4)$. This scales poorly with data size, rendering the direct method impractical for large-scale signal or image processing tasks. Consequently, its primary role in this study is to serve as a precision benchmark for evaluating the accuracy of the proposed fast algorithms.

\subsection{Fast Implementations via Orthogonal Symplectic Matrix Decomposition}

To overcome the $O(N^4)$ computational bottleneck of the direct method, we leverage the specific algebraic structure of the 2D NSFRFT. The core principle is to decompose the dense orthogonal symplectic matrix $\mathbf{X}$ into a product of simpler elementary matrices. This factorization allows the high-complexity 2D NSFRFT integral to be converted into a cascade of low-complexity 2D discrete operators.

We adapt two established matrix decomposition frameworks from the general 2D NSLCT \cite{46,47,48,49}. By substituting the general parameter blocks in these frameworks with the specific rotation parameters of the orthogonal symplectic matrix $\mathbf{X}$, the 2D NSFRFT is efficiently instantiated. The fundamental operators involved in these cascades include:
\begin{itemize}
	\item \textbf{2D Chirp Multiplication (CM):} Point-wise quadratic phase modulation in the spatial domain.
	\item \textbf{2D Chirp Convolution (CC):} Convolution with a quadratic phase kernel, implementable via 2D FFTs with complexity $\mathcal{O}(N^2 \log N)$.
	\item \textbf{2D Affine Transform (AT):} Geometric warping requiring spatial interpolation.
\end{itemize}
\noindent Detailed mathematical definitions and computational complexity analyses of these operators can be found in \cite{46,47,48,49} and are not repeated here.

Based on different decomposition strategies for $\mathbf{X}$, we present two distinct algorithms:
\begin{enumerate}
	\item \textbf{Algorithm \RomanNumeral{1} (CM-FT-AT-CM):} This method decomposes $\mathbf{X}$ into a sequence involving a 2D AT. While efficient, the inclusion of the 2D AT stage necessitates 2D interpolation, which may introduce numerical aliasing.
	\item \textbf{Algorithm \RomanNumeral{2} (Interpolation-Free):} To mitigate interpolation errors, this method employs a purely arithmetic decomposition using only 2D CM and 2D CC operators. This approach preserves the discrete grid structure, resulting in higher numerical accuracy.
\end{enumerate}

A summary of their decomposition forms, operator sequences, and computational complexities is provided in Table~\ref{tab:1}. Detailed parameter selection strategies and conditions are available in \cite{48,49}.

% Table 1
\begin{table}[H]
	\renewcommand{\arraystretch}{1.3} % Increased slightly for better spacing
	\centering
	\caption{Comparison of Fast Algorithms \RomanNumeral{1} and \RomanNumeral{2} for 2D NSFRFT}
	\resizebox{\textwidth}{!}{
		\begin{tabular}{|c|c|c|c|c|}
			\hline
			\textbf{Algorithm} & \textbf{Matrix Decomposition of $\mathbf{X}$} & \textbf{Operator Sequence} & \textbf{Complexity} & \textbf{Ref.} \\
			\hline
			\makecell*[c]{Algorithm \RomanNumeral{1} \\ (CM-FT-AT-CM)} & 
			$\makecell*[c]{
				\underbrace{\begin{bmatrix} \mathbf{I} & \mathbf{0} \\ \mathbf{A}\mathbf{B}^{-1} & \mathbf{I} \end{bmatrix}}_{\text{2D CM}}
				\underbrace{\begin{bmatrix} \mathbf{0} & \mathbf{I} \\ -\mathbf{I} & \mathbf{0} \end{bmatrix}}_{\text{2D FT}}
				\underbrace{\begin{bmatrix} (\mathbf{B}^T)^{-1} & \mathbf{0} \\ \mathbf{0} & \mathbf{B} \end{bmatrix}}_{\text{2D AT}}
				\underbrace{\begin{bmatrix} \mathbf{I} & \mathbf{0} \\ \mathbf{B}^{-1}\mathbf{A} & \mathbf{I} \end{bmatrix}}_{\text{2D CM}}
			}$ & 
			$\makecell*[c]{
				\mathcal{O}_{\mathrm{CM}}^{\mathbf{A}\mathbf{B}^{-1}}
				\mathcal{F}_{x,y} 
				\mathcal{O}_{\mathrm{AT}}^{\mathbf{B}} 
				\mathcal{O}_{\mathrm{CM}}^{\mathbf{B}^{-1}\mathbf{A}}
			}$ & 
			$\makecell*[c]{O(N^2 \log N)}$ & 
			\cite{48} \\
			\hline
			\makecell*[c]{Algorithm \RomanNumeral{2}-1 \\ (Symmetric Case) \\ $\mathbf{B} = \mathbf{B}^T$} & 
			$\makecell*[c]{
				\underbrace{\begin{bmatrix} \mathbf{I} & \mathbf{0} \\ (\mathbf{A} - \mathbf{I})\mathbf{B}^{-1} & \mathbf{I} \end{bmatrix}}_{\text{2D CM}}
				\underbrace{\begin{bmatrix} \mathbf{I} & \mathbf{B} \\ \mathbf{0} & \mathbf{I} \end{bmatrix}}_{\text{2D CC}}
				\underbrace{\begin{bmatrix} \mathbf{I} & \mathbf{0} \\ \mathbf{B}^{-1}(\mathbf{A} - \mathbf{I}) & \mathbf{I} \end{bmatrix}}_{\text{2D CM}}
			}$ & 
			$\makecell*[c]{
				\mathcal{O}_{\mathrm{CM}}^{(\mathbf{A}-\mathbf{I})\mathbf{B}^{-1}} 
				\mathcal{O}_{\mathrm{CC}}^{\mathbf{B}} 
				\mathcal{O}_{\mathrm{CM}}^{\mathbf{B}^{-1}(\mathbf{A}-\mathbf{I})}
			}$ & 
			$\makecell*[c]{O(N^2 \log N)}$ & 
			\cite{49} \\
			\hline
			\makecell*[c]{Algorithm \RomanNumeral{2}-2 \\ (General Case) \\ $\mathbf{B} \neq \mathbf{B}^T$} & 
			$\makecell*[c]{
				\underbrace{\begin{bmatrix} \mathbf{I} & \mathbf{0} \\ (\mathbf{D}' - \mathbf{I})\mathbf{B}'^{-1} & \mathbf{I} \end{bmatrix}}_{\text{2D CM}}
				\underbrace{\begin{bmatrix} \mathbf{I} & \mathbf{B}' \\ \mathbf{0} & \mathbf{I} \end{bmatrix}}_{\text{2D CC}} 
				\underbrace{\begin{bmatrix} \mathbf{I} & \mathbf{0} \\ \mathbf{B}'^{-1}(\mathbf{A}' - \mathbf{I}) & \mathbf{I} \end{bmatrix}}_{\text{2D CM}}
				\underbrace{\begin{bmatrix} \mathbf{I} & \mathbf{H} \\ \mathbf{0} & \mathbf{I} \end{bmatrix}}_{\text{2D CC}}
			}$ & 
			$\makecell*[c]{
				\mathcal{O}_{\mathrm{CM}}^{(\mathbf{D}' - \mathbf{I})\mathbf{B}'^{-1}} 
				\mathcal{O}_{\mathrm{CC}}^{\mathbf{B}'} 
				\mathcal{O}_{\mathrm{CM}}^{\mathbf{B}'^{-1}(\mathbf{A}' - \mathbf{I})} 
				\mathcal{O}_{\mathrm{CC}}^{\mathbf{H}}
			}$ & 
			$\makecell*[c]{O(N^2 \log N)}$ & 
			\cite{49} \\
			\hline
		\end{tabular}
	}
	\label{tab:1}
\end{table}

% Remark 3

	\textit{Remark (Comparison with Classical Iwasawa Decomposition \cite{46}):}
	It is crucial to distinguish the proposed algorithms from the classical fast computation method for the 2D NSLCT presented in \cite{46}, which is based on the Iwasawa decomposition. 
	The Iwasawa decomposition typically factorizes the transformation matrix into a sequence involving {two} 2D ATs. Since 2D ATs on discrete grids require spatial interpolation, the method in \cite{46} inevitably introduces interpolation errors twice.
	In contrast, our adapted {Algorithm \RomanNumeral{1}} optimizes this structure to involve only {one} 2D AT, thereby reducing the interpolation error sources. 
	Furthermore, {Algorithm \RomanNumeral{2}} utilizes a pure 2D CM-CC decomposition that eliminates 2D ATs entirely. Since 2D CC can be computed exactly via 2D FFT without grid interpolation, Algorithm \RomanNumeral{2} is {interpolation-free}, theoretically yielding the highest numerical accuracy among the three methods.

\subsection{Comparisons between the Two Fast Algorithms and the Direct Method}

This subsection presents a comprehensive comparative evaluation of the three implemented algorithms: the direct method, Algorithm \RomanNumeral{1}, and Algorithm \RomanNumeral{2}. We analyze their performance across three critical dimensions: computational efficiency (execution time), numerical accuracy, and reversibility.

The direct method, introduced in Section \ref{sec4.1}, serves as the rigorous benchmark for evaluating the precision of the fast algorithms. To ensure a robust assessment, we perform simulations using various types of input functions and distinct orthogonal symplectic parameter matrices $\mathbf{X}$. This allows us to quantify the trade-off between computational speed and numerical error for each implementation path.

Specifically, two representative 2D Hermite-Gaussian functions are chosen as test signals due to their diversity and analytical tractability. The first is the second-order function g$_1 = 4\exp({-\frac{x^2+y^2}{2}})(4x^2y^2 - 2(x^2+y^2) + 1)$; the second is a combination of basis functions, g$_2 = \Psi_{1,2}(x,y) + \Psi_{3,1}(x,y)$, where $\Psi_{m,n}(x,y) = \psi_m(x)\psi_n(y)$ and $\psi_m(x)$ is the 1D $m$-th order Hermite-Gaussian function defined by
\begin{equation}
\psi_m(x) = \frac{1}{\sqrt{2^m m! \sqrt{\pi}}} \exp(-\frac{x^2}{2}) H_m(x),
\label{eq:31}
\end{equation}
with $H_m(x)$ denoting the Hermite polynomial. The normalized mean square error (NMSE) is used to quantify the accuracy of the results.

\noindent (1) Computational Complexity

The computational complexity of two fast algorithms is summarized in Table~\ref{tab:1}; see also~\cite{46,47,48,49} for details on the underlying 2D discrete operators.

\noindent {(2) Accuracy}

\noindent {1) Simulation for Different Input Functions}

We evaluate the accuracy and efficiency of the two fast algorithms using different input functions, with the parameter matrix fixed as $\mathbf{X}_{\mathrm{ac1}}$. Accuracy is measured by the NMSE, defined by
\begin{equation}
\mathrm{NMSE_{ac}} = 
\frac{
\displaystyle\sum_{p} \sum_{q} \left| G(p\Delta_u, q\Delta_v) - G_0(p\Delta_u, q\Delta_v) \right|^2
}{
\displaystyle\sum_{p} \sum_{q} \left| G_0(p\Delta_u, q\Delta_v) \right|^2
},
\label{eq:32}
\end{equation}
where $G$ denotes the output of Algorithm~\RomanNumeral{1} or \RomanNumeral{2} and $G_0$ is the result of the direct method.

The parameter set used in the 2D NSFRFT numerical simulations is  $P=(0.4033, 0.1555,0.2851, -0.8555, \frac{\pi}{8})$, so the corresponding parameter matrix $\mathbf{X}_{\mathrm{ac1}}$ is
\begin{equation}
\mathbf{X}_{\mathrm{ac1}}
=
\begin{bmatrix}0.2635&0.4710&0.4177&-0.7309\\0.1837 &0.4817&-0.8499&-0.1091\\-0.4177 &0.7309 &0.2635&0.4710\\0.8499 &0.1091&0.1837&0.4817
\end{bmatrix}.
\label{eq:33}
\end{equation}

\begin{figure}[!htp]
	\centering
	\subfloat[]{\includegraphics[width=0.18\textwidth]{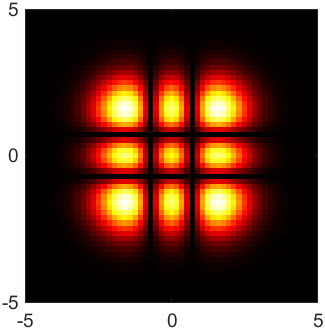}}
	\subfloat[]{\includegraphics[width=0.18\textwidth]{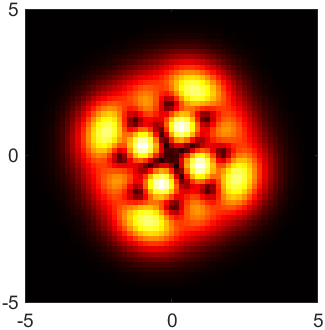}}
	\subfloat[]{\includegraphics[width=0.18\textwidth]{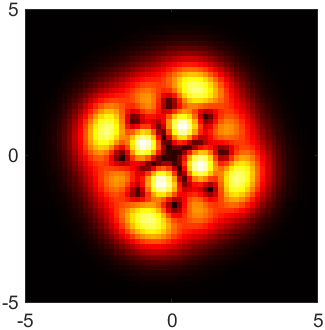}}
	\subfloat[]{\includegraphics[width=0.18\textwidth]{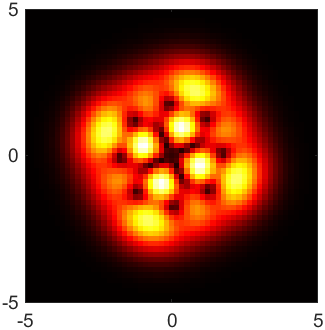}}
	\caption{2D NSFRFT under the condition  of input function g$_1$ and parameter matrix  $\mathbf{X}_{\mathrm{ac1}}$. (a) Input function; (b) Output of the direct method; (c) Output of Algorithm I; (d) Output of Algorithm II.}
 \label{fig:3}
\end{figure}

Fig.~\ref{fig:3} shows the simulation results of the 2D NSFRFT. Fig.~\ref{fig:3}(a) displays the input function g$_1$, while (b)--(d) present the outputs of the direct method, Algorithm~\RomanNumeral{1}, and Algorithm~\RomanNumeral{2}, respectively. All methods use $\Delta_x = \Delta_y = \Delta_u = \Delta_v = 0.1772$ with $M = N = 200$ sampling points.

Fig.~\ref{fig:3} confirms that both fast algorithms achieve results comparable to the direct method. Table~\ref{tab:2} further shows that they offer significant speed improvements, with Algorithm~\RomanNumeral{2} achieving the highest accuracy  because it is interpolation-free.

\begin{table}[h!]
\centering
\caption{The numerical calculation time and accuracy of the three methods under g$_1$ and $\mathbf{X}_{\mathrm{ac1}}$}
\label{tab:2}
\small 
\begin{tabular}{lcccc}
\toprule
\textbf{} & \textbf{Direct method} & \textbf{Algorithm~\RomanNumeral{1}} & \textbf{Algorithm~\RomanNumeral{2}}  \\ 
\midrule
Time  & 166.1043s  & 0.0101s & 0.0068s  \\ 
NMSE &  /   &  $2.2293 \times 10^{-3}$  & $3.6689 \times 10^{-8}$  \\ 
\bottomrule
\end{tabular}
\end{table}

\begin{figure}[htp]
	\centering
	\subfloat[]{\includegraphics[width=0.18\textwidth]{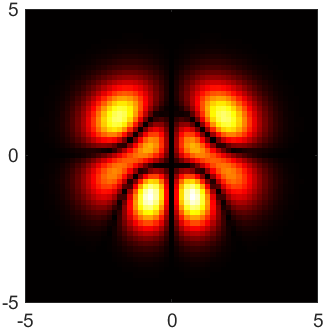}}
	\subfloat[]{\includegraphics[width=0.18\textwidth]{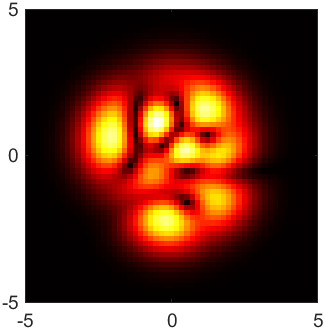}} 
	\subfloat[]{\includegraphics[width=0.18\textwidth]{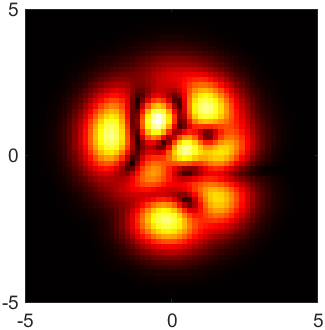}}
	\subfloat[]{\includegraphics[width=0.18\textwidth]{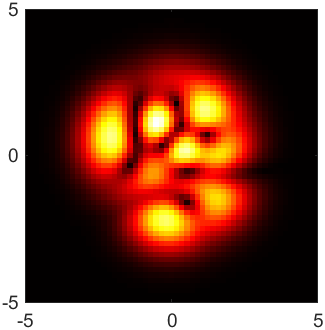}}
	\caption{2D NSFRFT under the condition  of input function g$_2$ and parameter matrix  $\mathbf{X}_{\mathrm{ac1}}$. (a) Input function; (b) Output of the direct method; (c) Output of Algorithm I; (d) Output of Algorithm II.}
 \label{fig:4}
\end{figure}

In the second simulation, the input function is changed to g$_2$ while keeping $\mathbf{X}_{\mathrm{ac1}}$ unchanged. As shown in Fig.~\ref{fig:4} and Table~\ref{tab:3}, the results are consistent with the previous case: both fast algorithms significantly reduce computation time, and Algorithm~\RomanNumeral{2} achieves the best accuracy.

\begin{table}[h!]
\centering
\caption{The numerical calculation time and accuracy of the three methods under g$_2$ and $\mathbf{X}_{\mathrm{ac1}}$}
\label{tab:3}
\small 
\begin{tabular}{lcccc}
\toprule
\textbf{} & \textbf{Direct method} & \textbf{Algorithm~\RomanNumeral{1}} & \textbf{Algorithm~\RomanNumeral{2}}  \\ 
\midrule
Time  & 162.1484s  & 0.0080s & 0.0078s   \\ 
NMSE &  /   &  $2.2407 \times 10^{-3}$  & $3.3724 \times 10^{-8}$  \\ 
\bottomrule
\end{tabular}
\end{table}

\noindent 2) Simulation with Different Parameter Matrices

We conduct a new round of numerical
simulation comparison experiments by changing different parameter matrices. The input function is still g$_1$ and a new parameter set $P = (0.1745, 0.5951, -0.7329, 0.2798, \frac{\pi}{9})$, yielding the matrix
\begin{equation}
\mathbf{X}_{\mathrm{ac2}} =
\begin{bmatrix}
0.4146 & 0.4635 & -0.6291 & 0.4664 \\
-0.6549 & -0.0867 & 0.0594 & 0.7484 \\
0.6291 & -0.4664 & 0.4146 & 0.4635 \\
-0.0594 & -0.7484 & -0.6549 & -0.0867
\end{bmatrix}.
\label{eq:34}
\end{equation}

\begin{figure}[!htp]
	\centering
	\subfloat[]{\includegraphics[width=0.18\textwidth]{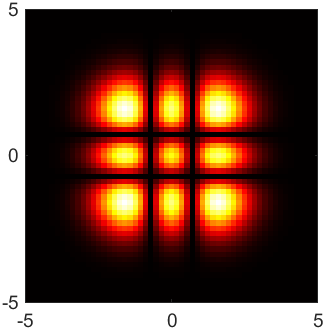}}
	\subfloat[]{\includegraphics[width=0.18\textwidth]{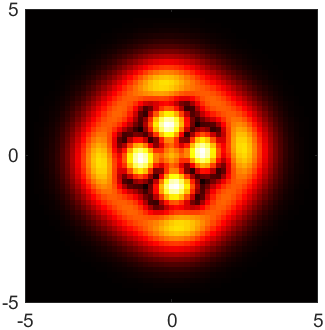}}
	\subfloat[]{\includegraphics[width=0.18\textwidth]{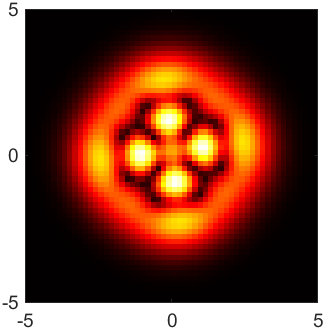}}
	\subfloat[]{\includegraphics[width=0.18\textwidth]{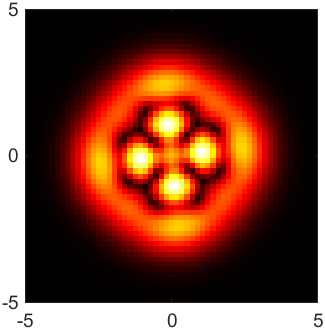}}
	\caption{2D NSFRFT under the condition  of input function g$_1$ and parameter matrix  $\mathbf{X}_{\mathrm{ac2}}$. (a) Input function; (b) Output of the direct method; (c) Output of Algorithm I; (d) Output of Algorithm II.}
 \label{fig:5}
\end{figure}

The simulation results are shown in Fig.~\ref{fig:5}, and the corresponding time and NMSE values are listed in Table~\ref{tab:4}. As before, both fast algorithms significantly outperform the direct method in efficiency, with Algorithm~\RomanNumeral{2} again achieving superior accuracy.

\begin{table}[h!]
\centering
\caption{The numerical calculation time and accuracy of the three methods under g$_1$ and $\mathbf{X}_{\mathrm{ac2}}$}
\label{tab:4}
\small 
\begin{tabular}{lcccc}
\toprule
\textbf{} & \textbf{Direct method} & \textbf{Algorithm~\RomanNumeral{1}} & \textbf{Algorithm~\RomanNumeral{2}}  \\ 
\midrule
Time  & 163.5826s  & 0.0090s & 0.0059s   \\ 
NMSE &  /   &  $2.5662 \times 10^{-2}$  & $3.671 \times 10^{-8}$  \\ 
\bottomrule
\end{tabular}
\end{table}

From the results of the three experiments, it can be concluded that the proposed two algorithms are highly efficient and exhibit satisfactory accuracy.

\noindent (3) Reversibility

Reversibility is a fundamental property of the 2D NSFRFT and underpins many of its applications. To evaluate this, we conduct a simulation using the input function g$_2$ and the parameter set $P = (-0.1601, 0.6966, 0.2625, 0.6483, \\ \frac{\pi}{6})$, yielding the matrix
\begin{equation}
\mathbf{X}_{\mathrm{re}}=
\begin{bmatrix}
-0.2699 & 0.2791 & 0.1473 & 0.9097\\
-0.9274 & -0.0074 & 0.2131 & -0.3074\\
-0.1473 & -0.9097 & -0.2699 & 0.2791\\
-0.2131 & 0.3074 & -0.9274 & -0.0074
\end{bmatrix}.
\label{eq:35}
\end{equation}

Reversibility is assessed by computing the NMSE between the original signal $g$ and its reconstruction $g'$ obtained through forward and inverse 2D NSFRFT:
\begin{equation}
\mathrm{NMSE}_{\mathrm{re}} = 
\frac{
\displaystyle\sum_{m} \sum_{n} \left| 
g'(m\Delta_x, n\Delta_y) - g(m\Delta_x, n\Delta_y) 
\right|^2
}{
\displaystyle\sum_{m} \sum_{n} \left| g(m\Delta_x, n\Delta_y) \right|^2
},
\label{eq:36}
\end{equation}
where \( g'(x,y) = (\mathcal{F}_{P})^{-1}[\mathcal{F}_{P}[g]](x,y) \).

\begin{figure}[htp]
	\centering
	\subfloat[]{\includegraphics[width=0.18\textwidth]{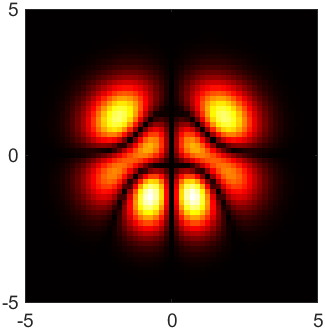}}
	\subfloat[]{\includegraphics[width=0.18\textwidth]{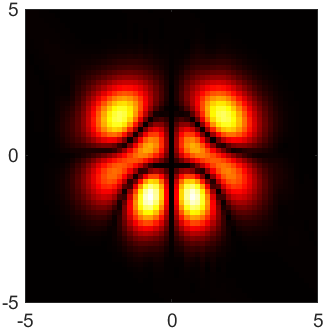}}
	\subfloat[]{\includegraphics[width=0.18\textwidth]{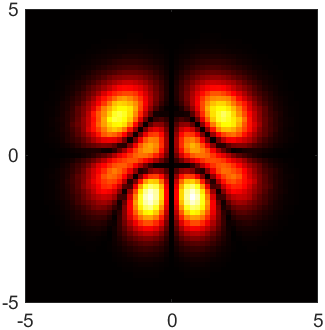}}
	\subfloat[]{\includegraphics[width=0.18\textwidth]{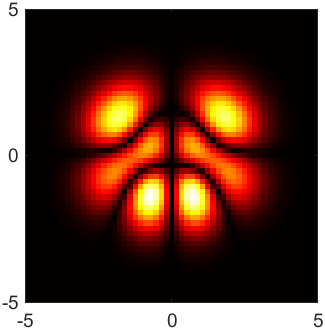}}
	\caption{ The reversibility of the 2D NSFRFT under the condition of input function g$_2$ and  parameter matrix $\mathbf{X}_{\mathrm{re}}$. (a) Input function; (b) Output of the direct method; (c) Output of Algorithm I; (d) Output of Algorithm II.}
\label{fig:6}
\end{figure}

Fig.~\ref{fig:6} presents the reversibility simulation results for g$_2$, while Table~\ref{tab:5} summarizes the corresponding computation time and NMSE values. Both Algorithms~\RomanNumeral{1} and \RomanNumeral{2} effectively preserve reversibility with significantly reduced computation time compared to the direct method. Notably, Algorithm~\RomanNumeral{2} achieves the highest accuracy, confirming its superior performance in reversible 2D NSFRFT implementation.

\begin{table}[h!]
\centering

\begin{threeparttable} % 开始环境
	\caption{The numerical calculation time and accuracy of the reversibility experiment under g$_2$ and  $\mathbf{X}_{\mathrm{re}}$ by three methods}
	\label{tab:5}
	\small 
	\begin{tabular}{lcccc}
		\toprule
		\textbf{} & \textbf{Direct method} & \textbf{Algorithm~\RomanNumeral{1}} & \textbf{Algorithm~\RomanNumeral{2}}  \\ 
		\midrule
		Time  & 328.7254s  & 0.0729s & 0.0250s  \\ 
		NMSE & $6.8233 \times 10^{-1}$ &  $4.6038 \times 10^{-3}$  & $7.989 \times 10^{-31}$  \\ 
		\bottomrule
	\end{tabular}
	
	% tablenotes 部分
	\begin{tablenotes}
		\footnotesize
		\item \textit{Note:} The direct method exhibits large errors due to truncation error accumulation in the $O(N^4)$ process.
	\end{tablenotes}
\end{threeparttable}

\end{table}

The four simulations collectively highlight the advantages of the proposed algorithms, which exhibit both high computational efficiency and accuracy. In particular, for applications requiring reversibility, the fast algorithms achieve superior performance in terms of both precision and speed.

\subsection{Transformation of Typical Signals}
\label{Section 4.4}
Having established the superior accuracy and efficiency of Algorithm \RomanNumeral{2} in the previous subsection, we utilize this algorithm to {empirically demonstrate} the generalized rotational characteristics of the 2D NSFRFT. We analyze its effect on three representative 2D signals: a smooth Gaussian function, a discontinuous rectangular aperture, and a geometric cross-shaped aperture. These examples visually demonstrate how the transform parameters reshape the signal energy distribution in the joint space-frequency domain through numerical simulation.
\subsubsection{Experimental Setup}
In this simulation, the parameters of the 2D NSFRFT are set as $P  = (0.1019, -0.2525, 0.8944, -0.3548, \frac{\pi}{7})$. This specific configuration ensures a non-trivial case where all degrees of freedom are active, effectively {inducing} coupled rotations in the phase space. The three input signals are defined as follows:
\begin{itemize}
	\item \textbf{Gaussian Signal ($f_1$):} An isotropic 2D Gaussian function defined as $f_1(x,y) = \exp\left(-\frac{x^2+y^2}{2\sigma^2}\right)$ with $\sigma=5$.
	\item \textbf{Rectangular Aperture ($f_2$):} A square aperture defined as $f_2(x,y) = 1$ for $|x|\le 4, |y|\le 4$, and 0 otherwise.
	\item \textbf{Cross-shaped Aperture ($f_3$):} A composite geometric structure defined as the union of two orthogonal bars: $f_3(x,y) = 1$ if $(|x|\le 6 \land |y|\le 1) \lor (|x|\le 1 \land |y|\le 6)$, and 0 otherwise (corresponding to length $l=12$ and width $w=2$).
\end{itemize}

\subsubsection{Analysis of Simulation Results}
\begin{figure}[h!]
	\centering
	\subfloat[]{\includegraphics[width=0.2\textwidth]{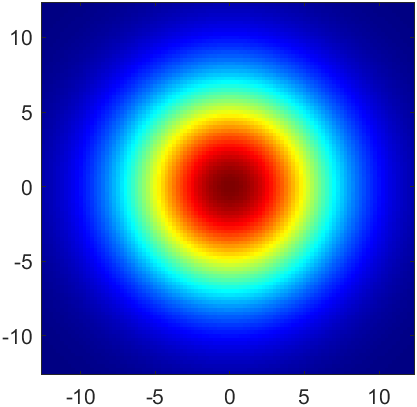}}
	\subfloat[]{\includegraphics[width=0.2\textwidth]{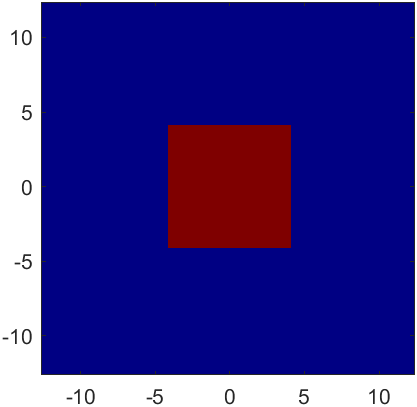}}
	\subfloat[]{\includegraphics[width=0.2\textwidth]{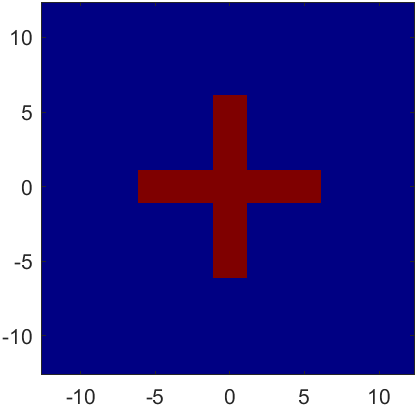}}\\
	\subfloat[]{\includegraphics[width=0.2\textwidth]{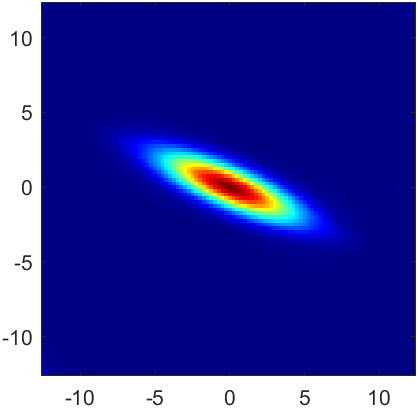}}
	\subfloat[]{\includegraphics[width=0.2\textwidth]{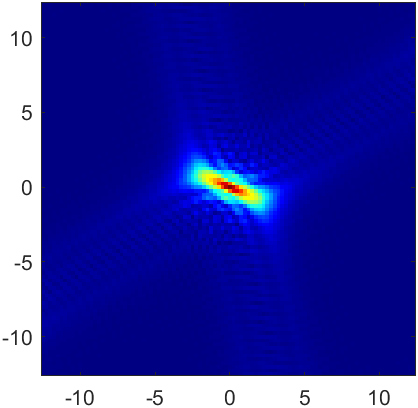}}
	\subfloat[]{\includegraphics[width=0.2\textwidth]{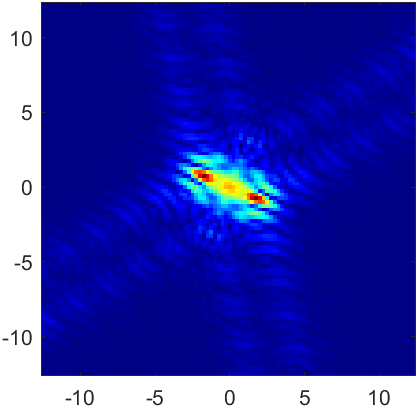}}
	\caption{Amplitude distributions of typical signals under the proposed 2D NSFRFT computed via Algorithm \RomanNumeral{2}. Top row: Input signals in the spatial domain $(x, y)$. Bottom row: Corresponding output amplitudes in the transform domain $(u, v)$. (a, d) Gaussian signal exhibiting rotation and reshaping. (b, e) Rectangular aperture exhibiting skewed diffraction patterns. (c, f) Cross-shaped aperture exhibiting complex coupled interference.}
	\label{fig:transform_results}
\end{figure}

The amplitude distributions of the input signals and their corresponding 2D NSFRFT outputs are presented in Fig. \ref{fig:transform_results}.

\begin{itemize}
	\item \textbf{Gaussian Signal (Figs. \ref{fig:transform_results}(a) and (d) ):} 
	The input is a circular Gaussian function with an isotropic energy distribution. The amplitude of the transformed signal (Fig. \ref{fig:transform_results}(d)) retains a Gaussian-like profile but exhibits a distinct geometric rotation and reshaping in the output $(u, v)$ plane. This visually confirms that the 2D NSFRFT acts as a generalized linear canonical operator, imparting quadratic phase modulation while strictly reorienting the energy distribution in the phase space.
	
	\item \textbf{Rectangular Aperture (Figs. \ref{fig:transform_results}(b) and (e)):} 
	The input is a square function with sharp boundaries. Unlike separable transforms (e.g., 2D SFRFT), which would produce orthogonal Sinc-function diffraction patterns aligned with the canonical axes, the 2D NSFRFT produces a skewed diffraction pattern (Fig. \ref{fig:transform_results}(e)). The diffraction lobes are rotated into non-orthogonal directions, exhibiting significant cross-term interference. This provides direct visual evidence of the spatial coupling capability of the proposed transform, demonstrating that spatial frequencies along the $x$ and $y$ axes are intrinsically mixed during the transformation.
	
	\item \textbf{Cross-shaped Aperture (Figs. \ref{fig:transform_results}(c) and  (f)):} 
	The output of the cross-shaped signal (Fig. \ref{fig:transform_results}(f)) displays a complex interference pattern where the diffraction arms of the two bars are rotated into non-orthogonal directions. This further validates the transform's ability to manipulate the geometry of composite structures, highlighting its potential for processing {complex wavefields characterized by} nonseparable structured interference.
\end{itemize}

\section{Applications}
\label{Section 5}

The proposed 2D NSFRFT not only generalizes the separable case but also introduces four additional degrees of freedom (DOF), unlocking unique capabilities for processing coupled multidimensional signals. In this section, we validate the practical utility of the transform in two key domains: signal filtering and medical image encryption.

We first focus on filter design, where the advantages of the nonseparable structure are most pronounced. We implement two distinct filtering paradigms: optimal filtering for general noise reduction, and specialized bandpass/bandstop filtering for separating coupled chirp signals. The latter represents a specific task where conventional 2D FRFTs with rotational interpretations (such as the 2D SFRFT, GT, and CFRFT) often fail, highlighting the unique advantage of the proposed method.

Subsequently, we apply the transform to double random phase encryption and decryption (DRPED) for medical images. We demonstrate that the expanded parameter space of the 2D NSFRFT significantly enhances the key space and security level, offering a robust solution for protecting sensitive medical data.

\subsection{Filter Design}
The multiplicative filter in the 2D NSFRFT domain is mathematically defined as
\begin{equation}
	\hat{f}(x, y) = (\mathcal{F}_{P})^{-1}\big[H(u,v)\mathcal{F}_{P}[f]\big](x,y),
	\label{eq:38}
\end{equation}
where $H(u,v)$ denotes the filter transfer function. Since the performance of this operation is strictly determined by the accuracy of the transform domain, the tractability of parameter estimation is a prerequisite for effective filtering. Therefore, before designing specific filters, we first demonstrate that the optimal parameters of the proposed transform can be located efficiently, avoiding the computational pitfalls common in higher-dimensional generalizations.

\subsubsection{Parameter Search Efficiency and Optimization Landscape}
A key practical advantage of the 2D NSFRFT over the general 2D NSLCT lies in its optimization tractability. While the general 2D NSLCT (corresponding to the symplectic group $\mathrm{Sp}(4, \mathbb{R})$) offers maximum flexibility with 10 DOFs, it suffers from the curse of dimensionality in blind parameter estimation tasks. In contrast, the 2D NSFRFT is constrained to the generalized rigid rotation subgroup with only 4 DOFs, striking a balance between flexibility and searchability.

To quantify this advantage, we conducted a Monte Carlo simulation comparing the parameter search efficiency of both transforms. The objective was to maximize the focusing quality (defined by the normalized peak-to-energy ratio) of a coupled 2D chirp signal $g(x,y) = \exp[\mathrm{j}(0.8x^2 + 0.6xy + 0.4y^2)] \cdot \exp(-\frac{x^2+y^2}{10})$. Both transforms utilized a random search strategy over 1,000 independent trials.

\begin{figure}[htp]
	\centering
	\subfloat[]{\includegraphics[width=0.8\textwidth]{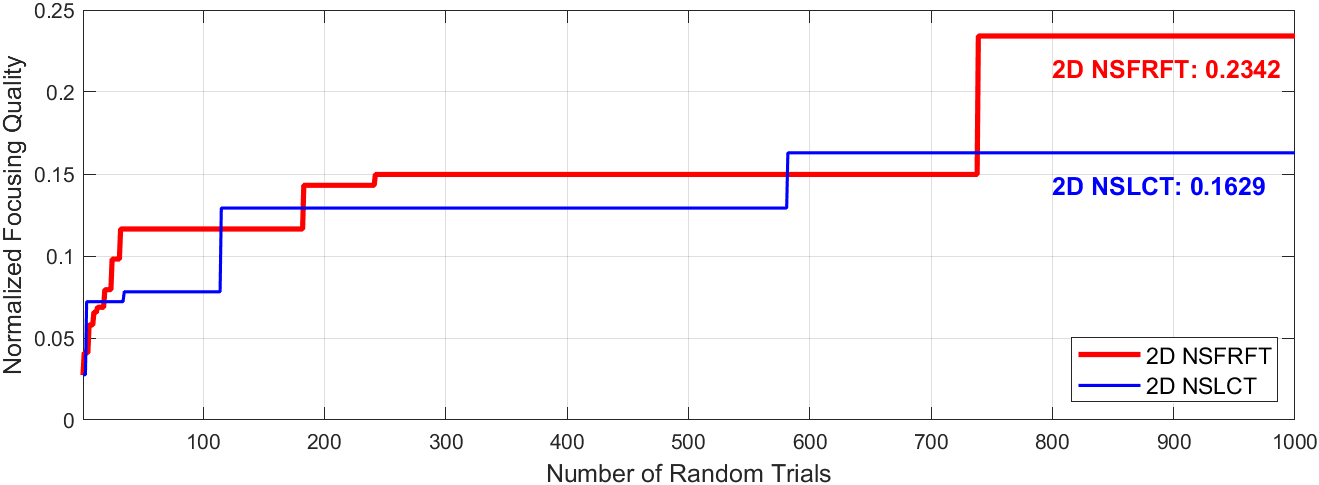}}\\
	\subfloat[2D NSFRFT Best Result \\ Score: 0.2342]{\includegraphics[width=0.4\textwidth]{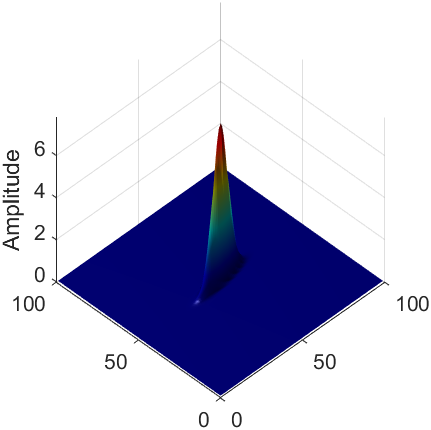}}
	\subfloat[2D NSLCT Best Result \\
	Score: 0.1629]{\includegraphics[width=0.4\textwidth]{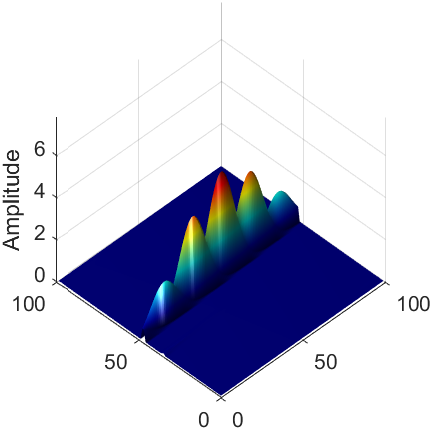}}
	\caption{Comparison of parameter search efficiency and focusing capability between the proposed 2D NSFRFT (4 DOFs) and the general 2D NSLCT (10 DOFs). (a) Convergence curves over 1,000 Monte Carlo trials. The proposed transform (Red) converges to a higher normalized score, while the general baseline (Blue) suffers from the curse of dimensionality. (b) and (c) 3D visualization of the best transformation results. The 2D NSFRFT achieves a sharp, single-peak impulse (perfect decoupling), whereas the 2D NSLCT yields a dispersed ridge structure (partial focusing), demonstrating the specific advantage of the proposed constrained framework.}
	\label{fig:5.2.1}
\end{figure}

Fig. \ref{fig:5.2.1} illustrates the results. The comparison reveals two critical findings:
\begin{itemize}
	\item \textbf{Search Efficiency:} The 2D NSFRFT (Red curve) exhibits rapid convergence, identifying a global optimum with a high focusing score ($\approx 0.2342$) within limited iterations. The general 2D NSLCT (Blue curve), despite its broader theoretical solution space, fails to converge effectively, stagnating at a significantly lower score ($\approx 0.1629$). Theoretical analysis suggests that this is because the parameter space of the 2D NSFRFT is geometrically compact (comprising the unit hypersphere $S^3$ and a bounded angular interval), whereas the 2D NSLCT operates in an unbounded non-compact space, making global search significantly more difficult.
	\item \textbf{Focusing Quality:} The 3D magnitude distributions of the best-found solutions demonstrate the physical difference. The proposed 2D NSFRFT result Fig. \ref{fig:5.2.1}(b) forms a sharp, solitary impulse, indicating that the coupled quadratic phase terms have been perfectly compensated. Conversely, the 2D NSLCT result Fig. \ref{fig:5.2.1}(c) forms a dispersed energy ridge. This ridge artifact suggests that the high-dimensional optimization algorithm became trapped in a local minimum, achieving only partial focusing without full decoupling.
\end{itemize}
Therefore, while the 2D NSLCT is more general, the proposed 2D NSFRFT offers a specific advantage in engineering applications by providing a search-friendly, geometrically consistent parameter space that ensures robust signal decoupling.

\subsubsection{Optimal Filter Design}
Consider the observation model $f(x,y) = g(x,y) + n(x,y)$, where $g$ is the desired signal and $n$ is additive noise. Assuming known correlation functions ($R_{gf}, R_{gg}, R_{ff}$), the optimal filter in the 2D NSFRFT domain minimizes the MSE. Due to the orthogonality of the transform, the optimal transfer function $H_{\text{opt}}(u, v)$ is derived as \cite{12}
\begin{equation}
H_{\text{opt}}(u, v) = 
\frac{R_{{G_{P}} {F_{P}}}(u, v, u, v)}{R_{{F_{P}} {F_{P}}}(u, v, u, v)},
\label{eq:39}
\end{equation}
where \( R_{G_P F_P} \) and \( R_{F_P F_P} \) denote the cross-correlation and auto-correlation functions in the 2D NSFRFT domain, respectively, defined by
\begin{align}
R_{{G_{P}} {F_{P}}}(u, v, u, v) &= 
\int_{\mathbb{R}^4} K_{P}(x, y, u, v) K^*_{P}(\sigma, \tau, u, v) R_{gf}(x, y, \sigma, \tau) \, \mathrm{d}x \, \mathrm{d}y \, \mathrm{d}\sigma \, \mathrm{d}\tau, \label{eq:40} \\
R_{{F_{P}} {F_{P}}}(u, v, u, v) &= 
\int_{\mathbb{R}^4} K_{P}(x, y, u, v)  K^*_{P}(\sigma, \tau, u, v) R_{ff}(x, y, \sigma, \tau) \, \mathrm{d}x \, \mathrm{d}y \, \mathrm{d}\sigma \, \mathrm{d}\tau, \label{eq:41}
\end{align}
and the MSE for the optimal filter is
\begin{equation}
\begin{aligned}
\mathrm{MSE} = \int_{\mathbb{R}^2} &\; R_{{G_{P}} {G_{P}}}(u, v, u, v)- 2 \, \operatorname{Re} \left( H_{\text{opt}}^*(u, v) R_{{F_{P}} {G_{P}}}(u, v, u, v) \right) + \left| H_{\text{opt}}(u, v) \right|^2 R_{{F_{P}} {F_{P}}}(u, v, u, v) \, \mathrm{d}u \, \mathrm{d}v,
\end{aligned}
\label{eq:42}
\end{equation}
where 
\begin{equation}
\begin{aligned}
R_{{G_{P}} {G_{P}}}(u, v, u, v) &= 
\int_{\mathbb{R}^4} K_{P}(x, y, u, v)  K^*_{P}(\sigma, \tau, u, v) R_{gg}(x, y, \sigma, \tau) \, \mathrm{d}x \, \mathrm{d}y \, \mathrm{d}\sigma \, \mathrm{d}\tau.
\end{aligned}
\label{eq:43}
\end{equation}
In practice, the optimal parameter set $P_{\text{opt}}$ minimizing Eq.~(\ref{eq:42}) is determined using a genetic algorithm (GA) to enhance the global search capability and mitigate the tendency to converge to local minima.  The design procedure is summarized in Algorithm~\ref{alg:opt-filter}.

\begin{algorithm}[htbp]
\caption{Optimal Filter Design in the 2D NSFRFT Domain}
\label{alg:opt-filter}
\begin{algorithmic}[1]
\Require Observed signal \( f(x,y) \), known correlation functions \( R_{gf}, R_{gg}, R_{ff} \)
\Ensure Optimal filter \( H_{\text{opt}}(u,v) \) and optimal parameter set \( P_{\text{opt}} = (a,b,c,d,\theta)_{\text{opt}} \)

\For{each candidate parameter set \( P = (a,b,c,d,\theta) \)}
    \State Compute 2D NSFRFT kernel \( K_P(x,y,u,v) \)
    \State Calculate correlation functions and construct \( H_{\text{opt}}(u,v) \) using Eqs.~(\ref{eq:39})--(\ref{eq:41})
    \State Evaluate MSE using Eq.~(\ref{eq:42})
    \If{current MSE $<$ minimum recorded MSE}
        \State Update \( P_{\text{opt}} \leftarrow P \)
        \State Store current \( H_{\text{opt}}(u,v) \)
    \EndIf
\EndFor
\State \textbf{return} \( H_{\text{opt}}(u,v) \), \( P_{\text{opt}} \), MSE
\end{algorithmic}
\end{algorithm}

We evaluate the performance using three metrics: MSE (log-scale), structural similarity index measure (SSIM), and peak signal-to-noise ratio (PSNR). Comparisons are made against the 2D FT, 2D SFRFT, GT, and CFRFT.

\textbf{1) Synthetic Gaussian Signals} \\
We test two scenarios: a separable signal $g_1=\exp\left[{-(x^2+y^2)}\right]$ and a coupled signal $g_2=\exp\left[{-(x^2+1.5xy+y^2)}\right]$, both corrupted by white Gaussian noise.
The results across various SNRs are detailed in Table \ref{tab:6} ($g_1$) and Table \ref{tab:7} ($g_2$).
\begin{itemize}
    \item Observation: The 2D NSFRFT-based filter consistently achieves the lowest MSE and highest PSNR.
    \item Analysis: The advantage is particularly evident for $g_2$, where the coupled term $1.5xy$ challenges separable transforms. The 2D NSFRFT utilizes its cross-parameters to better compact the signal energy, validating its superiority in handling nonseparable terms.
\end{itemize}

\begin{table*}[h]
\renewcommand{\arraystretch}{2.1}
\centering
\caption{Optimal filtering performance for $g_1$ in different transform domains under various SNRs}
\resizebox{\textwidth}{!}{
    \begin{tabular}{|c|c|c|c|c|c|c|c|}
    \hline
     \multicolumn{1}{|c|}{\multirow{2}[2]{*}{\text{Transform}}} & \multicolumn{1}{c}{ \makecell[c]{MSE / PSNR \\ (Optimal parameters)}} & \multicolumn{6}{c|}{SNR} \\
\cline{2-8}
& $-15$ & $-10$ & $-5$ & $0$ & $5$ & $10$ & $15$ \\ 
\hline
{\text{FT}}    & $-3.4353 / 34.3586$ & $-3.7328  / 37.3480$ & $-4.0858  / 40.9039$ & $-4.4749  / 44.8011$ & $-4.8870  / 48.8938$ & $-5.3140  / 53.1459$ & $-5.7541  / 57.5504$ \\
\hline
\multirow{2}{*}{\text{2D SFRFT}} & $-3.8918 / 40.6031$ & $-4.2336 / 43.6547$ & $-4.6013 / 47.1122$ & $-4.9589 / 50.2362$ & $-5.3797 / 54.5295$ & $ -5.8127 / 58.7046$ & $-6.2568 / 64.0169$ \\
 & $(0.0182, -3.1234)$ & $(-0.0182, 3.1234)$ & $(0.0182, -3.1234)$ & $(-0.0126, 3.1290)$ & $(-0.0199, 0.0199)$ & $(-0.0169, 0.0169) $ & $(-0.0731, 0.0731)$ \\ \hline
\multirow{2}{*}{\text{GT}} & $-3.8331 / 41.2619$ & $-4.1792 / 44.0800$ & $-4.5579 / 47.4684$ & $-4.9581 / 49.9247$ & $-5.3788 / 54.1122$ & $-5.8122 / 58.4178$ & $-6.2560 / 62.7937$ \\
 & $(0.0755)$ & $(-0.0629)$ & $(-0.0619)$ & $(0.0005)$ & $(0.0002)$ & $(-0.0007)$ & $(-0.0010)$ \\ 
 \hline
\multirow{2}{*}{\text{CFRFT}} & $-3.8278 / 39.0811$ & $-4.1785 / 44.1336$ & $-4.5534 / 45.9228$ & $-4.9064 / 49.8208$ &$ -5.3482 / 54.6990$ & $-5.8318 / 59.5837$ & $-6.2724 / 63.8273$ \\
 & $(-2.6833, 2.6779)$ & $(1.2614, -1.0307)$ & $(2.0792, -2.0796)$ & $(1.7481, -2.3325)$ & $(1.4409, -1.7960)$ & $(-0.4212, 0.5674)$ & $(0.2855, -0.2082)$ \\ 
 \hline
\multirow{3}{*}{\text{2D NSFRFT}} 
& $\mathbf{-4.0156 / 42.2726}$ 
& $\mathbf{-4.2940 / 44.2647}$ 
& $\mathbf{-4.8726 / 49.9987}$ 
& $\mathbf{-5.0204 / 50.9161}$ 
& $\mathbf{-5.4116 / 55.5659}$ 
& $\mathbf{-5.8449 / 59.7583}$ 
& $\mathbf{-6.2762 / 64.2091}$ \\
& \begin{tabular}[c]{@{}c@{}}$(\mathbf{-0.0029}, \mathbf{-0.0068},$\\ $\mathbf{0.0008}, \mathbf{0.9999}, \mathbf{1.5774})$\end{tabular}
& \begin{tabular}[c]{@{}c@{}}$(\mathbf{0.0059}, \mathbf{-0.0056},$\\ $\mathbf{-0.0037}, \mathbf{0.9999}, \mathbf{4.7181})$\end{tabular}
& \begin{tabular}[c]{@{}c@{}}$(\mathbf{-0.0483}, \mathbf{0.0006},$\\ $\mathbf{-0.9979}, \mathbf{-0.0442}, \mathbf{4.7329})$\end{tabular}
& \begin{tabular}[c]{@{}c@{}}$(\mathbf{0.0000}, \mathbf{0.0000},$\\ $\mathbf{-0.9999}, \mathbf{0.0029}, \mathbf{4.6973})$\end{tabular}
& \begin{tabular}[c]{@{}c@{}}$(\mathbf{0.1087}, \mathbf{0.0417},$\\ $\mathbf{0.0487}, \mathbf{0.9920}, \mathbf{1.4767})$\end{tabular}
& \begin{tabular}[c]{@{}c@{}}$(\mathbf{0.1092}, \mathbf{0.0413},$\\ $\mathbf{0.0488}, \mathbf{0.9920}, \mathbf{1.4768})$\end{tabular}
& \begin{tabular}[c]{@{}c@{}}$(\mathbf{0.9401}, \mathbf{-0.3196},$\\ $\mathbf{0.1181}, \mathbf{0.0078}, \mathbf{6.2832})$\end{tabular} \\
\hline
\end{tabular}
}
\label{tab:6}
\end{table*}

\begin{table*}[h]
\renewcommand{\arraystretch}{2.1}
\centering
\caption{Optimal filtering performance for $g_2$ in different transform domains under various SNRs}
\resizebox{\textwidth}{!}{
    \begin{tabular}{|c|c|c|c|c|c|c|c|}
    \hline
     \multicolumn{1}{|c|}{\multirow{2}[2]{*}{\text{Transform}}} & \multicolumn{1}{c}{ \makecell[c]{MSE / PSNR \\ (Optimal parameters)}} & \multicolumn{6}{c|}{SNR} \\
\cline{2-8}
& $-15$ & $-10$ & $-5$ & $0$ & $5$ & $10$ & $15$ \\ 
\hline
{\text{FT}}    & $-3.3981 / 33.9949 $ & $-3.7100  / 37.1453$ & $-4.0643  / 40.7306$ & $-4.4554  / 44.6337$ & $-4.8721  / 48.7444$ & $-5.3038  / 53.0488$ & $-5.7449  / 57.4687$ \\
\hline
\multirow{2}{*}{\text{2D SFRFT}} & $-3.5313 / 36.6216$ & $-3.8531 / 39.8226$ & $-4.2173 / 43.4839$ & $-4.6215 / 47.5699$ & $-5.0483 / 51.7635$ & $-5.4855 / 56.0606$ & $-5.9309 / 60.4652$ \\
 & $(0.2841, -0.2841)$ & $(0.2808, -0.2808)$ & $(-0.2670, 0.2670)$ & $(0.2420, -0.2420)$ & $(0.2395, -0.2395)$ & $(-0.2360, 0.2360)$ & $(-0.2295, 0.2295)$ \\ 
 \hline
\multirow{2}{*}{\text{GT}} & $-3.4985 / 35.8262$ & $-3.8308 / 38.9588$ & $-4.2034 / 42.5642$ & $-4.6034 / 47.9651$ & $-5.0256 / 51.8688$ & $-5.4600 / 55.9760$ & $-5.9050 / 60.1697$ \\
 & $(-0.0005)$ & $(-0.0010)$ & $(-0.0011)$ & $(0.0760)$ & $(0.0775)$ & $(0.0784)$ & $(0.0713)$ \\ 
 \hline
\multirow{2}{*}{\text{CFRFT}} & $-3.5290 / 37.4070$ & $-3.8589 / 40.0224$ & $-4.2309 / 43.5807$ & $-4.6261 / 47.5823$ &$ -5.0433 / 51.6072$ & $-5.4730 / 55.6951$ & $-5.9162 / 60.0184$ \\
 & $(-0.2025, 0.2766)$ & $(0.2597, -0.1859)$ & $(0.2599, -0.1861)$ & $(0.2565, -0.1474)$ & $(0.2531, -0.1457)$ & $(0.2220, -0.1304)$ & $(0.2477, -0.1431)$ \\ 
 \hline
\multirow{3}{*}{\text{2D NSFRFT}} 
& $\mathbf{-3.8160 / 38.5729}$ 
& $\mathbf{-4.1812 / 42.1354}$ 
& $\mathbf{-4.5447 / 46.1486}$ 
& $\mathbf{-4.9270 / 49.6359}$ 
& $\mathbf{-5.3661 / 54.2214}$ 
& $\mathbf{-5.7803 / 58.4129}$ 
& $\mathbf{-6.2273 / 62.8833}$ \\
& \begin{tabular}[c]{@{}c@{}}$(\mathbf{-0.7117}, \mathbf{0.0033},$\\ $\mathbf{0.0207}, \mathbf{-0.7021}, \mathbf{2.3912})$\end{tabular}
& \begin{tabular}[c]{@{}c@{}}$(\mathbf{-0.0797}, \mathbf{0.6987},$\\ $\mathbf{-0.7059}, \mathbf{0.0846}, \mathbf{3.9318})$\end{tabular}
& \begin{tabular}[c]{@{}c@{}}$(\mathbf{-0.0550}, \mathbf{0.6788},$\\ $\mathbf{-0.7319}, \mathbf{-0.0227}, \mathbf{4.0098})$\end{tabular}
& \begin{tabular}[c]{@{}c@{}}$(\mathbf{-0.1536}, \mathbf{0.6829},$\\ $\mathbf{-0.6916}, \mathbf{0.1780}, \mathbf{3.9132})$\end{tabular}
& \begin{tabular}[c]{@{}c@{}}$(\mathbf{-0.0680}, \mathbf{0.6652},$\\ $\mathbf{-0.7430}, \mathbf{-0.0277}, \mathbf{4.0036})$\end{tabular}
& \begin{tabular}[c]{@{}c@{}}$(\mathbf{-0.1954}, \mathbf{0.6050},$\\ $\mathbf{-0.7584}, \mathbf{0.1435}, \mathbf{4.0417})$\end{tabular}
& \begin{tabular}[c]{@{}c@{}}$(\mathbf{-0.1627}, \mathbf{0.6338},$\\ $\mathbf{-0.7417}, \mathbf{0.1474}, \mathbf{4.0633})$\end{tabular} \\
\hline
\end{tabular}
}
\label{tab:7}
\end{table*}

\textbf{2) Simulated Newton Fringe Patterns} 

Newton rings are modeled as $I(x,y) = I_0 + A \cos[\varphi(x,y)]$, with phase $\varphi(x,y) = \frac{2\pi}{\lambda R} [(x - x_0)^2 + (y - y_0)^2] + \pi$.
Parameters are set to $I_0 = 2, A = 2, \lambda = 600 \text{nm}, R = 0.4$. Table \ref{tab:8} summarizes the denoising results.
\begin{itemize}
    \item Observation: The proposed method yields the highest SSIM scores.
    \item Analysis: The high SSIM indicates superior preservation of the delicate fringe structures. The flexible rotational capability of the 2D NSFRFT adapts to the curvature of the rings more accurately than rigid geometries like the GT.
\end{itemize}

\begin{table*}[h]
\renewcommand{\arraystretch}{2.2}
\centering
\caption{Optimal filtering performance for simulated newton fringe patterns in different transform domains under various SNRs}
\resizebox{\textwidth}{!}{
    \begin{tabular}{|c|c|c|c|c|c|c|c|}
    \hline
     \multicolumn{1}{|c|}{\multirow{2}[2]{*}{\text{Transform}}} & \multicolumn{1}{c}{ \makecell[c]{MSE / PSNR / SSIM \\ (Optimal parameters)}} & \multicolumn{6}{c|}{SNR} \\
\cline{2-8}
& $-15$ & $-10$ & $-5$ & $0$ & $5$ & $10$ & $15$ \\ 
\hline
{\text{FT}}    & $-0.1047  / 1.0775 / 0.5034$ & $-0.3649  / 3.7245  / 0.6443$ & $-0.6780 / 6.8984 / 0.7592$ & $-1.0209 / 10.3504  / 0.8460$ & $-1.3776 /  13.9087  / 0.9053$ & $ -1.7390 / 17.5144 / 0.9433$ & $-2.1082 / 21.1903 / 0.9665$ \\
\hline
\multirow{2}{*}{\text{2D SFRFT}} & $-0.1407 / 1.4778 / 0.5170$ & $-0.4213 /  4.2745 / 0.6584$ & $-0.6978 / 7.1205 / 0.7645$ & $-1.0347 / 10.4952 / 0.8481$ & $-1.3933 / 14.0603 / 0.9059$ & $-1.7563  / 17.6698 / 0.9457$ & $-2.1243 / 21.3455 / 0.9682$ \\
 & $(1.5729, 1.5848)$ & $(1.5692, 1.6790)$ & $(1.5725, 1.5830)$ & $(1.5674, 1.5611)$ & $(1.5653, 1.5660)$ & $(1.5651, 1.5662)$ & $(1.5620, 1.5833)$ \\ 
 \hline
\multirow{2}{*}{\text{GT}} & $-0.1229  / 1.2737 / 0.5189$ & $-0.3791 / 3.8713 / 0.6526$ & $-0.6856 / 6.9696 / 0.7623$ & $-1.0219 / 10.3488 / 0.8460$ & $-1.3731 / 13.8587 / 0.9039 $ & $-1.7327 / 17.4532 / 0.9425$ & $-2.1114  / 21.2195 / 0.9672$ \\
 & $(1.5699)$ & $(1.5706)$ & $(1.5699)$ & $(1.5698)$ & $(1.5694)$ & $(-1.5708)$ & $(-1.5703)$ \\ 
 \hline
\multirow{2}{*}{\text{CFRFT}} & $-0.1522 / 1.6037 / 0.5341$ & $-0.4224 / 4.4022 / 0.6806$ & $-0.7257 / 7.3949 / 0.7765$ & $-1.0281 / 10.4210 / 0.8485$ &$ -1.3923 / 14.0512 / 0.9057$ & $-1.7523 / 17.6367 / 0.9445$ & $-2.1235 / 21.3377 / 0.9683$ \\
 & $(1.5876, 1.5852)$ & $(1.5912, 1.5909)$ & $(1.5637, 1.5638)$ & $(1.5630, 1.5630)$ & $(1.5666, 1.5666)$ & $(1.5630, 1.5630)$ & $(1.5632, 1.5628)$ \\ 
 \hline
\multirow{3}{*}{\text{2D NSFRFT}} 
& $\mathbf{-0.1654 / 1.7150  /0.5349}$ 
& $\mathbf{-0.5355 / 5.4978  / 0.7318}$ 
& $\mathbf{-0.7374 / 7.5119 / 0.7806}$ 
& $\mathbf{-1.0515 / 10.6716 / 0.8501}$ 
& $\mathbf{-1.4005 / 14.1423 / 0.9067}$ 
& $\mathbf{-1.7566 / 17.6766 / 0.9461}$ 
& $\mathbf{-2.1263 / 21.3642 / 0.9684}$ \\
& \begin{tabular}[c]{@{}c@{}}$(\mathbf{0.9999}, \mathbf{-0.0008},$\\ $\mathbf{-0.0064}, \mathbf{-0.0001}, \mathbf{1.5794})$\end{tabular}
& \begin{tabular}[c]{@{}c@{}}$(\mathbf{0.9999}, \mathbf{0.0004},$\\ $\mathbf{-0.0024}, \mathbf{0}, \mathbf{1.5804})$\end{tabular}
& \begin{tabular}[c]{@{}c@{}}$(\mathbf{0.9999}, \mathbf{-0.0005},$\\ $\mathbf{-0.0066}, \mathbf{0.0001}, \mathbf{1.5754})$\end{tabular}
& \begin{tabular}[c]{@{}c@{}}$(\mathbf{0.9999}, \mathbf{-0.0001},$\\ $\mathbf{0.0028}, \mathbf{0}, \mathbf{1.5637})$\end{tabular}
& \begin{tabular}[c]{@{}c@{}}$(\mathbf{0.9999}, \mathbf{-0.0003},$\\ $\mathbf{0.0030}, \mathbf{0}, \mathbf{1.5637})$\end{tabular}
& \begin{tabular}[c]{@{}c@{}}$(\mathbf{0.9999}, \mathbf{0.0001},$\\ $\mathbf{-0.0103}, \mathbf{-0.0001}, \mathbf{1.5724})$\end{tabular}
& \begin{tabular}[c]{@{}c@{}}$(\mathbf{0.9999}, \mathbf{-0.0001},$\\ $\mathbf{-0.0114}, \mathbf{-0.0001}, \mathbf{1.5725})$\end{tabular} \\
\hline
\end{tabular}
}
\label{tab:8}
\end{table*}

\textbf{3) Real-World Image Denoising} 

We tested images from six public datasets (e.g., SSID, O\_HAZE). An optimal filter was applied individually to RGB channels.
Table \ref{tab:9} shows that while performance gains on natural images are modest (due to the lack of specific nonseparable couplings), the 2D NSFRFT achieves consistent improvements across all datasets. This confirms its robustness as a generalized filtering tool. It is worth noting that the specific advantage of the proposed transform becomes much more pronounced when handling signals with structured nonseparable interference, as demonstrated in the bandstop filtering experiments in Section \ref{section 5.1.3}.

\begin{table*}[htbp]
\renewcommand{\arraystretch}{2.2}
  \centering
  \caption{Denoising performance for real-world images from different datasets in various transform domains}
  \resizebox{\textwidth}{!}{
    \begin{tabular}{|c|c|c|c|c|c|c|c|}
    \hline
    \multicolumn{1}{|c|}{\multirow{2}[2]{*}{Dataset}}  & \multicolumn{1}{c|}{\multirow{2}[2]{*}{Index}}  & \multicolumn{6}{c|}{MSE / PSNR / SSIM} \\
    \cline{3-8}
          &       & \multicolumn{1}{c|}{\centering \makecell[c]{Noisy Input}}  & \multicolumn{1}{c|}{FT} & \multicolumn{1}{c|}{2D SFRFT} & \multicolumn{1}{c|}{GT} & \multicolumn{1}{c|}{CFRFT} & \multicolumn{1}{c|}{2D NSFRFT} \\
    \hline
    \multirow{2}{*}{SSID} & $48$    & $-1.4381 / 14.3806 / 0.5981$ & $-2.4091 / 24.0907 / 0.7583$ &  $-2.4236 / 24.2357 / 0.7641$  &  $-2.3549 / 23.5491 / 0.7225$  &  $-2.4204 / 24.2037 / 0.7600$  & $ \mathbf{ -2.4236 / 24.2364 / 0.7642} $ \\
    \cline{2-8}
          &    $165$   &  $-1.8268 / 18.2681 / 0.2656$  &  $-2.8460 / 28.4603 / 0.8421$  &  $-2.8520 / 28.5202 / 0.8457$  &  $-2.7782 / 27.7822 / 0.8095 $ &  $-2.8537 / 28.5375 / 0.8453$  &  $\mathbf{ -2.8575 / 28.5746 / 0.8461}$ \\
    \hline
        \multirow{2}{*}{O\_HAZE} & $02 $   & $-1.4381 / 14.3806 / 0.5981$ & $-1.4169 / 14.1694 / 0.4858 $ &  $-2.0644 / 20.6435 / 0.8512$  &  $-2.0541 / 20.5410 / 0.8333$  &  $-2.0626 / 20.6263 / 0.8480$  & $ \mathbf{ -2.0645 / 20.6449 / 0.8513}$ \\
    \cline{2-8}
          &   $45$    &  $-1.2334 / 12.3337 / 0.2616$  &  $-1.2237 / 12.2367 / 0.2092$  &  $-2.5494 / 25.4939 / 0.8811$  &  $-2.5160 / 25.1601 / 0.8507$ &  $-2.5493 / 25.4935 / 0.8811$  &  $\mathbf{ -2.5506 / 25.5059 / 0.8814}$ \\
    \hline    \multirow{2}{*}{I\_HAZE} & $01$    & $-1.3918 / 13.9185 / 0.6265$ & $-1.3818 / 13.8181 / 0.5812$ &  $-2.1295 / 21.2950 / 0.8859$  &  $-2.0993 / 20.9927 / 0.8541$  &  $-2.1294 / 21.2936 / 0.8859$  & $ \mathbf{ -2.1523 / 21.5228 / 0.8873}$ \\
    \cline{2-8}
          &   $03$    &  $-1.5540 / 15.5395 / 0.4369$  &  $-1.5396 / 15.3957 / 0.3383$  &  $-2.1883 / 21.8826 / 0.8792$  &  $-2.1605 / 21.6045 / 0.8402 $ &  $-2.1882 / 21.8821 / 0.8793$  &  $\mathbf{ -2.1894 / 21.8937 / 0.8798}$ \\
    \hline    \multirow{2}{*}{OTS\_BETA} & $25$    & $-1.1757 / 11.7567 / 0.6017$ & $-1.1614 / 11.6144 / 0.5127$ &  $-2.4626 / 24.6255 / 0.8825$  &  $-2.4271 / 24.2709 / 0.8284$  &  $-2.4737 / 24.7370 / 0.8754 $  & $ \mathbf{ -2.5682 / 25.6818 / 0.9223}$ \\
    \cline{2-8}
          &    $58$   &  $-1.3298 / 13.2984 / 0.6882$  &  $-1.2972 / 12.9716 / 0.5306$  &  $-2.6792 / 26.7922 / 0.8973$  &  $-2.6054 / 26.0542 / 0.8601 $ &  $-2.6114 / 26.1139 / 0.8486$  &  $\mathbf{ -2.7838 / 27.8383 / 0.9486}$ \\
    \hline    \multirow{2}{*}{ITS\_V2} & $02$    & $-0.9895 / 9.8952 / 0.6692$  & $-0.9822 / 9.8220 / 0.5808 $ &  $-2.5187 / 25.1867 / 0.8708$  &  $-2.6329 / 26.3286 / 0.9169$  &  $-2.6327 / 26.3270 / 0.9166$  & $ \mathbf{ -2.6348 / 26.3484 / 0.9183} $ \\
    \cline{2-8}
          &   $04$    &  $-1.7656 / 17.6560 / 0.8159$  &  $-1.7527 / 17.5271 / 0.7940$  &  $-2.1703 / 21.7035 / 0.8777$  &  $-2.3566 / 23.5657 / 0.9137$ &  $-2.3927 / 23.9267 / 0.9346$  &  $\mathbf{ -2.4001 / 24.0008 / 0.9347} $ \\
    \hline    \multirow{2}{*}{Rain\_100H} & $08$    & $-0.9843 / 9.8433 / 0.3517$ & $-1.0543 / 10.5433 / 0.6086$ &  $-2.1245 / 21.2448 / 0.6915$  &  $-2.1438 / 21.4377 / 0.7185$  &  $-2.1559 / 21.5585 / 0.7266$  & $ \mathbf{ -2.1623 / 21.6228 / 0.7300} $ \\
    \cline{2-8}
          &   $77$    &  $-1.0006 / 10.0059 / 0.4461$  &  $-1.0727 / 10.7270 / 0.5617$  &  $-1.8512 / 18.5119 / 0.7503$  &  $-1.8780 / 18.7800 / 0.7624 $ &  $-1.8984 / 18.9840 / 0.7837$  &  $\mathbf{ -1.8988 / 18.9880 / 0.7840}$ \\
    \hline
    \end{tabular}
    }
  \label{tab:9}
\end{table*}

\subsubsection{Passband and Stopband Filters for Coupled Signals}
\label{section 5.1.3}

A distinct advantage of the 2D NSFRFT is its ability to focus 2D chirp signals with nonseparable terms, a capability lacking in the 2D FT, SFRFT, GT, and CFRFT. Since such signals are common in radar and interferometry, we design specialized passband (signal extraction) and stopband (artifact removal) filters. The specific filtering procedure is illustrated in Fig. \ref{fig:flowchart}. First, the optimal transform parameters are estimated to focus the target component into a compact impulse. Based on whether the focused component represents the signal of interest or an interference artifact, a multiplicative binary mask $H(u,v)$ (bandpass or bandstop) with bandwidth $B$ is designed in the transform domain. Finally, the filtered signal is recovered via the inverse 2D NSFRFT.

\begin{figure}[htbp]
	\centering
	% 使用 resizebox 确保图表自动适应栏宽，避免溢出
	\resizebox{0.8\textwidth}{!}{%
		\begin{tikzpicture}[node distance=2cm]
			
			% 定义样式
			\tikzstyle{startstop} = [rectangle, rounded corners, minimum width=5cm, minimum height=1cm,text centered, draw=black, fill=red!30]
			\tikzstyle{io} = [trapezium, trapezium left angle=70, trapezium right angle=110,
			trapezium stretches body,  % 让梯形宽度跟随 minimum width
			minimum width=5cm,       % 固定梯形整体宽度（可按需调整）
			text width=5cm,            % 文字区域宽度，确保居中
			minimum height=1cm, text centered, draw=black, fill=blue!30]
			\tikzstyle{process} = [rectangle, 
			minimum width=5cm,  % 与梯形(io)宽度一致，视觉更协调
			text width=4.8cm,   % 固定文字区域宽度（略小于最小宽度，留边距）
			minimum height=1cm, 
			text centered, 
			draw=black, 
			fill=green!30,
			align=center]       % 文字自动居中换行，适配固定宽度
			\tikzstyle{decision} = [diamond, minimum width=3cm, minimum height=1cm, text centered, draw=black, fill=orange!30]
			\tikzstyle{arrow} = [thick,->,>=stealth]
			
			% 节点
			\node (start) [startstop] {Start};
			\node (in1) [io, below of=start] {Input Signal $f(x,y)$, Parameters $P$};
			\node (trans) [process, below of=in1] {Compute 2D NSFRFT $F_P(u,v)$};
			\node (peak) [process, below of=trans] {Search Impulse Location};
			\node (dec) [decision, below of=peak, yshift=-0.5cm] {Is Signal Useful?};
			\node (bp) [process, left of=dec, xshift=-3.5cm] {Design Bandpass Filter $H(u,v)$};
			\node (bs) [process, right of=dec, xshift=3.5cm] {Design Bandstop Filter $H(u,v)$};
			\node (apply) [process, below of=dec, yshift=-1.5cm] {Apply Filter $F_{filtered} = F_P \cdot H$};
			\node (inv) [process, below of=apply] {Inverse 2D NSFRFT};
			\node (out) [io, below of=inv] {Filtered Image $f_{filtered}(x,y)$};
			\node (stop) [startstop, below of=out] {End};
			
			% 连接线
			\draw [arrow] (start) -- (in1);
			\draw [arrow] (in1) -- (trans);
			\draw [arrow] (trans) -- (peak);
			\draw [arrow] (peak) -- (dec);
			\draw [arrow] (dec) -- node[anchor=south] {Yes} (bp);
			\draw [arrow] (dec) -- node[anchor=south] {No} (bs);
			\draw [arrow] (bp) |- (apply);
			\draw [arrow] (bs) |- (apply);
			\draw [arrow] (apply) -- (inv);
			\draw [arrow] (inv) -- (out);
			\draw [arrow] (out) -- (stop);
			
		\end{tikzpicture}%
	}
	\caption{Flowchart of the proposed filtering strategy based on 2D NSFRFT. Depending on whether the focused energy peak corresponds to the target signal or noise, a specific mask (bandpass or bandstop) is constructed in the $(u,v)$ domain.}
	\label{fig:flowchart}
\end{figure}

\textbf{1) Signal Extraction (Passband Filtering)}

We constructed two synthetic coupled chirps embedded in noise:
\begin{itemize}
    \item $f_1 = \exp(-j(0.866x^2 - 2xy + 0.866y^2))$ (Strong coupling, SNR=-5dB). $P_1=(\frac{1}{2},\frac{1}{2},\frac{1}{2},\frac{1}{2},\frac{\pi}{3})$.
    \item $f_2 = \exp(j(0.2897x^2+0.0061xy+0.2877y^2))$ (Weak coupling, SNR=-10dB). $P_2=(-0.0046,0.0016,-0.0241, 0.9997,\frac{\pi}{6})$.
\end{itemize}

\begin{figure*}[htbp]
    \centering
    % 第一排图像，没有标题
    \begin{minipage}{0.14\textwidth}
        \centering
        \includegraphics[width=\textwidth]{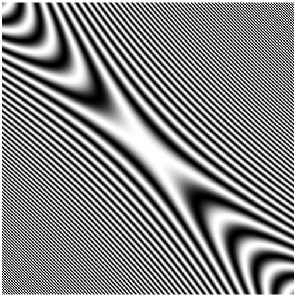} 
    \end{minipage}%
    \begin{minipage}{0.14\textwidth}
        \centering
        \includegraphics[width=\textwidth]{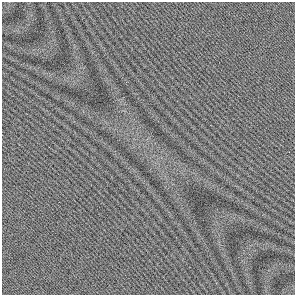}
    \end{minipage}%
    \begin{minipage}{0.14\textwidth}
        \centering
        \includegraphics[width=\textwidth]{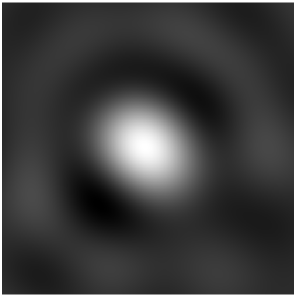} 
    \end{minipage}%
    \begin{minipage}{0.14\textwidth}
        \centering
        \includegraphics[width=\textwidth]{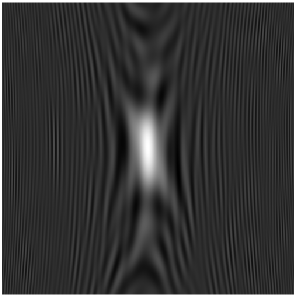} 
    \end{minipage}%
    \begin{minipage}{0.14\textwidth}
        \centering
        \includegraphics[width=\textwidth]{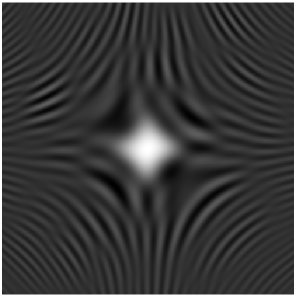}
    \end{minipage}%
    \begin{minipage}{0.14\textwidth}
        \centering
        \includegraphics[width=\textwidth]{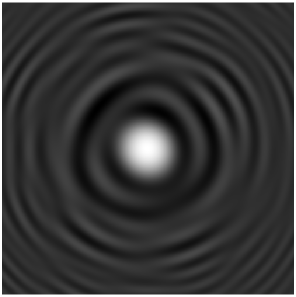}
    \end{minipage}%
    \begin{minipage}{0.14\textwidth}
        \centering
        \includegraphics[width=\textwidth]{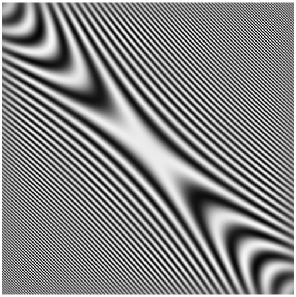}
    \end{minipage}
    
    \vspace{0.3cm} % 给两排之间增加垂直间距
    
    % 第二排图像
    \begin{minipage}{0.14\textwidth}
        \centering
        \includegraphics[width=\textwidth]{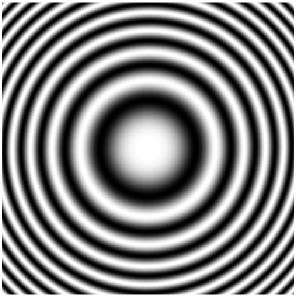} 
        \caption*{Original}
    \end{minipage}%
    \begin{minipage}{0.14\textwidth}
        \centering
        \includegraphics[width=\textwidth]{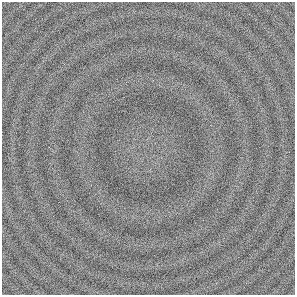}
        \caption*{Noisy}
    \end{minipage}%
    \begin{minipage}{0.14\textwidth}
        \centering
        \includegraphics[width=\textwidth]{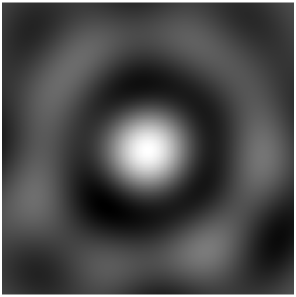} 
        \caption*{2D FT}
    \end{minipage}%
    \begin{minipage}{0.14\textwidth}
        \centering
        \includegraphics[width=\textwidth]{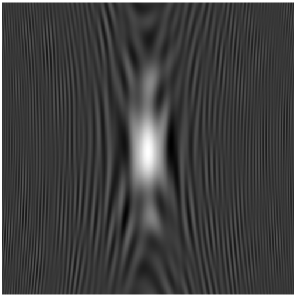} 
        \caption*{2D SFRFT}
    \end{minipage}%
    \begin{minipage}{0.14\textwidth}
        \centering
        \includegraphics[width=\textwidth]{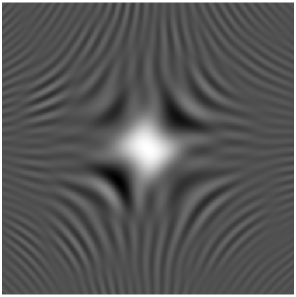}
        \caption*{GT}
    \end{minipage}%
    \begin{minipage}{0.14\textwidth}
        \centering
        \includegraphics[width=\textwidth]{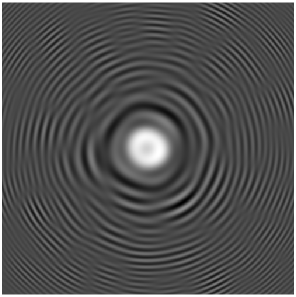}
        \caption*{CFRFT}
    \end{minipage}%
    \begin{minipage}{0.14\textwidth}
        \centering
        \includegraphics[width=\textwidth]{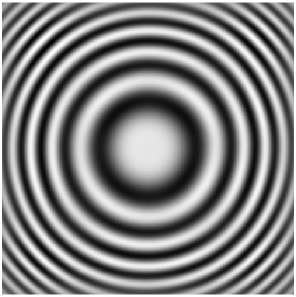}
        \caption*{2D NSFRFT}
    \end{minipage}
    
    \caption{Denoising synthetic 2D chirp signals with nonseparable terms in different transform domains.}
    \label{fig:10}
\end{figure*}

Results (Fig.~\ref{fig:10}): The 2D NSFRFT focuses these signals into sharp impulses, allowing easy extraction. Other transforms fail to focus the energy due to kernel mismatch, resulting in noisy reconstructions.

\textbf{2) Artifact Removal (Stopband Filtering)} 

Coupled chirps often appear as structured interference (artifacts). We simulated images corrupted by such artifacts ($f_3=\exp(j(0.2816x^2+0.1552xy+0.3064y^2+5.4319x-6.7898y)$ and $f_4=\exp(j(0.2639x^2+0.3552xy+0.2660y^2+6.7789x+0.0678y))$ with specific parameters $P_3=(-0.0692,0.0938,-0.8823,0.4560,\frac{\pi}{6})$ and $P_4=(0.1019,-0.2525,0.8944,-0.3548,\frac{\pi}{7})$).

\begin{figure}[htbp]
    \centering
    % 第一排图像，没有标题
    \begin{minipage}{0.14\textwidth}
        \centering
        \includegraphics[width=\textwidth]{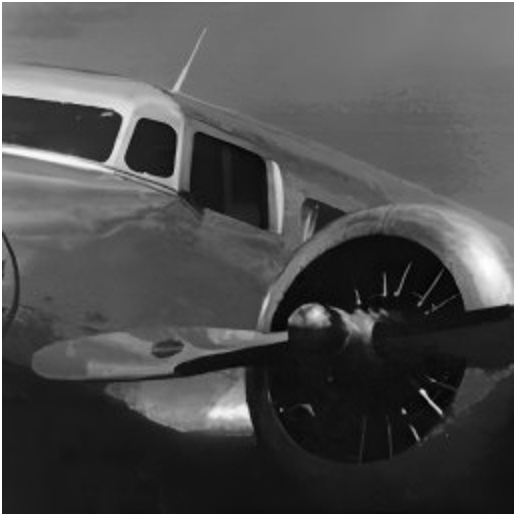} 
    \end{minipage}%
    \begin{minipage}{0.14\textwidth}
        \centering
        \includegraphics[width=\textwidth]{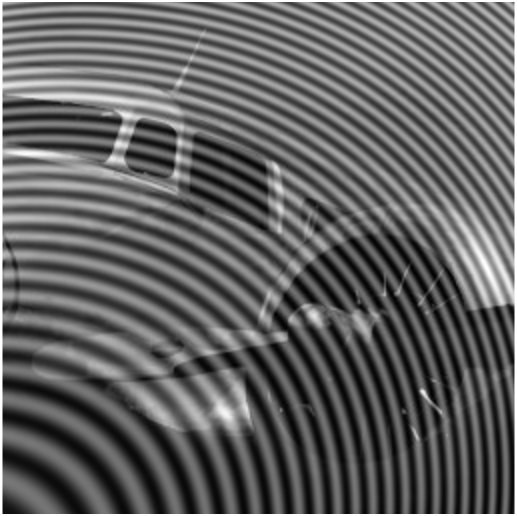}
    \end{minipage}%
    \begin{minipage}{0.14\textwidth}
        \centering
        \includegraphics[width=\textwidth]{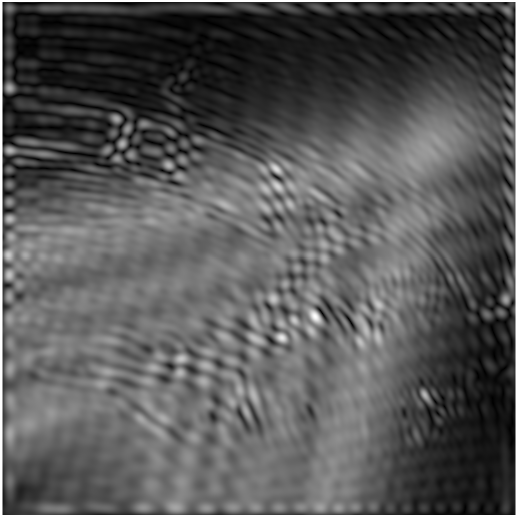} 
    \end{minipage}%
    \begin{minipage}{0.14\textwidth}
        \centering
        \includegraphics[width=\textwidth]{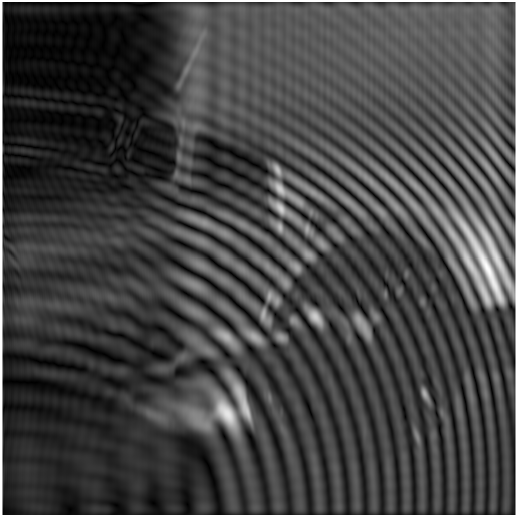} 
    \end{minipage}%
    \begin{minipage}{0.14\textwidth}
        \centering
        \includegraphics[width=\textwidth]{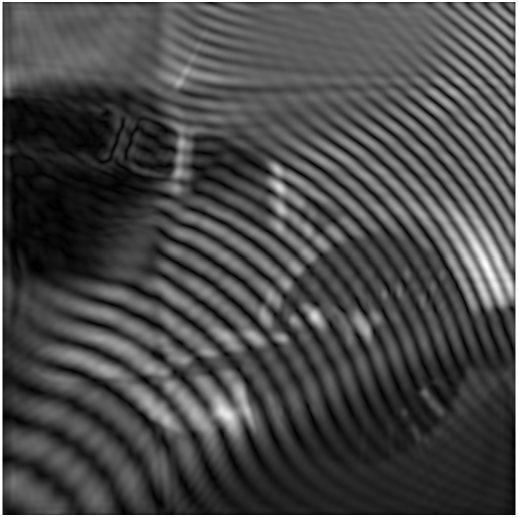}
    \end{minipage}%
    \begin{minipage}{0.14\textwidth}
        \centering
        \includegraphics[width=\textwidth]{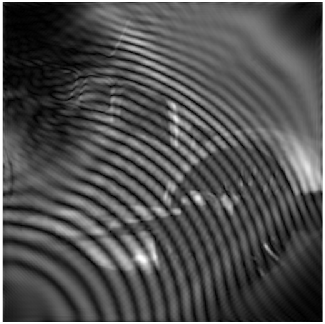}
    \end{minipage}%
    \begin{minipage}{0.14\textwidth}
        \centering
        \includegraphics[width=\textwidth]{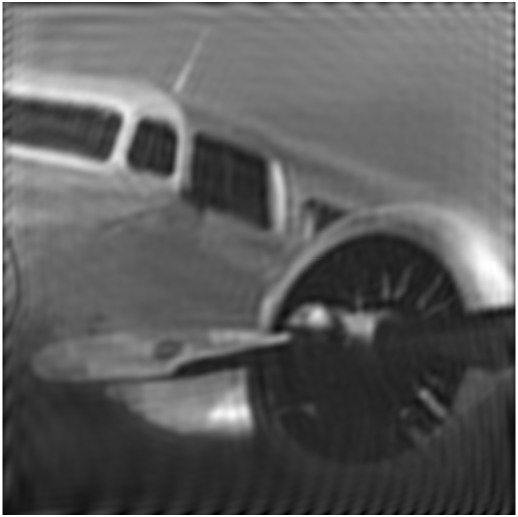}
    \end{minipage}
    
    \vspace{0.3cm} % 可选：给两排之间增加垂直间距
    
    % 第二排图像
    \begin{minipage}{0.14\textwidth}
        \centering
        \includegraphics[width=\textwidth]{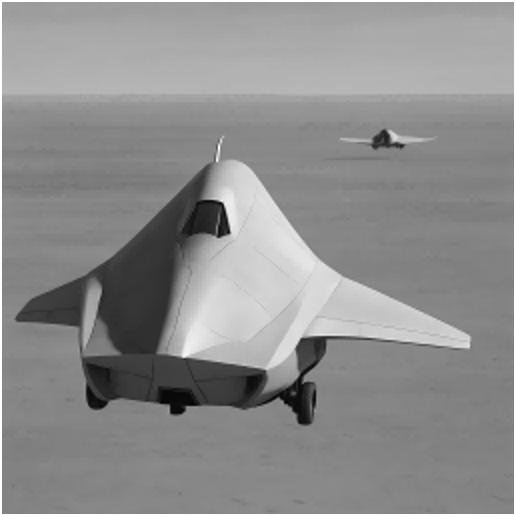} 
        \caption*{Original}
    \end{minipage}%
    \begin{minipage}{0.14\textwidth}
        \centering
        \includegraphics[width=\textwidth]{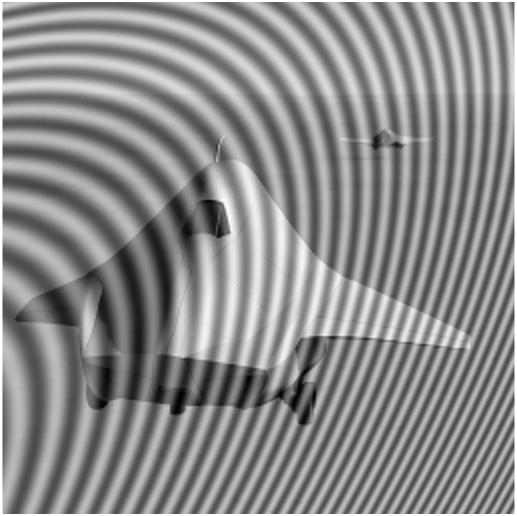}
        \caption*{Noisy}
    \end{minipage}%
    \begin{minipage}{0.14\textwidth}
        \centering
        \includegraphics[width=\textwidth]{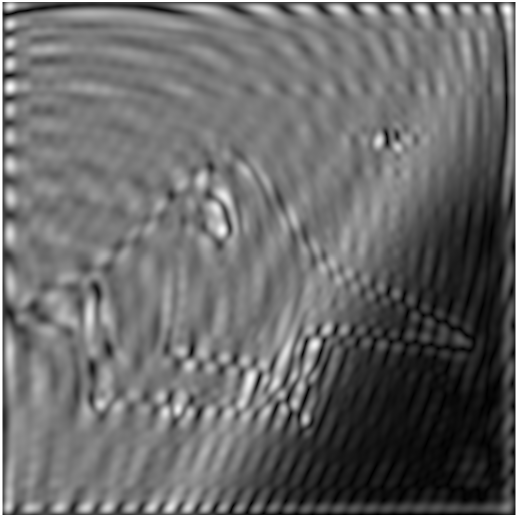} 
        \caption*{2D FT}
    \end{minipage}%
    \begin{minipage}{0.14\textwidth}
        \centering
        \includegraphics[width=\textwidth]{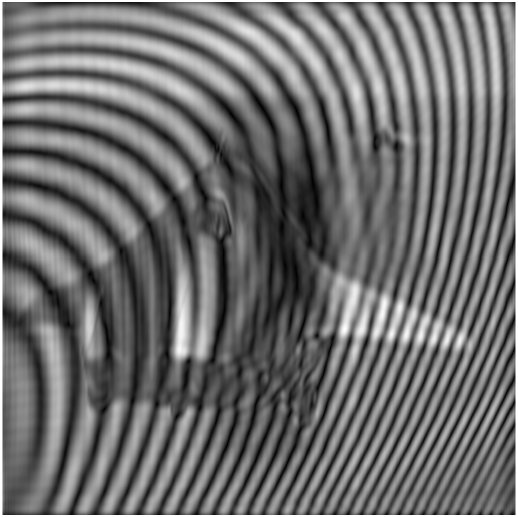} 
        \caption*{2D SFRFT}
    \end{minipage}%
    \begin{minipage}{0.14\textwidth}
        \centering
        \includegraphics[width=\textwidth]{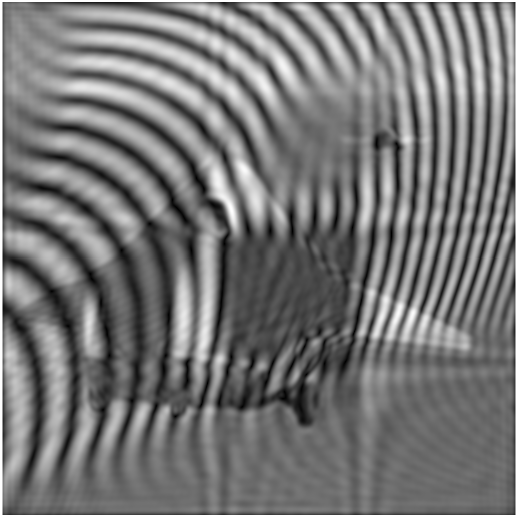}
        \caption*{GT}
    \end{minipage}%
    \begin{minipage}{0.14\textwidth}
        \centering
        \includegraphics[width=\textwidth]{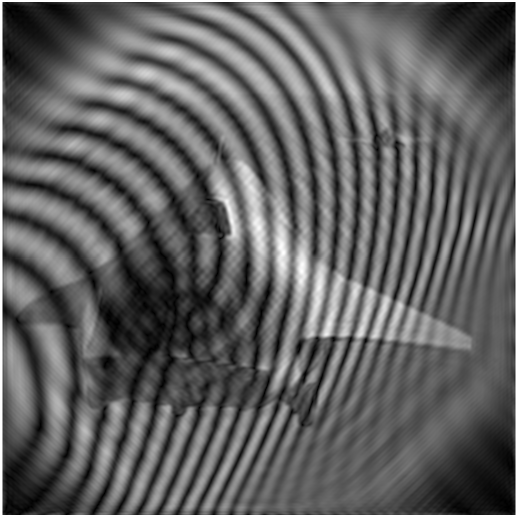}
        \caption*{CFRFT}
    \end{minipage}%
    \begin{minipage}{0.14\textwidth}
        \centering
        \includegraphics[width=\textwidth]{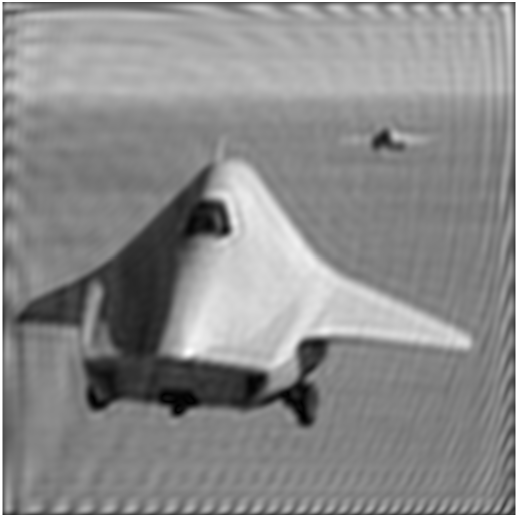}
        \caption*{2D NSFRFT}
    \end{minipage}
    
    \caption{Removal of 2D chirp artifacts acting as structured interference using different transform-domain filters.}
    \label{fig:11}
\end{figure}

Results (Fig.~\ref{fig:11}): By applying a stopband filter in the 2D NSFRFT domain, the interference is surgically removed. As shown in Fig.~\ref{fig:11}, comparative methods leave significant residual ripples or damage image details, whereas the proposed method effectively eliminates the artifacts while preserving the underlying structure.

\subsection{DRPED for Medical Images}
\begin{figure}[htbp]
	\centering
	\includegraphics[width=0.9\textwidth]{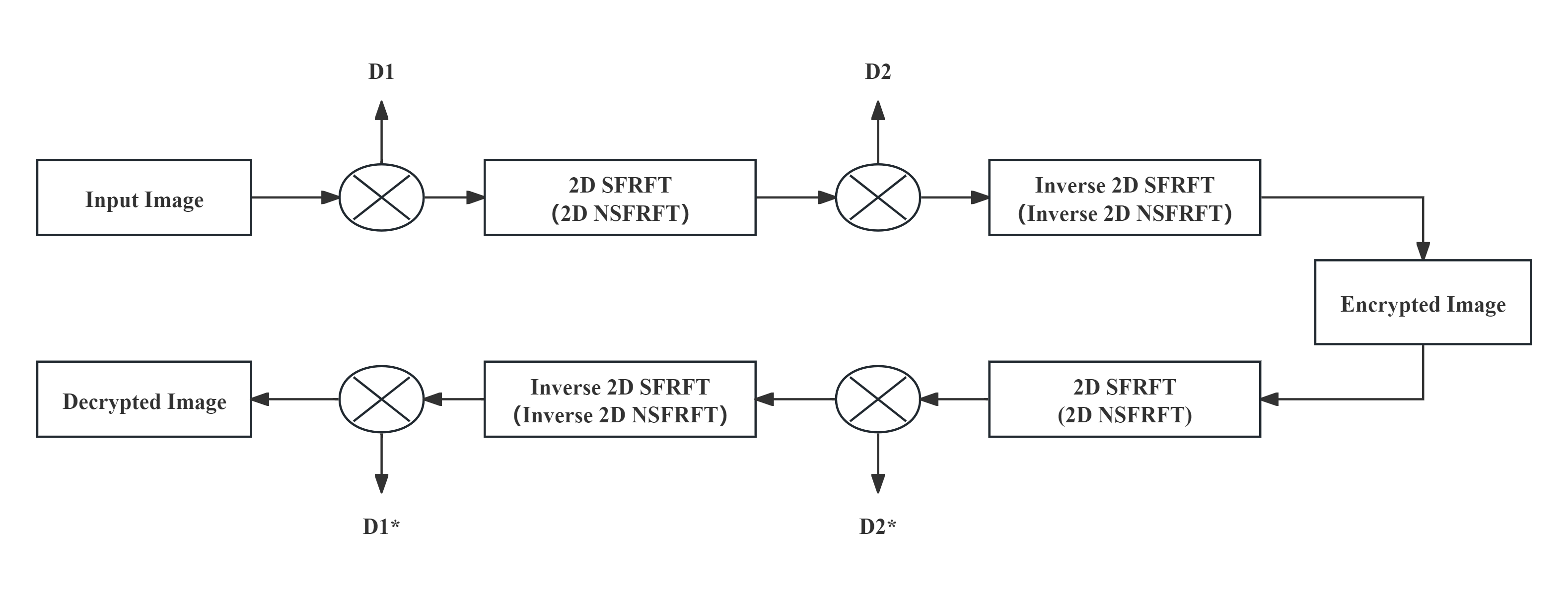}
	\caption{Schematic of DRPED algorithms based on the 2D SFRFT and the proposed 2D NSFRFT. D1 and D2 denote random phase masks.}
	\label{fig:7}
\end{figure}

The DRPED framework based on the 2D SFRFT has been widely studied as a benchmark for optical information security \cite{21,50,51,52,53,54,55}. Here, we extend this architecture to the 2D NSFRFT domain. Compared to the conventional separable case which is limited to 2 DOFs, the proposed 2D NSFRFT utilizes 4 DOFs. This expanded parameter space significantly enlarges the key space and enhances security against brute-force attacks, while maintaining a comparable computational complexity of $O(N^2\log N)$. The schematic of the encryption architecture is illustrated in Fig.~\ref{fig:7}.

\begin{figure}[htbp]
	\centering
	\subfloat[]{\includegraphics[width=0.18\textwidth]{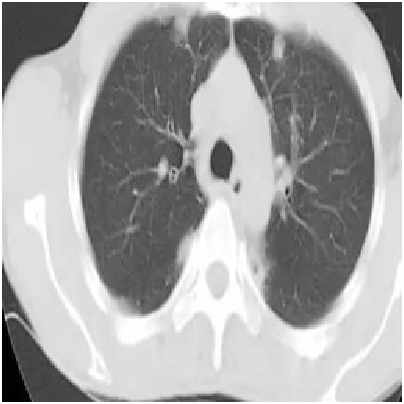}} \hfill
	\subfloat[]{\includegraphics[width=0.18\textwidth]{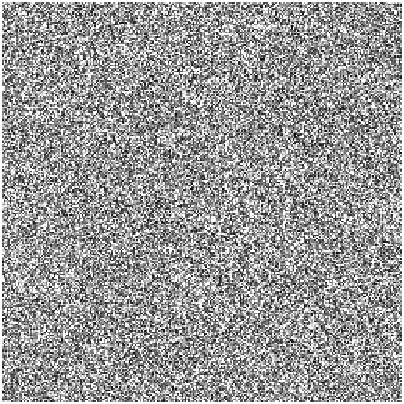}} \hfill
	\subfloat[]{\includegraphics[width=0.18\textwidth]{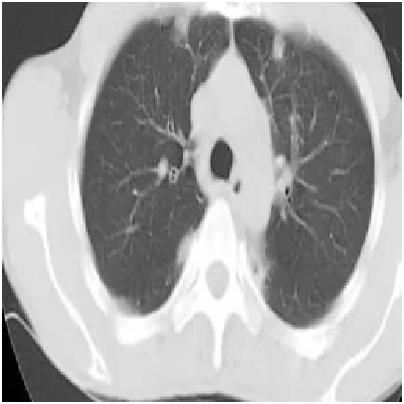}} \hfill
	\subfloat[]{\includegraphics[width=0.18\textwidth]{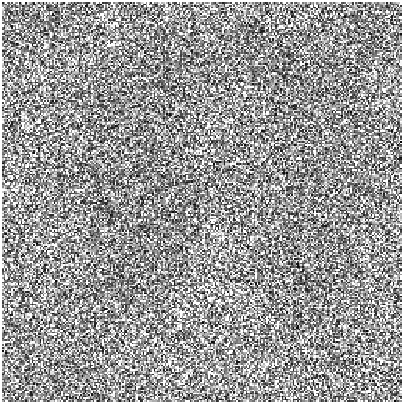}}
	\caption{Simulation results for medical image encryption using 2D NSFRFT. (a) Input medical image. (b) Encrypted cipher-text. (c) Decrypted image with correct keys ($P_{\text{key}}$). (d) Decrypted image with incorrect keys ($P_{\text{wrong}}$).}
	\label{fig:8}
\end{figure}

\textbf{Simulation Verification:}
To validate the proposed scheme for securing sensitive medical data, we utilize a $256\times256$ medical MRI image as the input (Fig.~\ref{fig:8}(a)). The 2D NSFRFT parameters are selected as the private keys: $P_{\text{key}} = (0.7548, 0.4147, -0.0442, -0.5063, \frac{\pi}{3})$. The resulting encrypted image (Fig.~\ref{fig:8}(b)) exhibits a visually noise-like distribution, leaking no structural information.

Decryption performance is evaluated using the Mean Squared Error (MSE), defined as
\begin{equation}
	\mathrm{MSE} = \frac{1}{M \times N} \sum_{x=0}^{M-1} \sum_{y=0}^{N-1} \left( |f_{\text{dec}}(x,y)| - |f_{\text{in}}(x,y)| \right)^2.
	\label{eq:37}
\end{equation}

The decryption results are presented in Fig.~\ref{fig:8}:
\begin{itemize}
	\item \textbf{Correct Decryption:} With the correct keys, the medical image is perfectly recovered (Fig.~\ref{fig:8}(c)). The MSE is $8.33 \times 10^{-26}$, indicating near-perfect numerical reconstruction.
	\item \textbf{Incorrect Decryption:} Decryption using a wrong key set $P_{\text{wrong}}=(\frac{1}{2}, \frac{1}{2}, \frac{1}{2}, \frac{1}{2}, \frac{\pi}{5})$ fails completely, yielding a noise-like image (Fig.~\ref{fig:8}(d)) with no recognizable features.
\end{itemize}

\begin{figure}[htbp]
	\centering
	\includegraphics[width=0.5\textwidth]{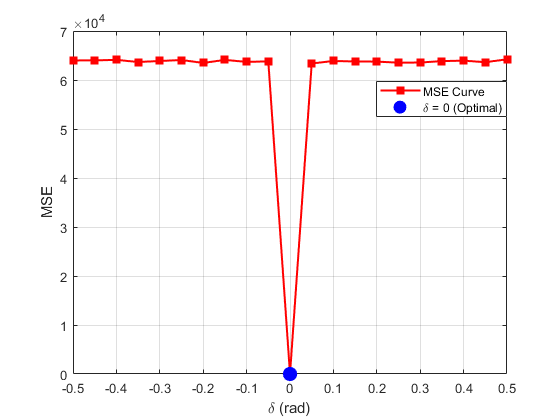}
	\caption{Key sensitivity analysis: Variation of MSE between the decrypted and original images with respect to deviation $\delta$ in the parameter $\theta$.}
	\label{fig:9}
\end{figure}

\textbf{Sensitivity Analysis:}
To evaluate the robustness of the system, we performed a key sensitivity test. Keeping the quaternion parameters $(a,b,c,d)$ fixed, we introduced a minute deviation $\delta$ to the rotation angle parameter $\theta$ (ranging from $-0.5$ to $0.5$). As shown in Fig.~\ref{fig:9}, the MSE curve exhibits a sharp V-shape. Even a minimal deviation of $\delta \approx 0.05$ causes the MSE to rise drastically, preventing meaningful decryption. This high sensitivity confirms that the 2D NSFRFT-based DRPED provides a rigorous security layer suitable for medical applications.

\section{Conclusions}
\label{Section 6}

In this paper, we have proposed the 2D NSFRFT, a generalized mathematical framework that fundamentally extends the classical 2D FRFT from separable operations to fully coupled {space-frequency} rotations. {From a group-theoretic perspective, by constructing the transform as the intersection of the symplectic group $\mathrm{Sp}(4, \mathbb{R})$ and the special orthogonal group $\mathrm{SO}(4)$ (parameterized by quaternions)}, we have established a unified theory that encompasses the 2D SFRFT, GT, and CFRFT as special cases, thereby providing a complete algebraic description of generalized rotations in the 4D phase space.

The contributions of this work are summarized as follows:
\begin{itemize}
	\item \textbf{Theoretical Generalization and Algorithm:} We derived properties of the 2D NSFRFT and developed two fast discretization algorithms. Theoretical analysis and runtime tests confirm that the fast algorithms maintain a computational complexity of $O(N^2 \log N)$, comparable to the classical 2D FFT and SFRFT, ensuring their feasibility for real-time signal processing.
	
	\item \textbf{Robust Signal Analysis and Filtering:} A critical advantage of the 2D NSFRFT is its capability to handle 2D signals with nonseparable coupling terms (e.g., coupled chirps). We demonstrated that, unlike the general 2D NSLCT which operates on the non-compact symplectic group, the 2D NSFRFT is constrained to a geometrically compact parameter space (isomorphic to the unitary group $\mathrm{U}(2)$), which makes global parameter search computationally feasible and stable. Based on this, we designed both minimum MSE optimal filters and specialized passband/stopband filters. Experimental results verify that the proposed method significantly outperforms existing transforms in extracting coupled signals and removing structured artifacts.
	
	\item \textbf{Enhanced Medical Image Security:} We extended the DRPED framework to the 2D NSFRFT domain specifically for medical imaging applications. The introduction of four independent DOFs significantly expands the key space and increases the system's sensitivity to key mismatches. Simulations on MRI images validate the effectiveness and high robustness of the proposed scheme, confirming its capability to securely protect sensitive medical data.
\end{itemize}

In summary, the 2D NSFRFT bridges the gap between theoretical generality and engineering applicability. It provides a powerful tool for analyzing non-stationary 2D signals exhibiting coupling characteristics, showing great potential in radar signal processing, optical information security, and medical imaging. 

\appendix

	\section{Group-Theoretic Derivation of the 2D NSFRFT}
	\label{Appendix C}
	
	In this appendix, we rigorously derive the definition of the 2D NSFRFT by determining the intersection of the special orthogonal group $\mathrm{SO}(4)$ and the symplectic group $\mathrm{Sp}(4, \mathbb{R})$. We demonstrate that the proposed 4-DOF kernel arises naturally from the maximal compact subgroup of the symplectic group, which is isomorphic to the unitary group $\mathrm{U}(2)$.
	
	\subsection{Geometric Decomposition in $\mathrm{SO}(4)$}
	Any 4D rotation matrix $\mathbf{Z} \in \mathrm{SO}(4)$ represents a rigid rotation in the phase space. Based on the theory of isoclinic rotations (often referred to as Cayley's factorization), $\mathbf{Z}$ can be uniquely factorized into a left-isoclinic rotation $\mathbf{M}_L$ and a right-isoclinic rotation $\mathbf{M}_R$, generated by two unit quaternions $q_L = a+bi+cj+dk$ and $q_R = p+qi+rj+sk$:
	\begin{equation}
		\mathbf{Z} = \mathbf{M}_L(q_L) \cdot \mathbf{M}_R(q_R),
		\label{eq:SO4_decomp}
	\end{equation}
	where $\mathbf{M}_L$ and $\mathbf{M}_R$ are defined as:
	\begin{equation}
		\mathbf{M}_L = \begin{bmatrix} 
			a & -b & -c & -d \\ 
			b & a & -d & c \\ 
			c & d & a & -b \\ 
			d & -c & b & a 
		\end{bmatrix}, \quad
		\mathbf{M}_R = \begin{bmatrix} 
			p & -q & -r & -s \\ 
			q & p & s & -r \\ 
			r & -s & p & q \\ 
			s & r & -q & p 
		\end{bmatrix}.
	\end{equation}
	This structure encompasses all 6 degrees of freedom of rigid rotations in 4D phase space.
	
	\subsection{Symplectic Intersection and Constraint Solving}
	For $\mathbf{Z}$ to represent a valid 2D NSLCT, it must preserve the symplectic form. Thus, we seek the intersection group $K = \mathrm{SO}(4) \cap \mathrm{Sp}(4, \mathbb{R})$. 
	A matrix $\mathbf{Z} \in \mathrm{SO}(4)$ is symplectic if and only if:
	\begin{equation}
		\mathbf{Z} \mathbf{J} \mathbf{Z}^T = \mathbf{J}, \quad \text{where } \mathbf{J} = \begin{bmatrix} \mathbf{0} & \mathbf{I} \\ -\mathbf{I} & \mathbf{0} \end{bmatrix}.
	\end{equation}
	Since $\mathbf{Z}$ is orthogonal ($\mathbf{Z}^T = \mathbf{Z}^{-1}$), this condition is equivalent to the commutation relation $\mathbf{Z}\mathbf{J} = \mathbf{J}\mathbf{Z}$. 
	
	Substituting Eq.~(\ref{eq:SO4_decomp}) into the commutation relation implies that the specific algebraic structure of $\mathbf{M}_R$ must be constrained. Solving the resulting linear system yields the unique necessary conditions for the right quaternion parameters:
	\begin{equation}
		q = 0, \quad s = 0.
	\end{equation}
	This implies that $q_R$ must degenerate to a complex exponential structure $q_R = \cos\theta + (\sin\theta) j$ (mapping to a planar rotation), reducing the 6 DOFs of $\mathrm{SO}(4)$ to 4 DOFs.
	
	\subsection{Final Algebraic Structure ($\mathrm{U}(2)$ Isomorphism)}
	With $q=s=0$, let $p=\cos\theta$ and $r=\sin\theta$. The resulting matrix $\mathbf{Z}$ (and its transpose used for the kernel $\mathbf{M}=\mathbf{Z}^T$) collapses to the specific block form characteristic of the unitary group representation:
	\begin{equation}
		\mathbf{M} = \begin{bmatrix} \mathbf{A} & \mathbf{B} \\ -\mathbf{B} & \mathbf{A} \end{bmatrix},
	\end{equation}
	where the sub-blocks are derived as:
	\begin{align}
		\mathbf{A} &= \begin{bmatrix}
			a\cos\theta - c\sin\theta & b\cos\theta - d\sin\theta  \\
			-b\cos\theta - d\sin\theta & a\cos\theta + c\sin\theta 
		\end{bmatrix}, \\
		\mathbf{B} &= \begin{bmatrix}
			a\sin\theta + c\cos\theta & b\sin\theta + d\cos\theta  \\
			-b\sin\theta + d\cos\theta & a\sin\theta - c\cos\theta
		\end{bmatrix}.
	\end{align}
	Substituting these blocks into the 2D NSLCT integral definition (Eq. (\ref{eq:1})) yields Eqs.~(\ref{eq:13})-(\ref{eq:18}).
	
	\textbf{Conclusion:}
	The derivation confirms that the proposed 2D NSFRFT is not an arbitrary construction but is mathematically isomorphic to the action of the {Unitary Group $\mathrm{U}(2)$}. It represents the unique subset of operations that are simultaneously rigid rotations (energy preserving) and 2D NSLCTs (symplectic area preserving).

	\section{Proofs of the 2D NSFRFT Properties}
	\label{Appendix B}

	\subsection*{B.1 Proof of Orthogonality of Kernel Functions}
	\begin{proof}
		\begin{align*}
			& \int_{\mathbb{R}^2} K_{P}(x,y,u,v) K_{P}^*(x,y,u',v') \, \mathrm{d}x\,\mathrm{d}y \\
			= & \, \int_{\mathbb{R}^2} \left[ \frac{1}{2\pi\sqrt{-\mathbf{T}}} \exp\left( \frac{-\mathrm{j}(p_1 x^2+p_2 xy+p_3 y^2)}{2\mathbf{T}} \right) \exp\left( \frac{-\mathrm{j}(m_1 ux+m_2 vx+m_3 uy+m_4 vy)}{\mathbf{T}} \right) \right. \\
			& \times \left. \exp\left( \frac{-\mathrm{j}(k_1 u^2+k_2 uv+k_3 v^2)}{2\mathbf{T}} \right) \right] \times \left[ \frac{1}{2\pi\sqrt{-\mathbf{T}}} \exp\left( \frac{-\mathrm{j}(p_1 x^2+p_2 xy+p_3 y^2)}{2\mathbf{T}} \right) \right. \\
			& \times \left. \exp\left( \frac{-\mathrm{j}(m_1 u'x+m_2 v'x+m_3 u'y+m_4 v'y)}{\mathbf{T}} \right) \exp\left( \frac{-\mathrm{j}(k_1 u'^2+k_2 u'v'+k_3 v'^2)}{2\mathbf{T}} \right) \right]^* \, \mathrm{d}x\,\mathrm{d}y \\
			= & \, \frac{1}{4\pi^2 \left| \mathbf{T} \right|} \exp\left[ \frac{-\mathrm{j}(k_1 u^2+k_2 uv+k_3 v^2)}{2\mathbf{T}} \right] \exp\left[ \frac{\mathrm{j}(k_1 u'^2+k_2 u'v'+k_3 v'^2)}{2\mathbf{T}} \right] \\
			& \times \int_{\mathbb{R}^2} {\exp\left[ \frac{-\mathrm{j}(p_1 x^2+p_2 xy+p_3 y^2)}{2\mathbf{T}} \right] \exp\left[ \frac{\mathrm{j}(p_1 x^2+p_2 xy+p_3 y^2)}{2\mathbf{T}} \right]} \\
			& \times \exp\left[ \frac{-\mathrm{j}}{\mathbf{T}} \big( m_1 ux + m_2 vx + m_3 uy + m_4 vy \big) \right] \exp\left[ \frac{\mathrm{j}}{\mathbf{T}} \big( m_1 u'x + m_2 v'x + m_3 u'y + m_4 v'y \big) \right] \, \mathrm{d}x\,\mathrm{d}y \\
			= & \, \frac{1}{4\pi^2 \left| \mathbf{T} \right|} \exp\left[ \frac{-\mathrm{j}(k_1 u^2+k_2 uv+k_3 v^2)}{2\mathbf{T}} \right] \exp\left[ \frac{\mathrm{j}(k_1 u'^2+k_2 u'v'+k_3 v'^2)}{2\mathbf{T}} \right] \\
			& \times \int_{\mathbb{R}^2} \exp\left\{ \frac{-\mathrm{j}}{\mathbf{T}} \left[ x \big( m_1(u-u') + m_2(v-v') \big) + y \big( m_3(u-u') + m_4(v-v') \big) \right] \right\} \, \mathrm{d}x\,\mathrm{d}y \\
			= & \, \frac{1}{4\pi^2 \left| \mathbf{T} \right|} \exp\left[ \frac{-\mathrm{j}(k_1 u^2+k_2 uv+k_3 v^2)}{2\mathbf{T}} \right] \exp\left[ \frac{\mathrm{j}(k_1 u'^2+k_2 u'v'+k_3 v'^2)}{2\mathbf{T}} \right] \\
			& \times 4\pi^2 \delta\left( \frac{m_1(u-u') + m_2(v-v')}{\mathbf{T}}, \frac{m_3(u-u') + m_4(v-v')}{\mathbf{T}} \right) \\
			= & \, \frac{1}{4\pi^2 \left| \mathbf{T} \right|} \times 4\pi^2 \left| \mathbf{T} \right| \delta(u-u', v-v') \times {\exp\left[ \frac{-\mathrm{j}(k_1 u^2+k_2 uv+k_3 v^2)}{2\mathbf{T}} \right] \exp\left[ \frac{\mathrm{j}(k_1 u^2+k_2 uv+k_3 v^2)}{2\mathbf{T}} \right]} \\
			= & \, \delta(u-u', v-v')
		\end{align*}
		This completes the proof. 
	\end{proof}

	\subsection*{B.2 Proof of Inverse Formula}

	\begin{proof}
		According to Eqs. (\ref{eq:13})-(\ref{eq:18}), we have
		\begin{align*}
			& \int_{\mathbb{R}^2} F_{P}(u,v) K_{P}^*(x,y,u,v) \, \mathrm{d}u\,\mathrm{d}v \\
			= & \, \left(\frac{1}{2\pi\sqrt{-\mathbf{T}}}\right)^* \int_{\mathbb{R}^2} F_{P}(u,v) \exp\left[\frac{-\mathrm{j}(k_1 u^2+k_2 uv+k_3 v^2)}{2\mathbf{T}}\right] \\
			& \times \exp\left[\frac{-\mathrm{j}(m_1 ux+m_2 vx+m_3 uy+m_4 vy)}{\mathbf{T}}\right] \exp\left[\frac{-\mathrm{j}(p_1 x^2+p_2 xy+p_3 y^2)}{2\mathbf{T}}\right] \, \mathrm{d}u\,\mathrm{d}v \\
			= & \, \frac{1}{4\pi^2 \left| \mathbf{T} \right|} \int_{\mathbb{R}^2} \left[ \int_{\mathbb{R}^2} f(s,t) K_{P}(s,t,u,v) \, \mathrm{d}s\,\mathrm{d}t \right] \exp\left[\frac{-\mathrm{j}(k_1 u^2+k_2 uv+k_3 v^2)}{2\mathbf{T}}\right] \\
			& \times \exp\left[\frac{-\mathrm{j}(m_1 ux+m_2 vx+m_3 uy+m_4 vy)}{\mathbf{T}}\right] \exp\left[\frac{-\mathrm{j}(p_1 x^2+p_2 xy+p_3 y^2)}{2\mathbf{T}}\right] \, \mathrm{d}u\,\mathrm{d}v \\
			= & \, \frac{1}{4\pi^2 \left| \mathbf{T} \right|} \int_{\mathbb{R}^2} \int_{\mathbb{R}^2} f(s,t) \exp\left[\frac{\mathrm{j}(p_1 s^2+p_2 st+p_3 t^2)}{2\mathbf{T}}\right] \exp\left[\frac{\mathrm{j}(m_1 us+m_2 vs+m_3 ut+m_4 vt)}{\mathbf{T}}\right] \\
			& \times {\exp\left[\frac{\mathrm{j}(k_1 u^2+k_2 uv+k_3 v^2)}{2\mathbf{T}}\right] \exp\left[\frac{-\mathrm{j}(k_1 u^2+k_2 uv+k_3 v^2)}{2\mathbf{T}}\right]} \\
			& \times \exp\left[\frac{-\mathrm{j}(m_1 ux+m_2 vx+m_3 uy+m_4 vy)}{\mathbf{T}}\right] \exp\left[\frac{-\mathrm{j}(p_1 x^2+p_2 xy+p_3 y^2)}{2\mathbf{T}}\right] \, \mathrm{d}s\,\mathrm{d}t\,\mathrm{d}u\,\mathrm{d}v \\
			= & \, \frac{1}{4\pi^2 \left| \mathbf{T} \right|} \int_{\mathbb{R}^2} f(s,t) \exp\left[\frac{\mathrm{j}(p_1 s^2+p_2 st+p_3 t^2)}{2\mathbf{T}}\right] \exp\left[\frac{-\mathrm{j}(p_1 x^2+p_2 xy+p_3 y^2)}{2\mathbf{T}}\right] \\
			& \times \left[ \int_{\mathbb{R}^2} \exp\left[\frac{\mathrm{j}\big(m_1 u(s-x)+m_2 v(s-x)+m_3 u(t-y)+m_4 v(t-y)\big)}{\mathbf{T}}\right] \, \mathrm{d}u\,\mathrm{d}v \right] \mathrm{d}s\,\mathrm{d}t \\
			= & \, \frac{1}{4\pi^2 \left| \mathbf{T} \right|} \int_{\mathbb{R}^2} f(s,t) \exp\left[\frac{\mathrm{j}(p_1 s^2+p_2 st+p_3 t^2)}{2\mathbf{T}}\right] \exp\left[\frac{-\mathrm{j}(p_1 x^2+p_2 xy+p_3 y^2)}{2\mathbf{T}}\right] \\
			& \times \left\{ 4\pi^2 |\mathbf{T}|^2 \delta\big(m_1(s-x)+m_3(t-y), m_2(s-x)+m_4(t-y)\big) \right\} \, \mathrm{d}s\,\mathrm{d}t \\
			= & \, \int_{\mathbb{R}^2} f(s,t) \exp\left[\frac{\mathrm{j}(p_1 s^2+p_2 st+p_3 t^2)}{2\mathbf{T}}\right] \exp\left[\frac{-\mathrm{j}(p_1 x^2+p_2 xy+p_3 y^2)}{2\mathbf{T}}\right] \delta(s-x, t-y) \, \mathrm{d}s\,\mathrm{d}t \\
			= & \, f(x,y)
		\end{align*}
		This completes the proof. 
	\end{proof}

	\subsection*{B.3 Proof of Energy Conservation Property}
	\begin{proof}
		\begin{align*}
			& \int_{\mathbb{R}^2} \left| F_{P}(u,v) \right|^2 \, \mathrm{d}u\,\mathrm{d}v \\
			= & \, \int_{\mathbb{R}^2} F_{P}(u,v) F_{P}^*(u,v) \, \mathrm{d}u\,\mathrm{d}v \\
			= & \, \int_{\mathbb{R}^2} \left[ \int_{\mathbb{R}^2} f(x,y) K_{P}(x,y,u,v) \, \mathrm{d}x\,\mathrm{d}y \right] \left[ \int_{\mathbb{R}^2} f^*(x',y') K_{P}^*(x',y',u,v) \, \mathrm{d}x'\,\mathrm{d}y' \right] \, \mathrm{d}u\,\mathrm{d}v \\
			= & \, \int_{\mathbb{R}^2} \int_{\mathbb{R}^2} f(x,y) f^*(x',y') \left[ \int_{\mathbb{R}^2} K_{P}(x,y,u,v) K_{P}^*(x',y',u,v) \, \mathrm{d}u\,\mathrm{d}v \right] \, \mathrm{d}x\,\mathrm{d}y\,\mathrm{d}x'\,\mathrm{d}y' \\
			= & \, \int_{\mathbb{R}^2} \int_{\mathbb{R}^2} f(x,y) f^*(x',y') \left\{ \frac{1}{4\pi^2 \left| \mathbf{T} \right|} \exp\left[\frac{-\mathrm{j}(p_1 x^2+p_2 xy+p_3 y^2)}{2\mathbf{T}}\right] \exp\left[\frac{\mathrm{j}(p_1 x'^2+p_2 x'y'+p_3 y'^2)}{2\mathbf{T}}\right] \right. \\
			& \times \int_{\mathbb{R}^2} {\exp\left[\frac{-\mathrm{j}(k_1 u^2+k_2 uv+k_3 v^2)}{2\mathbf{T}}\right] \exp\left[\frac{\mathrm{j}(k_1 u^2+k_2 uv+k_3 v^2)}{2\mathbf{T}}\right]} \\
			& \times \left. \exp\left[\frac{-\mathrm{j}(m_1 ux+m_2 vx+m_3 uy+m_4 vy)}{\mathbf{T}}\right] \exp\left[\frac{\mathrm{j}(m_1 ux'+m_2 vx'+m_3 uy'+m_4 vy')}{\mathbf{T}}\right] \, \mathrm{d}u\,\mathrm{d}v \right\} \, \mathrm{d}x\,\mathrm{d}y\,\mathrm{d}x'\,\mathrm{d}y' \\
			= & \, \int_{\mathbb{R}^2} \int_{\mathbb{R}^2} f(x,y) f^*(x',y') \left\{ \frac{1}{4\pi^2 \left| \mathbf{T} \right|} \exp\left[\frac{-\mathrm{j}(p_1 x^2+p_2 xy+p_3 y^2)}{2\mathbf{T}}\right] \exp\left[\frac{\mathrm{j}(p_1 x'^2+p_2 x'y'+p_3 y'^2)}{2\mathbf{T}}\right] \right. \\
			& \times \left. \int_{\mathbb{R}^2} \exp\left[\frac{-\mathrm{j}}{\mathbf{T}} \big( u(m_1(x-x') + m_3(y-y')) + v(m_2(x-x') + m_4(y-y')) \big) \right] \, \mathrm{d}u\,\mathrm{d}v \right\} \, \mathrm{d}x\,\mathrm{d}y\,\mathrm{d}x'\,\mathrm{d}y' \\
			= & \, \int_{\mathbb{R}^2} \int_{\mathbb{R}^2} f(x,y) f^*(x',y') \left\{ \frac{1}{4\pi^2 \left| \mathbf{T} \right|} \exp\left[\frac{-\mathrm{j}(p_1 x^2+p_2 xy+p_3 y^2)}{2\mathbf{T}}\right] \exp\left[\frac{\mathrm{j}(p_1 x'^2+p_2 x'y'+p_3 y'^2)}{2\mathbf{T}}\right] \right. \\
			& \times \left. 4\pi^2 |\mathbf{T}|^2 \delta\big(m_1(x-x')+m_3(y-y'), m_2(x-x')+m_4(y-y')\big) \right\} \, \mathrm{d}x\,\mathrm{d}y\,\mathrm{d}x'\,\mathrm{d}y' \\
			= & \, \int_{\mathbb{R}^2} \int_{\mathbb{R}^2} f(x,y) f^*(x',y') \exp\left[\frac{-\mathrm{j}(p_1 x^2+p_2 xy+p_3 y^2)}{2\mathbf{T}}\right] \exp\left[\frac{\mathrm{j}(p_1 x'^2+p_2 x'y'+p_3 y'^2)}{2\mathbf{T}}\right] \\
			& \times \delta(x-x', y-y') \, \mathrm{d}x\,\mathrm{d}y\,\mathrm{d}x'\,\mathrm{d}y' \\
			= & \, \int_{\mathbb{R}^2} f(x,y) f^*(x,y) \, \mathrm{d}x\,\mathrm{d}y \\
			= & \, \int_{\mathbb{R}^2} \left| f(x,y) \right|^2 \, \mathrm{d}x\,\mathrm{d}y
		\end{align*}
	\end{proof}
	
\subsection*{B.4 Proof of Conjugation Property}
	\begin{proof}
	The complex conjugate of the kernel $K_{P}(x,y,u,v)$ exhibits Hermitian symmetry, relating it to the kernel of the inverse transformation with swapped input/output variables. Specifically, we show that:
	\begin{equation}
		K_{P}^*(x,y,u,v) = \gamma \cdot K_{P_{inv}}(u,v,x,y),
	\end{equation}
	where $P_{inv} = (a, -b, -c, -d, -\theta)$ is the inverse parameter set defined in the main text.
	
	\textbf{1) Analysis of the Amplitude Term:}
	The amplitude coefficient is $A = \frac{1}{2\pi\sqrt{-\mathbf{T}}}$, the conjugation $A^*$ depends on the sign of $\mathbf{T}$:
	\begin{itemize}
		\item If $\mathbf{T} < 0$, $-\mathbf{T} > 0$, so $A \in \mathbb{R}$ and $A^* = A$.
		\item If $\mathbf{T} > 0$, $-\mathbf{T} < 0$, so $A \in i\mathbb{R}$ and $A^* = -A$.
	\end{itemize}
	This sign flip perfectly matches the definition of $\gamma$:
	\begin{equation}
		\gamma = \mathrm{sgn}(-\mathbf{T}) = \begin{cases} 
			1, & \text{if } \mathbf{T} < 0, \\
			-1, & \text{if } \mathbf{T} > 0.
		\end{cases}
	\end{equation}
	
	\textbf{2) Analysis of the Exponential Terms:}
	We verify that the phase terms of $K_{P}^*$ match those of $K_{P_{inv}}$ with variables $(u,v)$ and $(x,y)$ swapped.
	The exponential part of the conjugate kernel $K_{P}^*$ effectively reverses the sign of the imaginary unit $\mathrm{j}$.
	For the inverse kernel $K_{P_{inv}}$, we substitute the parameters from $P_{inv}$ into the standard definition. Using the quaternion relationships, the coefficients transform as:
	\begin{equation}
		\hat{m}_1 = -m_1, \quad \hat{m}_2 = -m_3, \quad \hat{m}_3 = -m_2, \quad \hat{m}_4 = -m_4.
	\end{equation}
	For example, substituting $\theta \to -\theta, b \to -b, d \to -d$ into $\hat{m}_2$:
	\begin{equation}
		\hat{m}_2 = (-b)\sin(-\theta) + (-d)\cos(-\theta) = b\sin\theta - d\cos\theta = -m_3.
	\end{equation}
	These sign reversals in the coefficients, combined with the common factor $\frac{\mathrm{j}}{\mathbf{T}}$, mathematically replicate the effect of the complex conjugation on the phase term.
	
	\textbf{3) Conclusion:}
	Since both the amplitude factor $\gamma$ and the exponential phase terms align perfectly with the definition of the inverse kernel, the property Eq.~(\ref{eq:conjugation_prop}) holds.
		
	\noindent This completes the proof. 
	\end{proof}

\subsection*{B.5 Proof of Space Reversal Property}
\begin{proof}
	Let $g(x, y) = f(-x, -y)$. According to the definition of the 2D NSFRFT, we have:
	\begin{equation}
		\mathcal{F}_{P}[g](u, v) = \int_{\mathbb{R}^2} f(-x,-y) K_{P}(x,y,u,v) \mathrm{d}x \, \mathrm{d}y.
	\end{equation}
	Perform the variable substitution $x' = -x$ and $y' = -y$. The Jacobian determinant of this transformation is $|\det(-\mathbf{I})| = 1$, so $\mathrm{d}x\,\mathrm{d}y = \mathrm{d}x'\,\mathrm{d}y'$. The integral limits remain $(-\infty, \infty)$. The expression becomes:
	\begin{equation}
		\mathcal{F}_{P}[g](u, v) = \int_{\mathbb{R}^2} f(x',y') K_{P}(-x',-y',u,v) \mathrm{d}x' \, \mathrm{d}y'.
		\label{eq:B2}
	\end{equation}
	Next, we analyze the symmetry of the kernel $K_{P}(-x', -y', u, v)$ based on its components:
	
	\textbf{1) Quadratic Phase Terms:} 
	The input and output chirp terms depend strictly on quadratic variables ($x^2, xy, y^2$ and $u^2, uv, v^2$). Since $(-x')^2 = x'^2$ and $(-x')(-y') = x'y'$, these terms are even functions of the coordinates and remain unchanged.
	
	\textbf{2) Linear Phase Kernel:} 
	The transformation kernel involves the exponential of linear products. Substituting the negated coordinates:
	\begin{align}
		& \exp\left( \frac{\mathrm{j}}{\mathbf{T}} [m_1 u(-x') + m_2 v(-x') + m_3 u(-y') + m_4 v(-y')] \right) \notag \\
		= & \exp\left( \frac{\mathrm{j}}{\mathbf{T}} [m_1 (-u)x' + m_2 (-v)x' + m_3 (-u)y' + m_4 (-v)y'] \right).
	\end{align}
	This expression effectively transfers the negative sign from the input coordinates $(x', y')$ to the output coordinates $(u, v)$.
	
	\textbf{Conclusion:}
	Combining these observations, we obtain the relationship:
	\begin{equation}
		K_{P}(-x', -y', u, v) = K_{P}(x', y', -u, -v).
	\end{equation}
	Substituting this back into Eq. (\ref{eq:B2}), we arrive at:
	\begin{equation}
		\mathcal{F}_{P}[g](u, v) = \int_{\mathbb{R}^2} f(x',y') K_{P}(x',y',-u,-v) \mathrm{d}x' \, \mathrm{d}y' = \mathcal{F}_{P}[f](-u, -v).
	\end{equation}
	This completes the proof.
	\end{proof}
	
	\subsection*{B.6. Proof of Differentiation Property}
	
	\begin{proof}
		The differentiation properties describe how the spatial derivative operators transform into the 2D NSFRFT domain. We derive this by exploiting the algebraic relation between the spatial and frequency gradients of the kernel.
		
		\textbf{1. Integration by Parts}
		
		Let $g(x,y) = \frac{\partial f(x,y)}{\partial x}$. The 2D NSFRFT of $g(x,y)$ is given by:
		\begin{equation}
			G_{P}(u,v) = \int_{\mathbb{R}^2} \frac{\partial f(x,y)}{\partial x} K_{P}(x,y,u,v) \,\mathrm{d}x\,\mathrm{d}y.
		\end{equation}
		Assuming $f(x,y)$ vanishes at infinity, integration by parts yields:
		\begin{equation} \label{eq:int_by_parts}
			G_{P}(u,v) = - \int_{\mathbb{R}^2} f(x,y) \frac{\partial K_{P}}{\partial x} \,\mathrm{d}x\,\mathrm{d}y.
		\end{equation}
		From the kernel definition in Eq. (14), the partial derivative with respect to $x$ is:
		\begin{equation} \label{eq:dk_dx}
			\frac{\partial K_{P}}{\partial x} = \frac{\mathrm{j}}{\mathbf{T}} \left( p_1 x + \frac{p_2}{2} y + m_1 u + m_2 v \right) K_{P}.
		\end{equation}
		Substituting Eq. (\ref{eq:dk_dx}) into Eq. (\ref{eq:int_by_parts}), we obtain:
		\begin{equation} \label{eq:G_intermediate}
			G_{P}(u,v) = -\frac{\mathrm{j}}{\mathbf{T}} \left[ p_1 \mathcal{F}_{P}[xf] + \frac{p_2}{2} \mathcal{F}_{P}[yf] + (m_1 u + m_2 v) F_{P}(u,v) \right],
		\end{equation}
		where $\mathcal{F}_{P}[xf]$ and $\mathcal{F}_{P}[yf]$ denote the transform of $xf(x,y)$ and $yf(x,y)$, respectively.
		
		\textbf{2. Operator Relation in Transform Domain}
		
		To express $\mathcal{F}_{P}[xf]$ and $\mathcal{F}_{P}[yf]$ in terms of frequency operators, we differentiate the kernel with respect to $u$ and $v$:
		\begin{subequations}
			\begin{align}
				\frac{\partial K_{P}}{\partial u} &= \frac{\mathrm{j}}{\mathbf{T}} \left( m_1 x + m_3 y + k_1 u + \frac{k_2}{2} v \right) K_{P}, \\
				\frac{\partial K_{P}}{\partial v} &= \frac{\mathrm{j}}{\mathbf{T}} \left( m_2 x + m_4 y + \frac{k_2}{2} u + k_3 v \right) K_{P}.
			\end{align}
		\end{subequations}
		Rearranging these equations, we establish a linear system linking the spatial multiplication operators to the frequency differential operators:
		\begin{equation} \label{eq:linear_system}
			\begin{pmatrix} m_1 & m_3 \\ m_2 & m_4 \end{pmatrix} 
			\begin{pmatrix} xK_{P} \\ yK_{P} \end{pmatrix} 
			= \frac{\mathbf{T}}{\mathrm{j}} 
			\begin{pmatrix} \frac{\partial K_{P}}{\partial u} \\ \frac{\partial K_{P}}{\partial v} \end{pmatrix} 
			- \begin{pmatrix} C_u \\ C_v \end{pmatrix} K_{P},
		\end{equation}
		where $C_u = k_1 u + \frac{k_2}{2} v$ and $C_v = \frac{k_2}{2} u + k_3 v$.
		
		\textbf{3. Solving for Spatial Variables}
		
		Let $\Delta = m_1 m_4 - m_2 m_3$. Provided $\Delta \neq 0$, we can invert the matrix in Eq. (\ref{eq:linear_system}) to solve for $xK_{P}$ and $yK_{P}$. Applying the integral operator $\int_{\mathbb{R}^2} f(x,y)(\cdot) \mathrm{d}x \, \mathrm{d}y$ to both sides, we convert the kernel relations into operator relations acting on $F_{P}(u,v)$:
		\begin{subequations}
			\begin{align}
				\mathcal{F}_{P}[xf] &= \frac{1}{\Delta} \left[ m_4 \left( \frac{\mathbf{T}}{\mathrm{j}} \frac{\partial F_{P}}{\partial u} - C_u F_{P} \right) - m_3 \left( \frac{\mathbf{T}}{\mathrm{j}} \frac{\partial F_{P}}{\partial v} - C_v F_{P} \right) \right], \label{eq:xf_op} \\
				\mathcal{F}_{P}[yf] &= \frac{1}{\Delta} \left[ -m_2 \left( \frac{\mathbf{T}}{\mathrm{j}} \frac{\partial F_{P}}{\partial u} - C_u F_{P} \right) + m_1 \left( \frac{\mathbf{T}}{\mathrm{j}} \frac{\partial F_{P}}{\partial v} - C_v F_{P} \right) \right]. \label{eq:yf_op}
			\end{align}
		\end{subequations}
		
		\textbf{4. Final Derivation and Simplification}
		
		Substituting the expressions for $\mathcal{F}_{P}[xf]$ and $\mathcal{F}_{P}[yf]$ back into Eq. (\ref{eq:G_intermediate}) yields the linear combination of operators acting on $F_{P}(u,v)$. While the initial derivation produces coefficients involving algebraic combinations of $\{p_i, m_i, k_i\}$, these can be significantly simplified by exploiting the symplectic property of the 2D NSFRFT parameter matrix $\mathbf{X}$.
		
		Specifically, observing that the determinant term $\Delta = m_1 m_4 - m_2 m_3$ is identical to $\mathbf{T} = \det(\mathbf{B})$, and applying the symplectic constraint $\mathbf{A}\mathbf{B}^T = \mathbf{B}\mathbf{A}^T$, the derived coefficients simplify to the elements of the parameter matrix $\mathbf{X} = \begin{pmatrix} \mathbf{A} & \mathbf{B} \\ \mathbf{C} & \mathbf{D} \end{pmatrix} = \begin{pmatrix} \mathbf{A} & \mathbf{B} \\ \mathbf{-B} & \mathbf{A} \end{pmatrix}$. The resulting Differentiation property is:
		\begin{equation}
			\mathcal{F}_{P}\left[ \frac{\partial f}{\partial x} \right](u,v) = \left( a_{11} \frac{\partial}{\partial u} + a_{21} \frac{\partial}{\partial v} - \mathrm{j} c_{11} u - \mathrm{j} c_{21} v \right) F_{P}(u,v)
		\end{equation}
		where $a_{11}, a_{21}, c_{11}, c_{21}$ are the elements of the sub-matrices $\mathbf{A}$ and $\mathbf{C} = \mathbf{-B} $ corresponding to the 2D NSFRFT. 
		
		Similarly, following the same derivation procedure, the differentiation property with respect to the $y$-axis is obtained as:
		\begin{equation}
			\mathcal{F}_{P}\left[ \frac{\partial f}{\partial y} \right](u,v) = \left( a_{12} \frac{\partial}{\partial u} + a_{22} \frac{\partial}{\partial v} - \mathrm{j} c_{12} u - \mathrm{j} c_{22} v \right) F_{P}(u,v),
		\end{equation}
		where the coefficients correspond to the second column of the parameter matrix $\mathbf{X}$, i.e., $a_{12} = b\cos\theta - d\sin\theta$, $a_{22} = a\cos\theta + c\sin\theta$, $c_{12} = -b\sin\theta + d\cos\theta$, and $c_{22} = a\sin\theta - c\cos\theta$.
		
		This result proves that the 2D NSFRFT satisfies the general operator transformation rule of the 2D NSLCT, confirming the correctness of the derivation.

\noindent This completes the proof.
	\end{proof}
	
	\subsection*{B.7. Proof of Integration Property}
	
	\begin{proof}
		The integration property determines the 2D NSFRFT of the spatial integral of a signal. Since integration is the inverse operation of differentiation, this property is derived by inverting the operator relation established in the differentiation property.
		
		Let $h(x,y)$ be the integral of $f(x,y)$ with respect to $x$:
		\begin{equation} \label{eq:int_def}
			h(x,y) = \int_{-\infty}^{x} f(\tau, y) \, \mathrm{d}\tau.
		\end{equation}
		By the fundamental theorem of calculus, differentiating Eq. (\ref{eq:int_def}) with respect to $x$ recovers the original signal:
		\begin{equation} \label{eq:fund_thm}
			\frac{\partial h(x,y)}{\partial x} = f(x,y).
		\end{equation}
		Taking the 2D NSFRFT of both sides of Eq. (\ref{eq:fund_thm}), and applying the Differentiation property derived in Appendix B.6, we obtain:
		\begin{equation}
			\mathcal{F}_{P}\left[ \frac{\partial h}{\partial x} \right](u,v) = F_{P}(u,v).
		\end{equation}
		Substituting the explicit differential operator for the $x$-axis derived previously:
		\begin{equation} \label{eq:int_pde_x}
			\left( a_{11} \frac{\partial}{\partial u} + a_{21} \frac{\partial}{\partial v} - \mathrm{j} c_{11} u - \mathrm{j} c_{21} v \right) H_{P}(u,v) = F_{P}(u,v),
		\end{equation}
		where $H_{P}(u,v)$ is the 2D NSFRFT of  $h(x,y)$, and $F_{P}(u,v)$ is the transform of $f(x,y)$.
		
		Similarly, for the integration with respect to $y$, let $g(x,y) = \int_{-\infty}^{y} f(x, \eta) \, \mathrm{d}\eta$. The transform $G_{P}(u,v) = \mathcal{F}_{P}[g(x,y)]$ satisfies the following linear partial differential equation:
		\begin{equation} \label{eq:int_pde_y}
			\left( a_{12} \frac{\partial}{\partial u} + a_{22} \frac{\partial}{\partial v} - \mathrm{j} c_{12} u - \mathrm{j} c_{22} v \right) G_{P}(u,v) = F_{P}(u,v).
		\end{equation}
		This completes the proof.
	\end{proof}

	\textit{Note on Derivation for Properties 8-9:} 
	The direct derivation of the time translation, frequency modulation, and WD properties from the integral definition involves lengthy and tedious algebraic manipulations of quadratic phase terms. To provide a concise and rigorous proof, we derive these properties by treating the 2D NSFRFT as a special case of the general 2D  NSLCT. We apply the established operator rules of the 2D NSLCT \cite{45} and substitute the specific symplectic parameter matrix $\mathbf{X}$ of the proposed 2D NSFRFT defined in Eq. (\ref{eq:22}).
	
	\subsection*{B.8. Proof of Shift and Modulation Property}
	\begin{proof}
		
		For a generalized 2D NSLCT parameterized by a symplectic matrix $\mathbf{M}$, the transformation of a signal that is simultaneously time-shifted by $(m_1, m_2)$ and frequency-modulated by $(n_1, n_2)$ is given by:
		\begin{equation}
			\mathcal{L}^\mathbf{M} \left\{ f(x - m_1, y - m_2) e^{\mathrm{j} (n_1 x + n_2 y)} \right\}(u, v) = \Phi \cdot e^{\mathrm{j} (s_1 u + s_2 v)} \mathcal{L}^\mathbf{M}[f](u - r_1, v - r_2),
			\label{eq:NSLCT_shift_mod}
		\end{equation}
		where $\Phi$ is a constant phase factor. The relationship between the input parameter vector and the output parameter vector is strictly linear and determined by the symplectic matrix $\mathbf{M}$:
		\begin{equation}
			\begin{bmatrix} r_1 \\ r_2 \\ s_1 \\ s_2 \end{bmatrix} 
			= \mathbf{M} 
			\begin{bmatrix} m_1 \\ m_2 \\ n_1 \\ n_2 \end{bmatrix}
			= \begin{bmatrix}
				\mathbf{A} & \mathbf{B} \\ \mathbf{C} & \mathbf{D}
			\end{bmatrix}\begin{bmatrix} m_1 \\ m_2 \\ n_1 \\ n_2 \end{bmatrix}.
		\end{equation}
		
		Substituting the specific $4 \times 4$ symplectic matrix $\mathbf{X}$ of the 2D NSFRFT (as defined in Eq. (\ref{eq:22})) into the above equation yields:
		\begin{equation}
			G_P(u,v) = \Phi \cdot e^{\mathrm{j}(s_1 u + s_2 v)} F_P(u - r_1, v - r_2),
			\label{eq:tf_shifting}
		\end{equation}
		where
		\begin{equation}
			\begin{bmatrix} r_1 \\ r_2 \\ s_1 \\ s_2 \end{bmatrix} 
			= \begin{bmatrix}
				\mathbf{A} & \mathbf{B} \\ \mathbf{-B} & \mathbf{A}
			\end{bmatrix}
			\begin{bmatrix} m_1 \\ m_2 \\ n_1 \\ n_2 \end{bmatrix}.
		\end{equation}
		
	\noindent This completes the proof.
	\end{proof}
	
	\subsection*{B.9. Proof of Relationship with the 2D WD}
	\begin{proof}
		Based on the properties of the general 2D NSLCT, the WD of the transformed signal $\mathcal{L}^\mathbf{M}[f]$ is related to the WD of the input signal $f$ by a linear coordinate transformation in the phase space:
		\begin{equation}
			W_{\mathcal{L}^\mathbf{M}[f]}(x, y, u, v) = W_f(\widetilde{x}, \widetilde{y}, \widetilde{u}, \widetilde{v}),
		\end{equation}
		where 
		\begin{equation}
			\begin{bmatrix} \widetilde{x} \\ \widetilde{y} \\ \widetilde{u} \\ \widetilde{v} \end{bmatrix} 
			= \mathbf{M}^{-1} 
			\begin{bmatrix} x \\ y \\ u \\ v \end{bmatrix}
			= \mathbf{M}^{T} 
			\begin{bmatrix} x \\ y \\ u \\ v \end{bmatrix}
			= \begin{bmatrix} \mathbf{D}^T & -\mathbf{B}^T \\ -\mathbf{C}^T & \mathbf{A}^T \end{bmatrix} 
			\begin{bmatrix} x \\ y \\ u \\ v \end{bmatrix}.
		\end{equation}
		
		Substituting the specific block structure of the proposed 2D NSFRFT (where $\mathbf{D}=\mathbf{A}$ and $\mathbf{C}=-\mathbf{B}$) into the above equation, we have
		\begin{equation}
			W_{F_P}(x, y, u, v) = W_f(\widetilde{x}, \widetilde{y}, \widetilde{u}, \widetilde{v}),
			\label{eq:wigner_relation}
		\end{equation}
		\begin{equation}
			\begin{bmatrix} \widetilde{x} \\ \widetilde{y} \\ \widetilde{u} \\ \widetilde{v} \end{bmatrix} =\mathbf{X}^{-1}\begin{bmatrix} x \\ y \\ u \\ v \end{bmatrix}
			= \mathbf{X}^{T} 
			\begin{bmatrix} x \\ y \\ u \\ v \end{bmatrix}
			= \begin{bmatrix} \mathbf{A}^T & -\mathbf{B}^T \\ \mathbf{B}^T & \mathbf{A}^T \end{bmatrix} 
			\begin{bmatrix} x \\ y \\ u \\ v \end{bmatrix}.
		\end{equation}
		
		\noindent This completes the proof.
	\end{proof}

\section{Proof of 2D NSFRFT for Processing 2D Chirp Signals with Nonseparable Terms}
\label{Appendix A}

\begin{proof}
	Considering the 2D chirp signal with a nonseparable term given by
	\begin{equation}
		f(x, y) = \exp\left[{-\frac{\mathrm{j}(p_1 x^2 + p_2 xy + p_3 y^2)}{2\mathbf{T}}}\right],
		\label{eq:44}
	\end{equation}
	by substituting Eq. (\ref{eq:44}) into the 2D NSFRFT defined in Eq. (\ref{eq:13}), and applying the kernel expression from Eq. (\ref{eq:14}), after some simplification, we have
	\begin{equation}    
		\begin{aligned}
			F_{P}(u,v) 
			&= \frac{1}{2\pi\sqrt{-\mathbf{T}}} \exp\left[{\frac{\mathrm{j}(k_1 u^2 + k_2 uv + k_3 v^2)}{2\mathbf{T}}}\right]
			\int_{\mathbb{R}^2}  \exp\left[{\frac{\mathrm{j}(m_1 ux + m_2 vx + m_3 uy + m_4 vy)}{\mathbf{T}}}\right] \mathrm{d}x \, \mathrm{d}y.
		\end{aligned}
		\label{eq:45}
	\end{equation}
	Next, we consider the remaining part of the integral, by utilizing the properties of the 2D FT and the Dirac delta function yields
	\begin{equation}
		\begin{aligned}
			\int_{\mathbb{R}^2} \exp\left[{\frac{\mathrm{j}(m_1 ux + m_2 vx + m_3 uy + m_4 vy)}{\mathbf{T}}}\right] \mathrm{d}x \, \mathrm{d}y 
			= (2\pi)^2 |\mathbf{T}|^2\delta(m_1 u + m_2 v)\delta(m_3 u + m_4 v).
		\end{aligned}
		\label{eq:46}
	\end{equation}
	Combining the coefficient terms, we obtain
	\begin{equation}
		\frac{(2\pi)^2|\mathbf{T}|^2}{2\pi\sqrt{-\mathbf{T}}} = \frac{2\pi \mathbf{T}^2}{\sqrt{-\mathbf{T}}}.
		\label{eq:47}
	\end{equation}
	Substituting Eqs. (\ref{eq:46}) and (\ref{eq:47}) into Eq. (\ref{eq:45}), the final result of the 2D NSFRFT of \( f(x, y) \) is
	\begin{align}
		F_{P}(u,v)  = \frac{2\pi \mathbf{T}^2}{\sqrt{-\mathbf{T}}}\exp\left[{\frac{\mathrm{j}(k_1 u^2 + k_2 uv + k_3 v^2)}{2\mathbf{T}}}\right] \delta(m_1 u + m_2 v) \delta(m_3 u + m_4 v). \label{eq:48}
	\end{align}
	This completes the proof.
\end{proof}

\textit{Remark (Validity Scope and Singularities):}
The derived integral representation in Eq. (\ref{eq:48}) is valid under the general condition $\mathbf{T} \neq 0$, as explicitly constrained in Eq. (\ref{eq:15}).
In cases where $\mathbf{T}=0$, the transform physically degenerates. A specific instance of this is the identity transform, which corresponds to the parameter set $P=(1,0,0,0,0)$ and is separately defined in Eq. (\ref{eq:20}).
For other parameter sets yielding $\mathbf{T}=0$, the corresponding transform expressions are not derived in this work. To avoid mathematical ambiguity, these undefined degenerate cases are strictly excluded from the scope of the proposed 2D NSFRFT framework.

\bibliographystyle{elsarticle-num-names}
\bibliography{reference}
\biboptions{numbers,sort&compress}

\end{document}